\documentclass[fleqn,usenatbib]{mnras}

\usepackage{newtxtext,newtxmath}

\usepackage[T1]{fontenc}

\DeclareRobustCommand{\VAN}[3]{#2}
\let\VANthebibliography\thebibliography
\def\thebibliography{\DeclareRobustCommand{\VAN}[3]{##3}\VANthebibliography}

\usepackage{graphicx}	
\usepackage{amsmath}	
\usepackage{amssymb}	
\usepackage{natbib}
\usepackage{microtype}
\usepackage{xcolor}

\newcommand{\bham}{Birmingham Institute for Gravitational Wave Astronomy and School of Physics and Astronomy, University of Birmingham, Birmingham B15 2TT, UK}
\newcommand{\edinburgh}{Institute for Astronomy, University of Edinburgh, Royal Observatory, Blackford Hill, EH9 3HJ, UK}
\newcommand{\cambridge}{Institute of Astronomy, University of Cambridge, Madingley Road, Cambridge, CB3 0HA}
\newcommand{\northwestern}{Center for Interdisciplinary Exploration and Research in Astrophysics and Department of Physics and Astronomy, Northwestern University, 2145 Sheridan Road, Evanston, IL 60208-3112, USA}
\newcommand{\telaviv}{The School of Physics and Astronomy, Tel Aviv University, Tel Aviv 69978, Israel}
\newcommand{\cifar}{CIFAR Azrieli Global Scholars program, CIFAR, Toronto, Canada}
\newcommand{\dtu}{DTU Space, National Space Institute, Technical University
of Denmark, Elektrovej 327, 2800 Kgs. Lyngby, Denmark}
\newcommand{\mpi}{Max-Planck-Institut f{\"u}r Astrophysik, Karl-Schwarzschild Str. 1, D-85748 Garching, Germany}
\newcommand{\stockholm}{Department of Astronomy, Stockholm University, The Oskar Klein Centre, AlbaNova, SE-106 91 Stockholm, Sweden}
\newcommand{\cfa}{Center for Astrophysics \textbar{} Harvard \& Smithsonian, 60 Garden Street, Cambridge, MA 02138-1516, USA}
\newcommand{\brera}{INAF–Osservatorio Astronomico di Brera, via Bianchi 46, I-23807 Merate (LC), Italy}
\newcommand{\sron}{SRON, Netherlands Institute for Space Research, Sorbonnelaan 2, NL-3584 CA Utrecht, the Netherlands}
\newcommand{\radboud}{Department of Astrophysics/IMAPP, Radboud University, PO Box 9010, NL-6500 GL Nijmegen, the Netherlands}
\newcommand{\lco}{Las Cumbres Observatory, 6740 Cortona Dr, Suite 102, Goleta, CA 93117-5575, USA}
\newcommand{\ucsb}{Department of Physics, University of California, Santa Barbara, CA 93106-9530, USA}
\newcommand{\eso}{European Southern Observatory, Alonso de Co\'rdova 3107, Vitacura, Casilla 190001, Santiago, Chile}
\newcommand{\soton}{School of Physics and Astronomy, University of Southampton, Southampton, SO17 1BJ, UK}
\newcommand{\mssl}{Mullard Space Science Laboratory, University College London, Holmbury St Mary, Dorking, Surrey RH5 6NT, UK}
\newcommand{\anu}{The Research School of Astronomy and Astrophysics, Australian National University, ACT 2601, Australia}
\newcommand{\granada}{Departamento de F\'isica Te\'orica y del Cosmos, Universidad de Granada, E-18071 Granada, Spain}
\newcommand{\weizmann}{Department of Particle Physics and Astrophysics, Weizmann Institute of Science, 234 Herzl St, 76100 Rehovot, Israel}
\newcommand{\bello}{Departamento de Ciencias Fisicas, Universidad Andr\'es Bello, Avda. Republica 252, Santiago, Chile}
\newcommand{\cardiff}{School of Physics \& Astronomy, Cardiff University, Queens Buildings, The Parade, Cardiff, CF24 3AA, UK}
\newcommand{\iaps}{Istituto di Astrofisica e Planetologia Spaziali (INAF), Via Fosso del Cavaliere 100, Roma, I-00133, Italy}
\newcommand{\warsaw}{Astronomical Observatory, University of Warsaw, Al. Ujazdowskie 4, 00-478 Warszawa, Poland}
\newcommand{\lisbon}{CENTRA-Centro de Astrof\'{\i}sica e Gravita\c{c}\~ao and Departamento de F\'{\i}sica, Instituto Superior T\'ecnico, Universidade de Lisboa, Avenida Rovisco Pais, 1049-001 Lisboa, Portugal}
\newcommand{\iucaa}{The Inter-University Centre for Astronomy and Astrophysics, Ganeshkhind, Pune - 411007, India}
\newcommand{\iap}{Institute of Astrophysics Paris (IAP), and Sorbonne University, 98bis Boulevard Arago, F-75014 Paris}
\newcommand{\monash}{Monash Centre for Astrophysics, School of Physics and Astronomy, Monash University, Clayton, Victoria 3800, Australia}
\newcommand{\ozgrav}{The ARC Center of Excellence for Gravitational Wave Discovery--OzGrav, Australia}
\newcommand{\qub}{Astrophysics Research Centre, School of Mathematics and Physics, Queens University Belfast, Belfast BT7 1NN, UK}

\def\M{M$_\odot$}
\def\ergs{erg\,s$^{-1}$}
\def\kms{km\,s$^{-1}$}

\title[Answers to the qiz]{An outflow powers the optical rise of the nearby, fast-evolving tidal disruption event AT2019qiz}

\author[M. Nicholl et al]{{M.~Nicholl$^{1,2}$}\thanks{Contact e-mail: \href{mailto:mnicholl@star.sr.bham.ac.uk}{mnicholl@star.sr.bham.ac.uk}}, 
{T.~Wevers$^{3}$},
{S.~R.~Oates$^{1}$},
{K.~D.~Alexander$^{4}$\thanks{Einstein Fellow}},
{G.~Leloudas$^{5}$},
{F.~Onori$^{6}$},
\newauthor 
{A.~Jerkstrand$^{7,8}$},
{S.~Gomez$^{9}$},
{S.~Campana$^{10}$},
{I.~Arcavi$^{11,12}$},
{P.~Charalampopoulos$^{5}$},
\newauthor 
{M.~Gromadzki$^{13}$},
{N.~Ihanec$^{13}$},
{P.~G.~Jonker$^{14,15}$},
{A.~Lawrence$^{2}$},
{I.~Mandel$^{16,17,1}$},
\newauthor 
{S.~Schulze$^{18}$},
{P.~Short$^{2}$},
{J.~Burke$^{19,20}$},
{C.~McCully$^{19,20}$},
{D.~Hiramatsu$^{19,20}$},
\newauthor 
{D.~A.~Howell$^{19,20}$},
{C.~Pellegrino$^{19,20}$},
{H.~Abbot$^{21}$},
{J.~P.~Anderson$^{22}$},
{E.~Berger$^{9}$},
\newauthor 
{P.~K.~Blanchard$^{4}$},
{G.~Cannizzaro$^{14,15}$},
{T.-W.~Chen$^{8}$},
{M.~Dennefeld$^{23}$},
{L.~Galbany$^{24}$},
\newauthor 
{S.~Gonz\'alez-Gait\'an$^{25}$},
{G.~Hosseinzadeh$^{9}$},
{C.~Inserra$^{26}$},
{I.~Irani$^{18}$},
{P.~Kuin$^{27}$},
\newauthor 
{T.~M\"uller-Bravo$^{28}$},
{J.~Pineda$^{29}$},
{N.~P.~Ross$^{2}$},
{R.~Roy$^{30}$},
{S.~J.~Smartt$^{31}$},
{K.~W.~Smith$^{31}$},
\newauthor 
{B.~Tucker$^{21}$},
{{\L}.~Wyrzykowski$^{13}$},
{D.~R.~Young$^{31}$}
\\
Affiliations at end of paper
}

\date{}

\pubyear{2020}

\begin{document}
\label{firstpage}
\pagerange{\pageref{firstpage}--\pageref{lastpage}}
\maketitle

\begin{abstract}
At 66\,Mpc, AT2019qiz is the closest optical tidal disruption event (TDE) to date, with a luminosity intermediate between the bulk of the population and the faint-and-fast event iPTF16fnl. Its proximity allowed a very early detection and triggering of multiwavelength and spectroscopic follow-up well before maximum light. The velocity dispersion of the host galaxy and fits to the TDE light curve indicate a black hole mass $\approx 10^6$\,\M, disrupting a star of $\approx1$\,\M. By analysing our comprehensive UV, optical and X-ray data, we show that the early optical emission is dominated by an outflow, with a luminosity evolution $L\propto t^2$, consistent with a photosphere expanding at constant velocity ($\gtrsim 2000$\,\kms), and a line-forming region producing initially blueshifted H and He~II profiles with $v=3000-10000$\,\kms. The fastest optical ejecta approach the velocity inferred from radio detections {(modelled in a forthcoming companion paper from K.~D.~Alexander et al.)}, thus the same outflow {may be} responsible for both the fast optical rise and the radio emission -- the first time this connection has been {observed} in a TDE. The light curve rise begins $29\pm2$ days before maximum light, peaking when the photosphere reaches the radius where optical photons can escape. The photosphere then undergoes a sudden transition, first cooling at constant radius then contracting at constant temperature. At the same time, the blueshifts disappear from the spectrum and Bowen fluorescence lines (N~III) become prominent, implying a source of far-UV photons, while the X-ray light curve peaks at $\approx10^{41}$\,\ergs. {Assuming that these X-rays are from prompt accretion,} the size and mass of the outflow are consistent with the reprocessing layer needed to explain the large optical to X-ray ratio in this and other optical TDEs, {possibly favouring accretion-powered over collision-powered outflow models}.
\end{abstract}




\section{Introduction}\label{sec:intro}


An unfortunate star in the nucleus of a galaxy can find itself on an orbit that intersects the tidal radius of the central supermassive black hole (SMBH), where $R_t\approx R_*(M_\bullet/M_*)^{1/3}$ for a black hole of mass $M_\bullet$ and a star of mass $M_*$ and radius $R_*$ \citep{Hills1975}. This encounter induces a spread in the specific orbital binding energy across the star that is orders of magnitude greater than the mean binding energy \citep{Rees1988}, sufficient to tear the star apart in a `tidal disruption event' (TDE). The stellar debris, confined in the vertical direction by self-gravity \citep{Kochanek1994,Guillochon2014}, is stretched into a long, thin stream, roughly half of which remains bound to the SMBH \citep{Rees1988}. As the bound debris orbits the SMBH, relativistic apsidal precession causes the stream to self-intersect and dissipate energy {\citep{Shiokawa2015,Dai2015,Bonnerot2020}.}

This destruction can power a very luminous flare, up to or exceeding the Eddington luminosity, either when the intersecting streams circularise and form an accretion disk \citep{Rees1988,Phinney1989}, or even earlier if comparable radiation is produced directly from the stream collisions \citep{Piran2015,Jiang2016}. Such flares are now regularly discovered, at a rate exceeding a few per year, by the various wide-field time-domain surveys \citep[e.g.][]{Gezari2012,Holoien2014,Arcavi2014,vanVelzen2020}. 

Observed TDEs are bright in the UV, with characteristic temperatures $\sim2-5\times10^4$\,K and luminosities $\sim10^{44}$\,\ergs. They are classified according to their spectra, generally exhibiting broad, low equivalent width\footnote{Compared to other nuclear transients such as active galactic nuclei} emission lines of hydrogen, neutral and ionised helium, and Bowen fluorescence lines of doubly-ionised nitrogen and oxygen \citep[e.g.][]{Arcavi2014,Leloudas2019}. This prompted \citet{vanVelzen2020} to suggest three sub-classes labelled TDE-H, TDE-He and TDE-Bowen, though some TDEs defy a consistent classification by changing their apparent spectral type as they evolve \citep{Nicholl2019}. 

TDE flares were initially predicted to be brightest in X-rays, due to the high temperature of an accretion disk, and indeed this is the wavelength where the earliest TDE candidates were identified \citep{Komossa2002}. However, the optically-discovered TDEs have proven to be surprisingly diverse in their X-ray properties. Their X-ray to optical ratios at maximum light range from $\gtrsim10^{3}$ to $<10^{-3}$ \citep{Auchettl2017}. Producing such luminous optical emission without significant X-ray flux can be explained in one of two ways: either X-ray faint TDEs are powered primarily by stream collisions rather than accretion, or the accretion disk emission is reprocessed through an atmosphere \citep{Strubbe2009,Guillochon2014,Roth2016}.

Several lines of evidence have indicated that accretion disks do form promptly even in X-ray faint TDEs: Bowen fluorescence lines that require excitation from far-UV photons \citep{Blagorodnova2018,Leloudas2019}; low-ionisation iron emission appearing shortly after maximum light \citep{Wevers2019b}; and recently the direct detection of double-peaked Balmer lines that match predicted disk profiles \citep{Short2020,Hung2020}. Thus a critical question is to understand the nature and origin of the implied reprocessing layer. {\citet{Guillochon2014} showed that the unbound debris stream cannot be the site of reprocessing, because its apparent cross-section is too low to intercept a significant fraction of the TDE flux.}

Inhibiting progress is the messy geometry of the debris. Colliding streams, inflowing and outflowing gas, and a viewing-angle dependence on both the broad-band \citep{Dai2018} and spectroscopic \citep{Nicholl2019} properties all contribute to a messy knot that must be untangled. One important clue comes from radio observations: although only a small (but growing) sample of TDEs have been detected in the radio, in such cases we can measure the properties (energy, velocity, and density) of an outflow directly \citep[see recent review by][]{Alexander2020}. In some TDEs this emission is from a relativistic jet \citep{Zauderer2011,Bloom2011,Burrows2011,Cenko2012,Mattila2018}, which does not appear to be a common feature of TDEs, but other radio TDEs have launched sub-relativistic outflows \citep{vanvelzen2016,Alexander2016,Alexander2017,Anderson2019}. 

A number of radio-quiet TDEs have exhibited indirect evidence for slower outflows in the form of blueshifted optical/UV emission and absorption lines \citep{Roth2018,Hung2019,Blanchard2017}, suggesting that outflows may be common. This is important, as the expanding material offers a promising means to form the apparently ubiquitous reprocessing layer required by the optical/X-ray ratios. Suggested models include an Eddington envelope \citep{Loeb1997}, possibly inflated by radiatively inefficient accretion or an optically thick disk wind \citep{Metzger2016,Dai2018}; or a collision-induced outflow \citep{Lu2020}.

Understanding whether the optical reprocessing layer is connected to the non-relativistic outflows seen in some radio TDEs is therefore a crucial, and as yet under-explored, step towards a pan-chromatic picture of TDEs. {Multi-colour photometry at early times can reveal whether (or how quickly) the optical photosphere grows with time. Early radio detections can determine the time at which material is launched, to check for consistency with the optical flare \citep{Alexander2016}. Spectroscopy and X-ray observations can be used to search for signatures of accretion. In principle this could} allow us to distinguish whether outflows are launched \emph{by} accretion, or \emph{before} accretion (i.e., in the collisions that ultimately enable disk formation).

In this paper, we present a detailed study of the UV, optical and X-ray emission from AT2019qiz: the closest TDE discovered to date, and the first optical TDE at $z<0.02$ that has been detected in the radio. We place a particular focus on the spectroscopic evolution, finding clear evidence of an outflow launched well before the light curve maximum. By studying the evolution of the photosphere and line velocities, we infer a roughly homologous structure, with the fastest optically-emitting material (the reprocessing layer) likely also responsible for the radio emission (which is analysed in detail in a forthcoming companion paper; K.~D.~Alexander et al., in preparation). This event suggests a closer connection between the optical and radio TDE outflows than has been appreciated to date, while a peak in the X-ray light curve and the detection of Bowen lines indicates that accretion began promptly in this event and likely drives the outflow. The rapid rise and decline of the light curve suggests that the properties of outflows may be key to understanding the fastest TDEs.

We detail the discovery and classification of AT2019qiz in section \ref{sec:class}, and describe our observations and data reduction in section \ref{sec:obs}. We analyse the host galaxy, including evidence for a pre-existing AGN, in section \ref{sec:host}, and study the photometric and spectroscopic evolution of the TDE emission in sections \ref{sec:phot} and \ref{sec:spec}. This is then brought together into a coherent picture, discussed in section \ref{sec:dis}, before we conclude in section \ref{sec:conc}. All data in this paper will be made publicly available via WISeREP \citep{Yaron2012}.

\begin{figure}
  \centering
  \includegraphics[width=6cm]{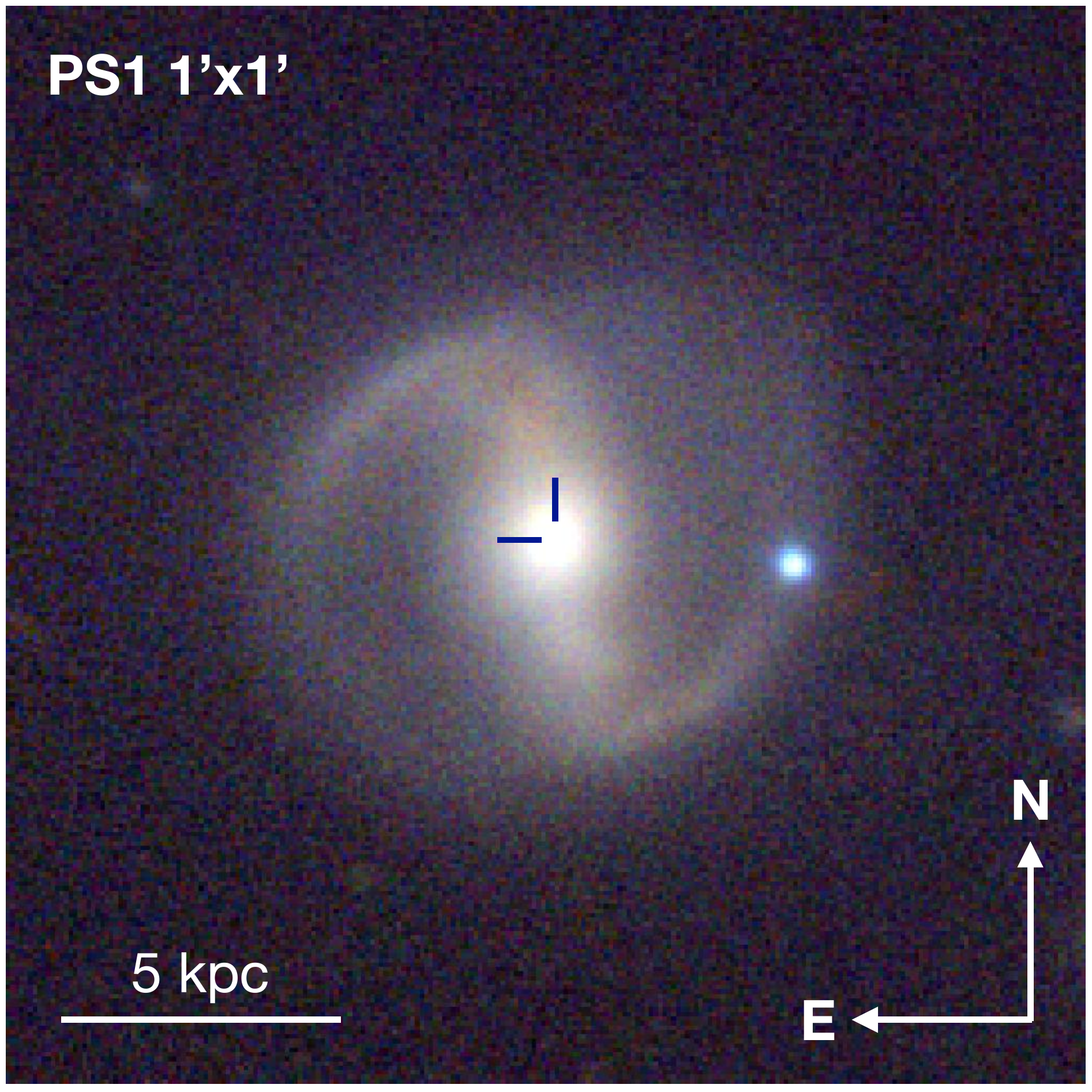}
  \caption{Pre-disruption $g,i,z$ colour image of the host of AT2019qiz, 2MASX\,J04463790-1013349, obtained from the PanSTARRS image server. Comparing to our highest S/N LCO image of the transient, we measure an offset of $15\pm46$\,pc from the centre of the host.}
  \label{fig:ps1}
\end{figure}

\section{Discovery and background}\label{sec:class}

AT2019qiz was discovered in real-time alerts from the Zwicky Transient Facility \citep[ZTF;][]{Bellm2019,Masci2019,Patterson2019}, at coordinates RA$=$04:46:37.88, Dec$=-$10:13:34.90, and was given the survey designation ZTF19abzrhgq. It is coincident with the centre of the galaxy 2MASX\,J04463790-1013349 (a.k.a.~WISEA\,J044637.88-101334.9). The transient was first identified by the ALeRCE broker, who reported it to the Transient Name Server on 2019-09-19 UT \citep{Forster2019}.
It was also reported by the AMPEL broker a few days later. In subsequent nights the same transient was independently reported by the Asteroid Terrestrial impact Last Alert System \citep[ATLAS;][]{Tonry2018} as ATLAS19vfr, by \textit{Gaia} Science Alerts \citep{Hodgkin2013} as Gaia19eks, and by the Panoramic Survey Telescope And Rapid Response System (PanSTARRS) Survey for Transients \citep[PSST;][]{Huber2015} as PS19gdd. The earliest detection is from ATLAS on {2019-09-18} UT.

We have been running the Classification Survey for Nuclear Transients with Liverpool and Lasair (C-SNAILS; \citealt{Nicholl2019csnails}) to search for TDEs in the public ZTF alert stream. AT2019qiz passed our selection criteria (based on brightness and proximity to the centre of the host galaxy\footnote{The source could have been triggered 5 days earlier but for missing offset information in the alert packet.}) on 2019-09-25 UT, and we triggered spectroscopy with the Liverpool Telescope. On the same day, \citet{Siebert2019} publicly classified AT2019qiz as a TDE using spectroscopy from Keck I.

Their reported spectrum, and our own data obtained over the following nights, showed broad He~II and Balmer emission lines superposed on a very blue continuum, characteristic of UV-optical TDEs. Follow-up observations from other groups showed that the source was rising in the UV \citep{Zhang2019}, but was not initially detected in X-rays \citep{Auchettl2019}. Radio observations with the Australia Telescope Compact Array (ATCA) revealed rising radio emission, reaching 2.6\,mJy at 21.2\,GHz on 2019-12-02 \citep{OBrien2019}, placing AT2019qiz among the handful of TDEs detected at radio wavelengths.

The spectroscopic redshift of AT2019qiz, as listed for the host galaxy in the NASA Extragalactic Database (NED) and measured from narrow absorption lines in the TDE spectrum, is $z=0.01513$. This corresponds to a distance of 65.6\,Mpc, assuming a flat cosmology with $H_0=70$\,km\,s$^{-1}$\,Mpc$^{-1}$ and $\Omega_\Lambda=0.7$. This makes AT2019qiz the most nearby TDE discovered to date.

AT2019qiz was included in the sample of 17 TDEs from ZTF studied by \citet{vanVelzen2020}. Their work focused primarily on the photometric evolution around peak, and correlations between TDE and host galaxy properties. In this paper, we analyse a rich dataset for AT2019qiz including densely sampled spectroscopy, very early and late photometric observations, and the first detection of the source in X-rays. We also examine the host galaxy in detail. While we link the optical properties of the TDE to its behaviour in the radio, the full radio dataset and analysis will be presented in a companion paper (K.~D.~Alexander et al., in preparation).

\begin{figure}
  \centering
  \includegraphics[width=\columnwidth]{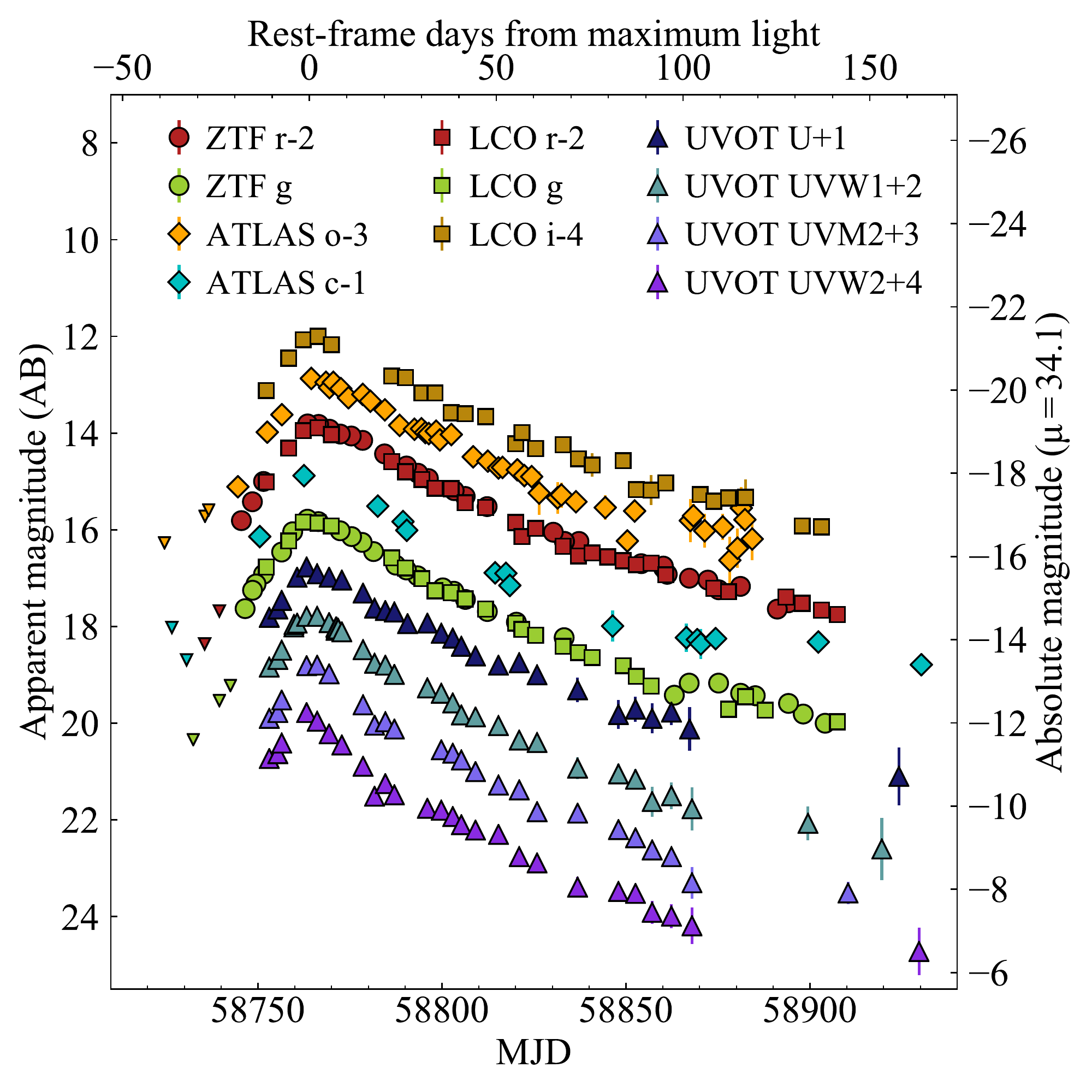}
  \caption{Optical and UV light curves of AT2019qiz. The host contribution has been removed using image differencing (ZTF, ATLAS, LCO) or subtraction of fluxes estimated from the galaxy SED (UVOT).}
  \label{fig:phot}
\end{figure}

\section{Observations}\label{sec:obs}

\subsection{Ground-based imaging}\label{sec:ground}

Well-sampled host-subtracted light curves of AT2019qiz were obtained by the ZTF public survey, in the $g$ and $r$ bands, and ATLAS in the $c$ and $o$ bands (effective wavelengths 5330 and 6790\,\AA). The ZTF light curves were accessed using the Lasair alert broker\footnote{\url{https://lasair.roe.ac.uk}} \citep{Smith2019}.

We triggered additional imaging with a typical cadence of four days in the $g,r,i$ bands using the Las Cumbres Observatory (LCO) global network of 1-m telescopes \citep{Brown2013}. We retrieved the reduced (de-biased and flat-fielded) images from LCO and carried out point spread function (PSF) fitting photometry using a custom wrapper for \textsc{daophot}. The zeropoint in each image was calculated by comparing the instrumental magnitudes of field stars with their catalogued magnitudes from PanSTARRS data release 1 \citep{Flewelling2016}.

\begin{figure}
  \centering
  \includegraphics[width=6.5cm]{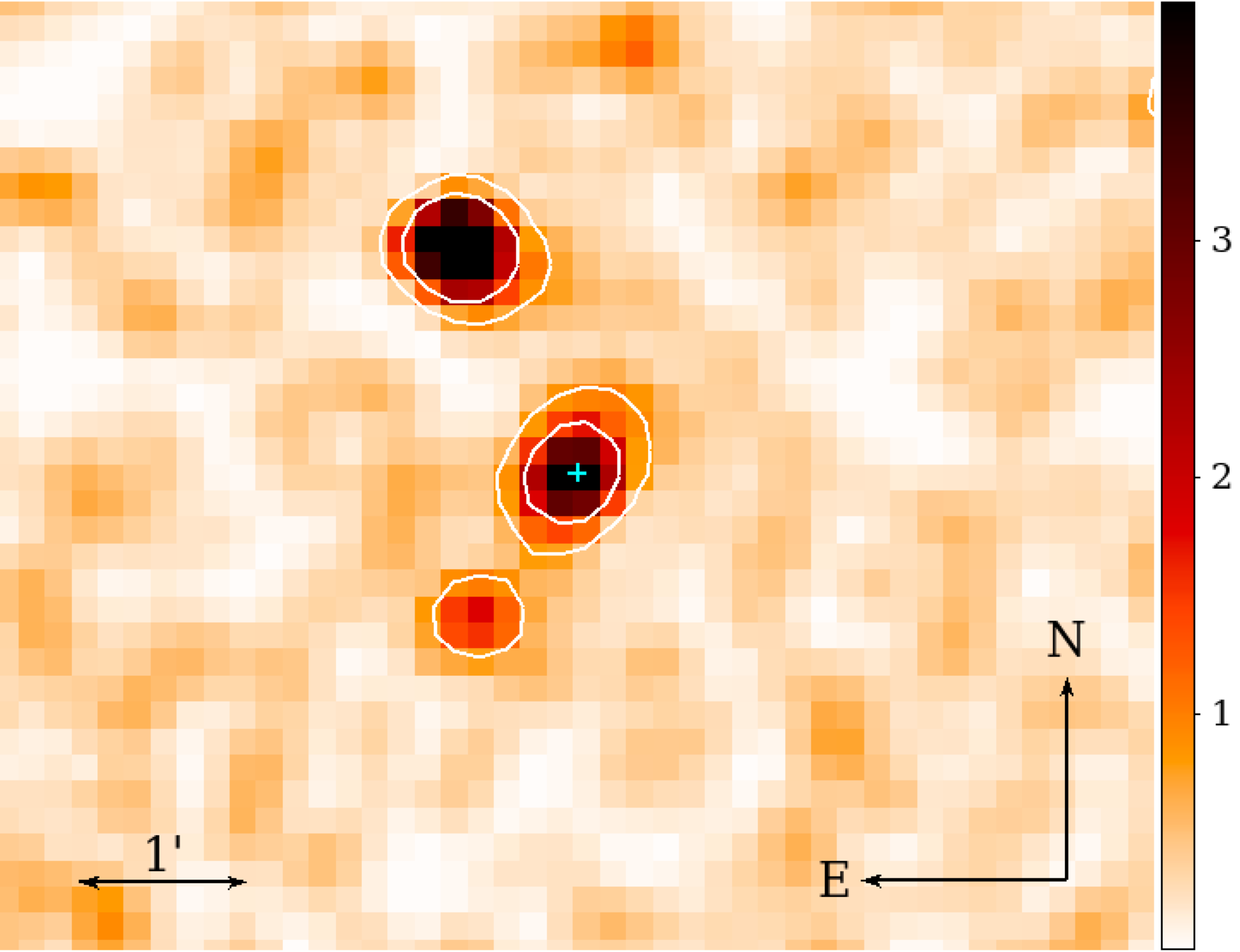}
  \caption{Stacked \textit{Swift} XRT image (total exposure time 30\,ks) centred at the position of AT2019qiz, marked by a cyan cross. An X-ray source is detected at $5.6\sigma$ significance. The image has been blocked $4\times4$ and smoothed using a Gaussian kernel for display. The colour bar gives the counts per pixel.}
  \label{fig:x}
\end{figure}

AT2019qiz resides in a bright galaxy, 2MASX\,J04463790-1013349, with $m_r=14.2$\,mag (Section \ref{sec:host}). A three-colour ($gri$) image of the host, comprised of deep stacks from PanSTARRS, is shown in Figure \ref{fig:ps1}. To isolate the transient flux, we aligned each LCO image with the PanSTARRS image in the corresponding filter using the \textsc{geomap} and \textsc{geotran} tasks in \textsc{pyraf}, computing the transformation from typically $> 50$ stars, before convolving the PanSTARRS reference image to match the PSF of the science image and subtracting with \textsc{hotpants} \citep{Becker2015}.

\subsection{Astrometry}

Measuring the TDE position in an LCO $r$ band image obtained on 2019-10-10 (at the peak of the optical light curve), and the centroid of the galaxy in the aligned PanSTARRS $r$ band image, we measure an offset of $0.12\pm0.37$ pixels (or $0.047''\pm0.144$) between the transient and the host nucleus, where the uncertainty is dominated by the root-mean-square error in aligning the images. This corresponds to a physical offset of $15\pm46$\,pc at this distance; the transient is therefore fully consistent with a nuclear origin.

An alternative astrometric constraint can be obtained using the {\it Gaia} Science Alerts (GSA) detections \citep{Hodgkin2013}.
Gaia19eks was discovered at a separation of 38 milli-arcseconds (mas) from the reported location of its host galaxy in {\it Gaia} data release 2 (GDR2; \citealt{Gaia2018}).
The estimated astrometric uncertainty of GSA is $\sim$100 mas \citep{Fabricius2016}, and the coordinate systems of GSA and GDR2 are well aligned \citep{Kostrzewa2018, Wevers2019b}.
This measurement corresponds to an even tighter constraint on the offset of $12\pm32$\,pc.

\subsection{\textit{Swift} UVOT data}\label{sec:uvot}

Target-of-opportunity observations spanning 39 epochs (PIs Yu and Nicholl) were obtained with the UV-Optical Telescope (UVOT) and X-ray Telescope (XRT) on-board the Neil Gehrels Swift Observatory (\textit{Swift}). The UVOT light curves were measured using a $5''$ aperture. This is approximately twice the UVOT point-spread function, ensuring the measured magnitudes capture most of the transient flux while minimising the host contribution (the coincidence loss correction for the UVOT data is also determined using a $5''$ aperture, ensuring a reliable calibration of these magnitudes). The count rates were obtained using the \textit{Swift} \textsc{uvotsource} tools and converted to magnitudes using the UVOT photometric zero points \citep{Breeveld2011}. The analysis pipeline used software HEADAS 6.24 and UVOT calibration 20170922. We exclude the initial images in the $B$, $U$, $UVW1$, and $UVW2$ filters (OBSID 00012012001) due to trailing within the images. We also exclude 2 later $UVW1$ images and a $UVW2$ image due to the source being located on patches of the detector known to suffer reduced sensitivity. A correction has yet to be determined for these patches\footnote{https://heasarc.gsfc.nasa.gov/docs/heasarc/caldb/swift/docs/\\uvot/uvotcaldb\_sss\_01.pdf}.

No host galaxy images in the UV are available for subtraction. We estimated the host contribution using a spectral energy distribution (SED) fit to archival data for this galaxy (details in section \ref{sec:host}). {The host magnitude and its error in each UVOT band was estimated using the mean and standard deviation of SED samples drawn from the posterior of this fit.} We then scaled the predicted flux by a factor 0.2, i.e.~the fraction of the host light within a $5''$ aperture in the PanSTARRS $g$-band image of the galaxy, before subtracting from the transient photometry. We checked that this method provides a reliable correction for the host flux by re-extracting the UVOT light curve using a $30''$ aperture to fully capture both the transient and host flux, and subtracting the model host magnitudes with no scaling.

Comparing the $5''$ light curves to the $30''$ light curves, we find a good match in the $U$ and $UVW1$ bands. In the bluer $UVM2$ and $UVW2$ bands, where the host SED and light profile is less constrained, we find that scaling the host flux by a factor 0.1 before subtraction yields better agreement, and we adopt this as our final light curve. {The fractional uncertainty ($\approx 20\%$) in the host flux was combined in quadrature with that of the transient flux when calculating the photometric errors.} {We show in the appendix a comparison between the UVOT light curves obtained using the $5''$ and $30''$ apertures.} The complete, host-subtracted UV and optical light curves from \textit{Swift}, LCO, ZTF and ATLAS are shown in Figure \ref{fig:phot}.

\begin{figure}
  \centering
  \includegraphics[width=\columnwidth]{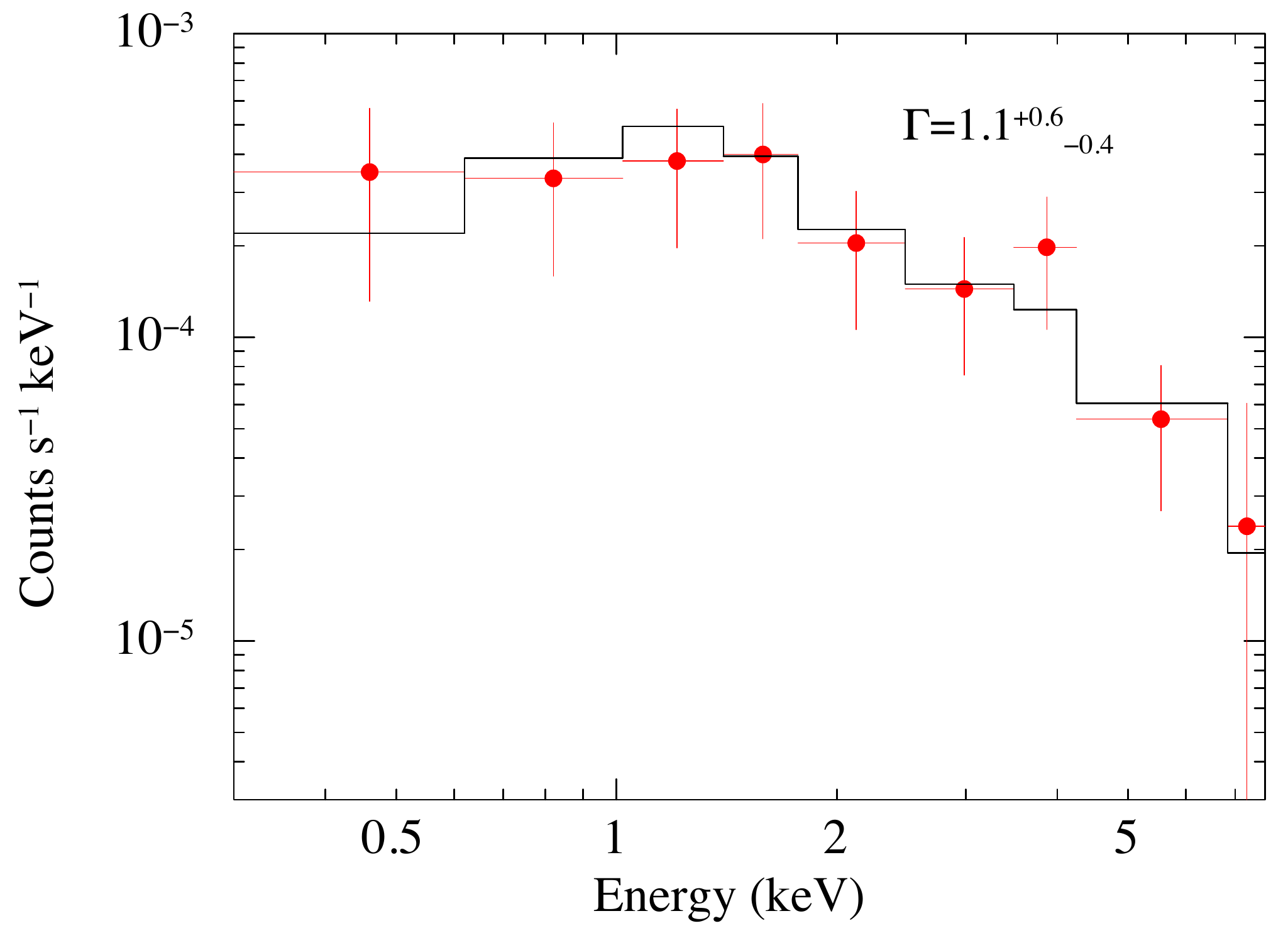}
  \caption{Time-averaged X-ray spectrum and best-fit absorbed power-law model, used to derive the counts-to-flux conversion.}
  \label{fig:xspec}
\end{figure}

\subsection{\textit{Swift} XRT data}

We processed the XRT data using the online analysis tools provided by the UK \textit{Swift} Science Data Centre \citep{Evans2007,Evans2009}. We first combined all of the data into a single deep stack (total exposure time 30\,ks), which we then downloaded for local analysis. The stacked image, shown in Figure \ref{fig:x}, clearly exhibits an X-ray source at the position of AT2019qiz.
Using a $50''$ aperture ($\sim2.5$ times the instrumental half-energy width) centered at the coordinates of the transient, we measure an excess $46.9 \pm 8.4$ counts above the background, giving a mean count rate of $(1.6\pm0.3)\times10^{-3}$ ct\,s$^{-1}$.

We then used the same tools to extract the mean X-ray spectrum, shown in Figure \ref{fig:xspec}, and light curve. Given the low number of counts we fit the spectrum using Cash statistics, and fixed the Galactic column density to $6.5\times 10^{20}$ cm$^{-2}$. The fit with a power law does not need an intrinsic column ($<2.8\times 10^{21}$ cm$^{-2}$, $90\%$ confidence level). The photon index of the fit is $\Gamma=1.1^{+0.6}_{-0.4}$. A blackbody model can give a comparable fit, but the inferred temperature and radius ($kT=0.9$\,keV, $R=6.0\times10^7$\,cm) are not consistent with other TDEs. {Such a radius is also much smaller than the Schwarzschild radius of a SMBH, however an apparently small emitting surface can also arise due to obscuration \citep{Wevers2019}.} The 0.3-10 keV unabsorbed flux from the power-law fit is $9.9^{+3.7}_{-3.4}\times 10^{-14}$ erg cm$^{-2}$ s$^{-1}$. At the distance of AT2019qiz, this corresponds to an X-ray luminosity $L_X=5.1\times10^{40}$\,erg\,s$^{-1}$.

\begin{figure}
  \centering
  \includegraphics[width=\columnwidth]{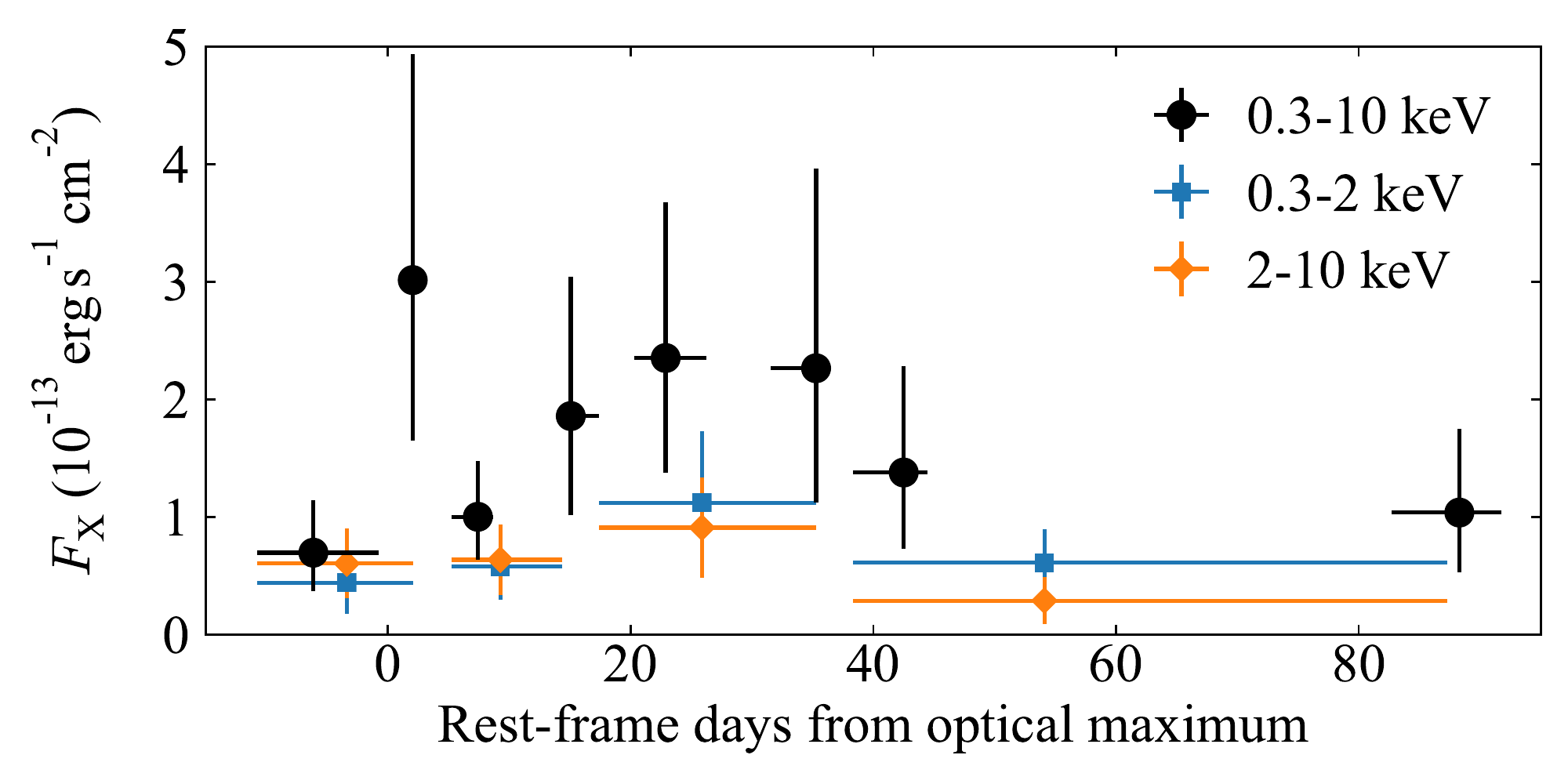}
  \includegraphics[width=\columnwidth]{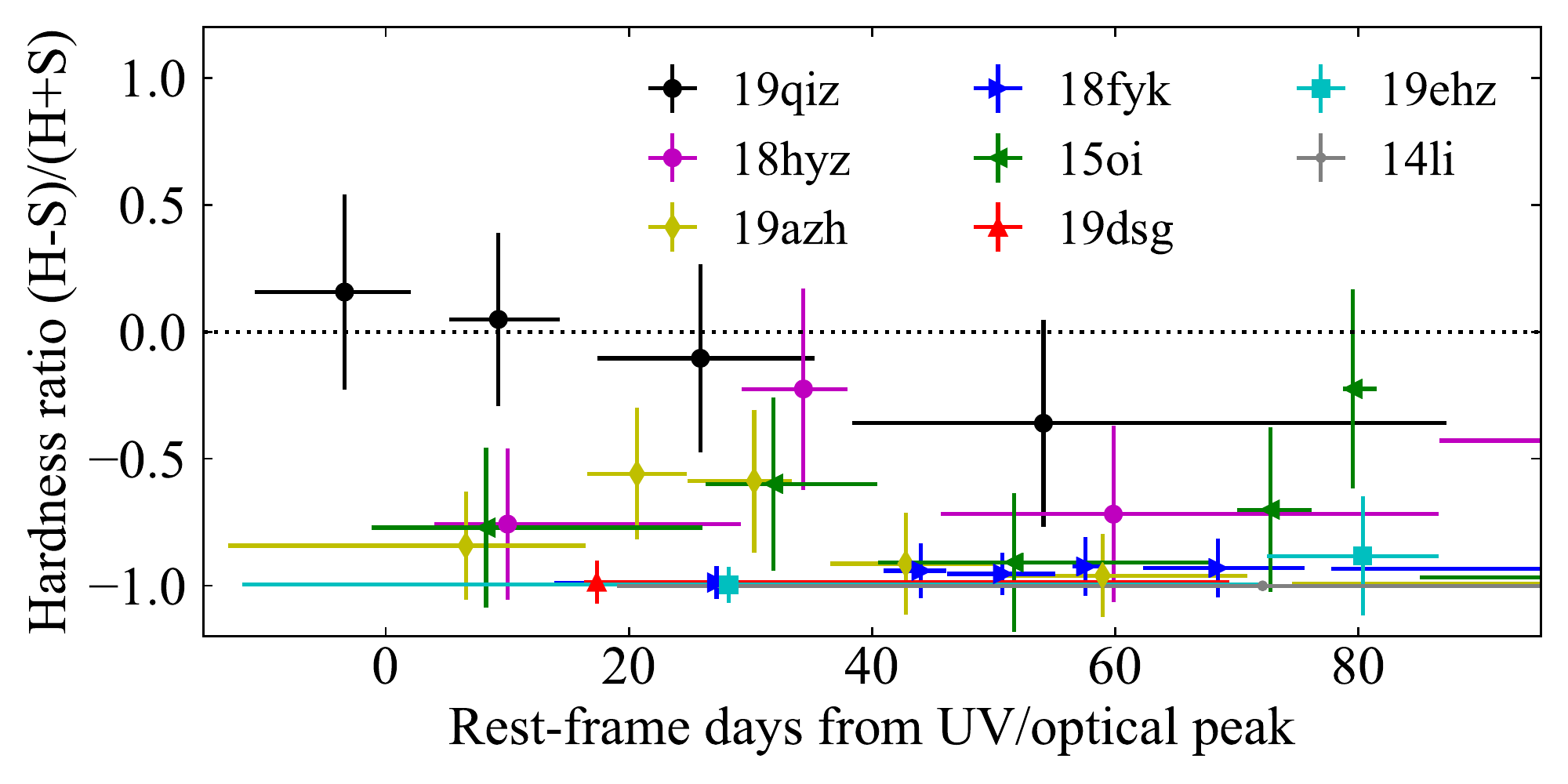}
  \caption{Top: XRT light curve (unabsorbed flux) in 0.3-10\,keV, 0.3-2\,keV (soft) and 2-10\,keV (hard) X-ray bands. The X-ray light curve peaks around 25 days after optical maximum.
  Bottom: evolution of the hardness ratio, defined as (hard\,$-$\,soft counts)/(hard\,$+$\,soft counts). The X-rays transition from hard to soft as the luminosity declines. This is in contrast to most X-ray TDEs that exhibit at all times a soft spectrum \citep{vanVelzen2020,Gomez2020,Holoien2016,Holoien2016b}.
  }
  \label{fig:hr}
\end{figure}

The X-ray light curve is shown in Figure \ref{fig:hr}. {We specify a target bin size of 10 counts above background, with a minimum of three counts to form a bin}. The 0.3-10\,keV light curve peaks around 25 days after optical maximum. We calculate the evolution of the hardness ratio as $(H-S)/(H+S)$, where $S$ is the count rate in the $0.3-2$\,keV band and $H$ the count rate in the $2-10$\,keV band; these ranges have been chosen to match \citet{Auchettl2017}. {The pipeline necessarily returns coarser temporal bins for the hardness ratio, due to the lower counts when dividing into the two bands}. {We compare this ratio to other TDEs with well-sampled XRT detections.} AT2019qiz exhibits an unusually hard ratio at early times, with $(H-S)/(H+S)=0.2\pm0.3$, but as the X-rays fade they also soften, reaching $(H-S)/(H+S)=-0.4\pm0.3$ by $\approx 50$ days after peak. This latter value is typical of the TDE sample studied by \citet{Auchettl2017}, whereas a positive ratio has only been seen previously in the jetted TDE J1644+57 \citep{Zauderer2011}.

\begin{figure*}
  \centering
  \includegraphics[width=\textwidth]{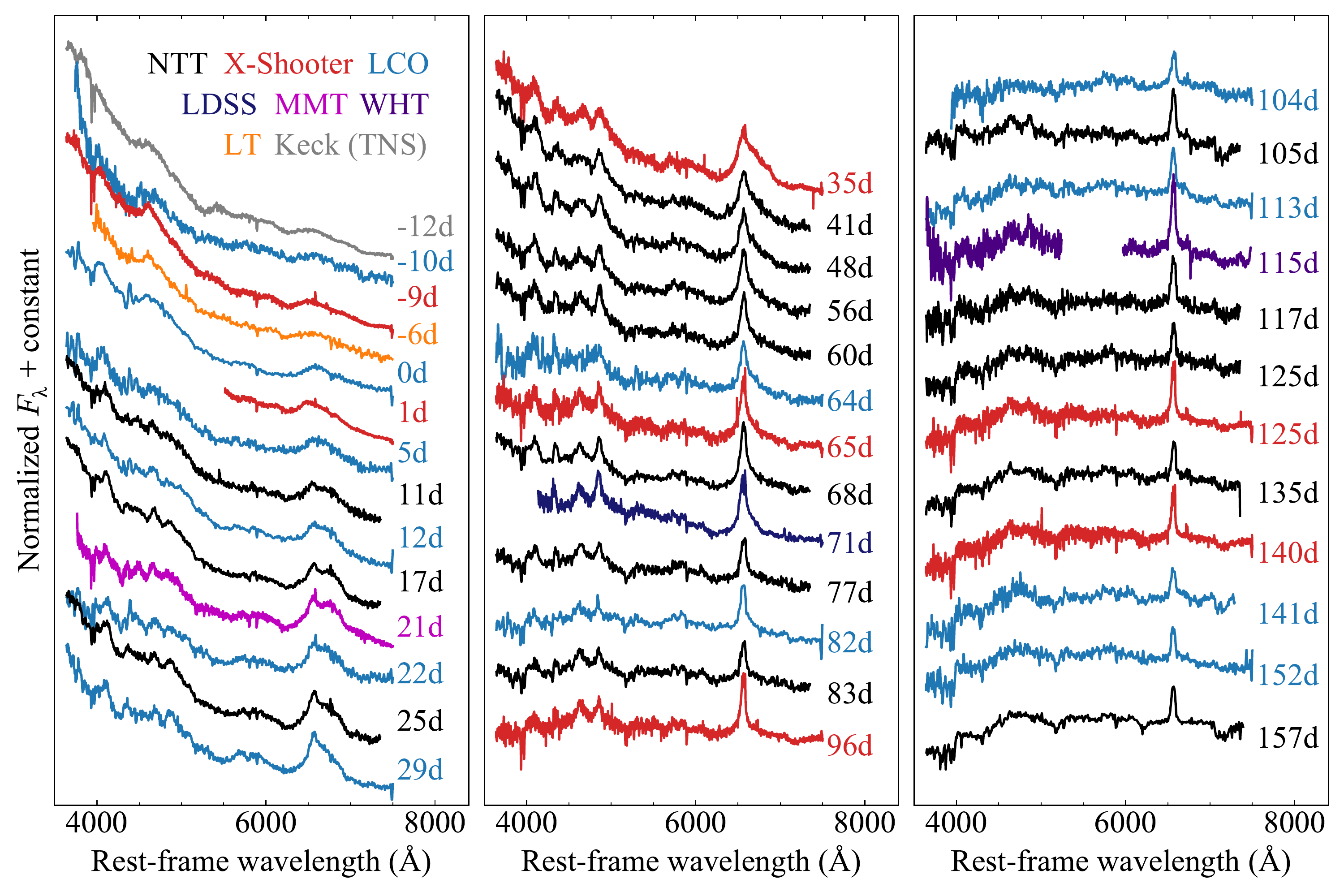}
  \caption{Complete series of spectra of AT2019qiz at phases from 12 days before until 157 days after the UV/optical peak. The phase of each spectrum is labelled, while colours individuate the telescopes/instruments used for the observations. The continuum transitions from TDE- to host-dominated over time, but broad TDE emission lines are visible throughout. A version of this figure with host light subtracted is given in the appendix.}
  \label{fig:spec}
\end{figure*}

\subsection{Optical spectroscopic data}

Spectra of AT2019qiz were obtained from the 3.6-m New Technology Telescope (NTT), using EFOSC2 with Grism\#11, through the advanced Public ESO Spectroscopy Survey of Transient Objects (ePESSTO+; \citealt{Smartt2015}); the LCO 2-m North and South telescopes with FLOYDS; the 2-m Liverpool Telescope (LT) with SPRAT \citep{Piascik2014} in the blue-optimised setting, as part of C-SNAILS \citep{Nicholl2019csnails}; the 6.5-m MMT telescope with Binospec \citep{Fabricant2019}; the 6.5-m Magellan Clay telescope with LDSS-3 and the VPH-ALL grism; the 4.2-m William Herschel Telescope with ISIS \citep{Jorden1990} and the R600 blue/red gratings; and the 8-m ESO Very Large Telescope using X-Shooter \citep{Vernet2011} in on-slit nodding mode, through our TDE target-of-opportunity program.

Reduction and extraction of these data were performed using instrument-specific pipelines or (in the cases of the LDSS-3 and ISIS data) standard routines in \textsc{iraf}. Reduced LCO and LT data were downloaded from the respective data archives, while we ran the pipelines \citep{Smartt2015,Freudling2013} locally for the EFOSC2 and X-Shooter data\footnote{The atmospheric dispersion corrector on X-Shooter failed on 2019-10-10, so we were unable to reduce the data in the UVB arm for this epoch.}. Typical reduction steps are de-biasing, flat-fielding and wavelength-calibration using standard lamps, cosmic-ray removal \citep{vanDokkum2012}, flux calibration using spectra of standard stars obtained with the same instrument setups, and variance-weighted extraction to a one-dimensional spectrum. We also retrieved the reduced classification spectrum obtained by \citet{Siebert2019} using the 10-m Keck-I telescope with LRIS \citep{Oke1995}, and made public via the Transient Name Server\footnote{\url{https://wis-tns.weizmann.ac.il}}.
All spectra are corrected for redshift and a foreground extinction of $E(B-V) = 0.0939$ using the dust maps of \citet{Schlafly2011} and the extinction curve from \citet{Cardelli1989}. All spectra are plotted in Figure \ref{fig:spec}. For host-subtracted spectra (section \ref{sec:host} and appendix), we apply these corrections after scaling and subtraction.

\section{Host galaxy properties}\label{sec:host}

\subsection{Morphology}\label{sec:morph}

The host of AT2019qiz is a face-on spiral galaxy. A large-scale bar is visible in the PanSTARRS image (Figure \ref{fig:ps1}). \citet{French2020} analysed \textit{Hubble Space Telescope} (\textit{HST}) images of four TDE hosts and identified bars in two. While central bars (on scales $\lesssim 100$\,pc) can increase the TDE rate by dynamically feeding stars towards the nucleus \citep{Merritt2004}, there is no evidence that large-scale bars increase the TDE rate \citep{French2020}. Given the proximity of AT2019qiz, this galaxy is an ideal candidate for \textit{HST} or adaptive optics imaging to resolve the structure of the nucleus.

\begin{figure}
  \centering
  \includegraphics[width=\columnwidth]{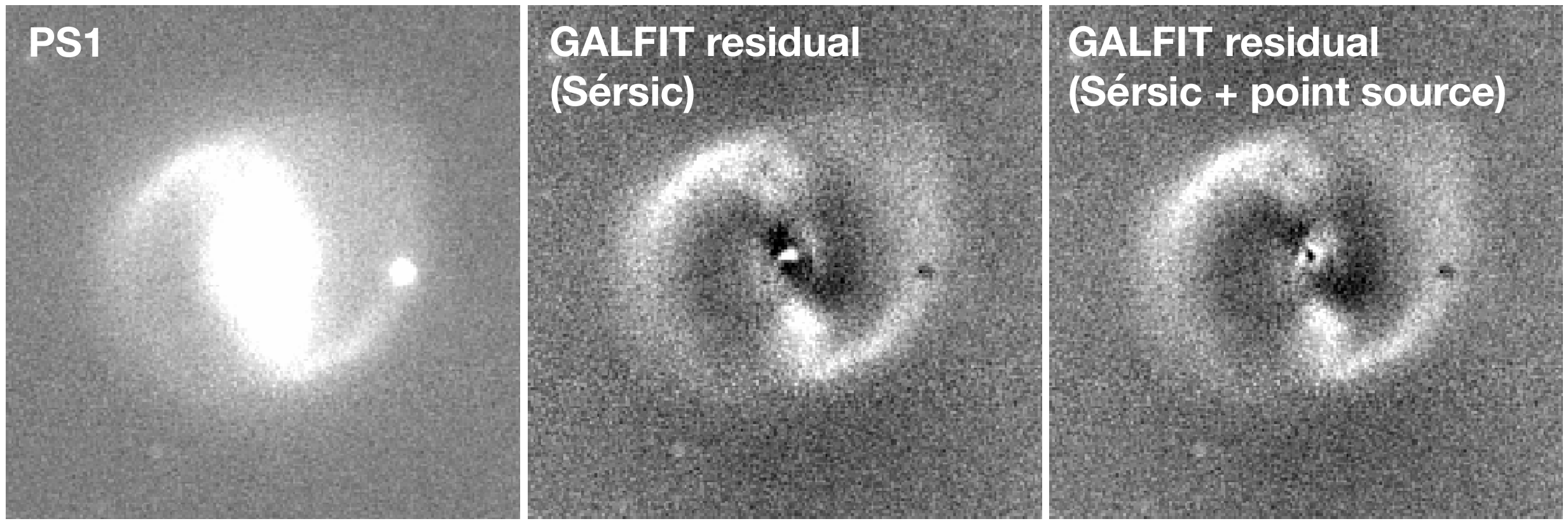}
  \caption{Left: PanSTARRS $g$-band image of the host galaxy. North is up, and east is towards the left of the image. We fit a S\'ersic function for the overall surface brightness profile using \textsc{galfit}, but make no attempt to model the spiral arms. The model also includes a point source for the nearby star to the west, and optionally a central point source for the galactic nucleus. Middle: Subtraction residuals without a central point source. Right: Subtraction residuals when including a central point source. We find visibly smoother residuals in this case.}
  \label{fig:galfit}
\end{figure}

Recent studies have shown that TDE host galaxies typically have a more central concentration of mass than the background galaxy population \citep{LawSmith2017,Graur2018}. The most recent compilation \citep{French2020rev} shows that the S\'ersic indices of TDE hosts range from $\approx 1.5-6$, consistent with the background distribution of quiescent galaxies but significantly higher than star-forming galaxies. We measure the S\'ersic index for the host of AT2019qiz by fitting the light distribution in the PanSTARRS $r$-band image in a $200\times400$ pixel box, centered on the nucleus, using \textsc{galfit} \citep{peng2002}. The residuals are shown in Figure \ref{fig:galfit}.
We do not fit for the spiral structure. Following \citet{French2020}, we investigate the effect of including an additional central point source (using the point-spread function derived from stars in the image as in section \ref{sec:ground}). The residuals appear flatter when including the point source, {however the change in reduced $\chi^2$ is minor} ($\chi^2= 1.36$ with the point source or 1.46 without). The best-fit S\'ersic index is 5.2 (with the point source) or 6.3 (without). In either case, this is consistent with the upper end of the observed distribution for TDE hosts.

\subsection{Velocity dispersion and black hole mass}

Following \citet{Wevers2017,Wevers2019}, we fit the velocity dispersion of stellar absorption lines with the code \textsc{ppxf} \citep{Cappellari2017} to estimate the mass of the central SMBH. We use a late-time spectrum obtained from X-shooter, resampled to a logarithmic spacing in wavelength and with the continuum removed via polynomial fits. We find a dispersion $\sigma=69.7\pm2.3$\,km\,s$^{-1}$.

Using relations between velocity dispersion and black hole mass (the $M_\bullet-\sigma$ relation), this gives a SMBH mass $\log (M_\bullet/M_\odot)=5.75\pm0.45$ in the calibration of \citet{McConnell2013}, or $\log (M_\bullet/M_\odot)=6.52\pm0.34$ in the calibration of \citet{Kormendy2013}.
The calibration of \citet{Gultekin2009} gives an intermediate value $\log (M_\bullet/M_\odot)=6.18\pm0.44$. The reason for the large spread in these estimates is that these relations were calibrated based on samples that comprised mostly black holes more massive than $10^7$\,\M. However, the estimates here are consistent, within the errors, with an independent mass measurement based on the TDE light curve (section \ref{sec:phot}).

\subsection{Host SED model}

Archival photometry of this galaxy is available from the PanSTARRS catalog in the $g,r,i,z,y$ filters, as well as in data releases from the 2 Micron All Sky Survey \citep[2MASS;][]{Skrutskie2006} in the $J,H,K$ filters, and the Wide-field Infrared Survey Explorer \citep[WISE;][]{Wright2010} in the WISE bands $W1-4$. We retrieved the Kron magnitudes from PanSTARRS, the extended profile-fit magnitudes (`m\_ext') from 2MASS, and the magnitudes in a $44''$ circular aperture (chosen to fully capture the galaxy flux) from WISE.

We fit the resultant spectral energy distribution (SED) with stellar population synthesis models in \textsc{Prospector} \citep{Leja2017} to derive key physical parameters of the galaxy. The free parameters in our model are stellar mass, metallicity, the current star-formation rate and the widths of five equal-mass bins for the star-formation history, and three parameters controlling the dust fraction and reprocessing (see \citealt{Leja2017} for details). {\citet{Leja2017} identify important degeneracies between age--metallicity--dust, and the dust mass--dust attenuation curve. \textsc{Prospector} is specifically designed to account for such degeneracies in parameter estimation using Markov chain Monte Carlo analysis to fully explore the posterior probability density.} \citet{vanVelzen2020} also used \textsc{Prospector} to model this galaxy (but only the PanSTARRS data); the mass and metallicity we find using the full SED are consistent with their results, within the uncertainties. A difference in our modelling is that we allow for a non-parametric star-formation history to better understand the age of the system.

\begin{figure}
  \centering
  \includegraphics[width=\columnwidth]{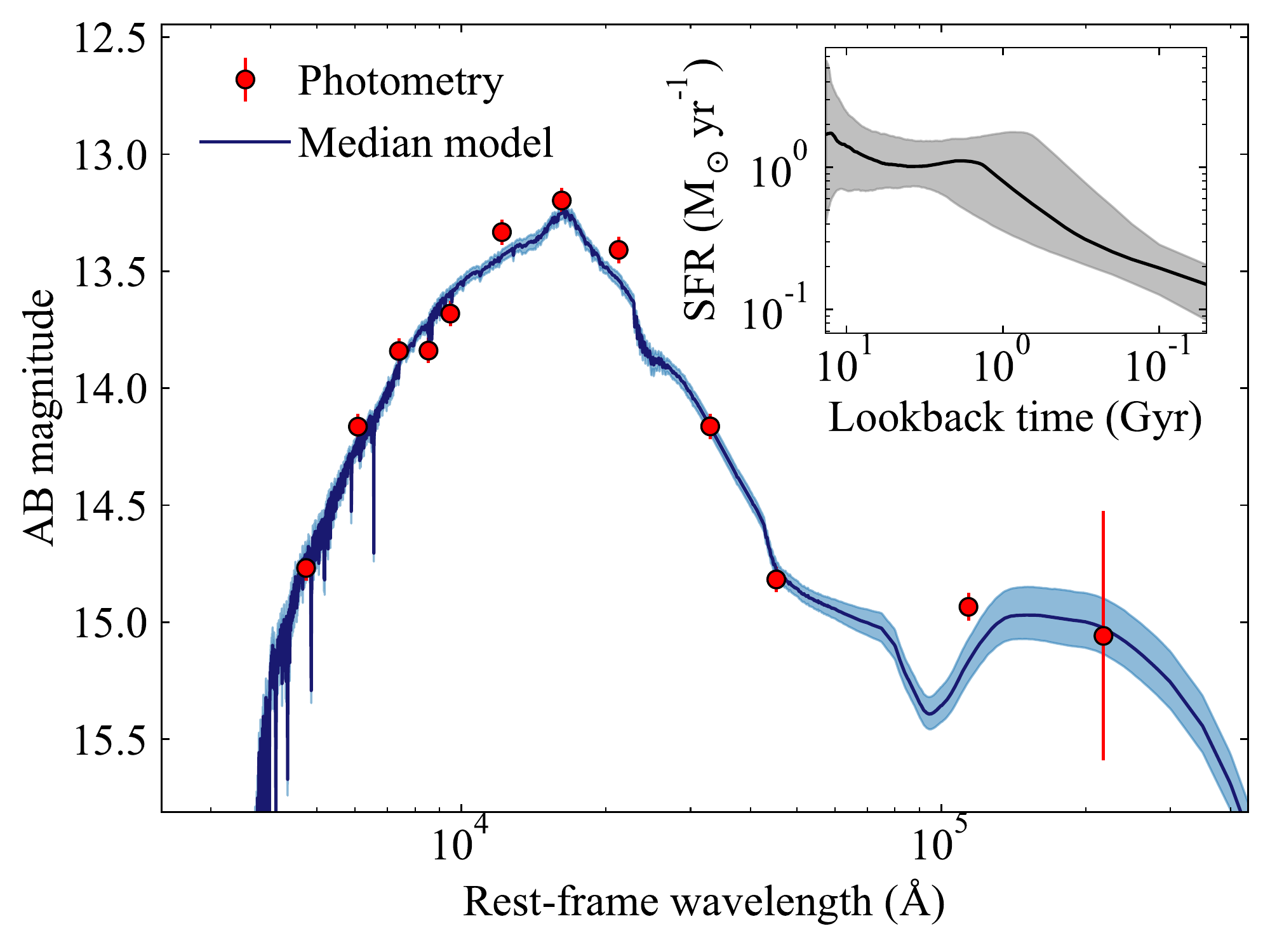}
  \caption{Archival photometry of the host galaxy, and SED fit using \textsc{prospector}. The best fit model, as well as the $1\sigma$ dispersion in model realisations, are shown. The inset shows the derived star-formation history, which is approximately flat at $1-2$\,\M\,yr$^{-1}$ prior to a steep drop in the last $\sim1$\,Gyr.}
  \label{fig:host}
\end{figure}

The best-fitting model is shown compared to the archival photometry in Figure \ref{fig:host}. We find stellar mass $\log (M_*/M_\odot) = 10.26^{+0.12}_{-0.15}$, a sub-solar metallicity $\log {Z/Z_\odot}=-0.84^{+0.28}_{-0.34}$ (but see section \ref{sec:bpt}), and a low specific star-formation rate $\log {\rm sSFR}=-11.21^{+0.23}_{-0.55}$ in the last 50\,Myr, where the reported values and uncertainties are the median and 16th/84th percentiles of the {marginalized} posterior distributions. The model also prefers a modest internal dust extinction, $A_V=0.16\pm0.04$\,mag. The stellar mass reported by \textsc{prospector} is the integral of the star-formation history, and {so includes stars and stellar remnants}. From our model we measure a `living' mass fraction {(i.e.~stars still undergoing nuclear burning)} of $0.58\pm0.02$.

In the same figure, we plot the median and $1\sigma$ uncertainty on the star-formation history {derived from the fit} versus lookback time since the Big Bang. We find a roughly constant star-formation rate of $\approx 2$\,\M\,yr$^{-1}$, prior to a sharp drop in the last $\approx 1.5$\,Gyr. A recent decline in star-formation is a common feature of TDE host galaxies, as evidenced by the over-representation of quiescent Balmer-strong galaxies (and the subset of post-starburst galaxies) among this population \citep{Arcavi2014,French2016,French2020rev}. Spectroscopy of the host after the TDE has completely faded will be required to confirm whether this galaxy is also a member of this class.

\begin{table}
  \centering
  \begin{tabular}{cc}
    \hline
  Line &  Flux ($10^{-16}$\,erg\,s$^{-1}$\,cm$^{-2}$) \\
  \hline
  H$\beta$ & $7.9 \pm 1.7$ \\
  {[O III]~$\lambda5007$} & $8.0 \pm 3.1$ \\
  {[O I]~$\lambda6300$}  & $<3.0$ \\
  H$\alpha$ & $14.2 \pm 3.8$ \\
  {[N II]~$\lambda6584$} & $7.6 \pm 1.9$ \\
  {[S II]~$\lambda\lambda6717,6731$}  & $6.3 \pm 1.0$ \\
    \hline
   & $\log$ ratio \\
  \hline
  {[O III]} / H$\beta$ & $0.04 \pm 0.25$ \\
  {[N II]} / H$\alpha$ & $-0.25 \pm 0.14$ \\
  {[S II]} / H$\alpha$ & $-0.32 \pm 0.16$ \\
  {[O I]} / H$\alpha$ & $\lesssim -0.7$ \\
  \hline
  \end{tabular}
  \caption{Host emission line fluxes and BPT line ratios \citep{Baldwin1981}, averaged over the X-shooter spectra.}
  \label{tab:bpt}
\end{table}

\subsection{Galaxy emission lines and evidence for an AGN}\label{sec:bpt}

The metallicity preferred by \textsc{prospector} would be very low for a galaxy of $\gtrsim10^{10}$\,\M, though \citet{vanVelzen2020} find similarly low metallicities for all TDE hosts in their sample, including AT2019qiz, from their SED fits. Spectroscopic line ratios provide a more reliable way to measure metallicity. The TDE spectra clearly show narrow lines from the host galaxy. We measure the fluxes of diagnostic narrow lines using Gaussian fits. Specifically, we measure H$\alpha$, H$\beta$, [O III]\,$\lambda5007$, [O I]\,$\lambda6300$, [N II]\,$\lambda6584$, and [S II]\,$\lambda\lambda6717,6731$. We report the mean of each of these ratios (averaged over the six X-shooter spectra) in Table \ref{tab:bpt}. No significant time evolution is seen in the narrow line fluxes. To estimate the metallicity, we use the N2 metallicity scale ([N II]\,$\lambda6584$/H$\alpha$), adopting the calibration from \citet{Pettini2004}, to find an oxygen abundance $12+\log({\rm O/H})=8.76\pm0.14$. This corresponds to a metallicity $Z/Z_\odot=1.17$, more in keeping with a typical massive galaxy. However, this [N~II]\,$\lambda6584$/H$\alpha$ ratio is outside the range used to calibrate the \citet{Pettini2004} relation, so this metallicity may not be reliable. Applying the calibration of \citet{Marino2013}, valid over a wider range, we find a slightly lower metallicity of $12+\log({\rm O/H})=8.63$, consistent with solar metallicity.

\begin{figure}
  \centering
  \includegraphics[width=\columnwidth]{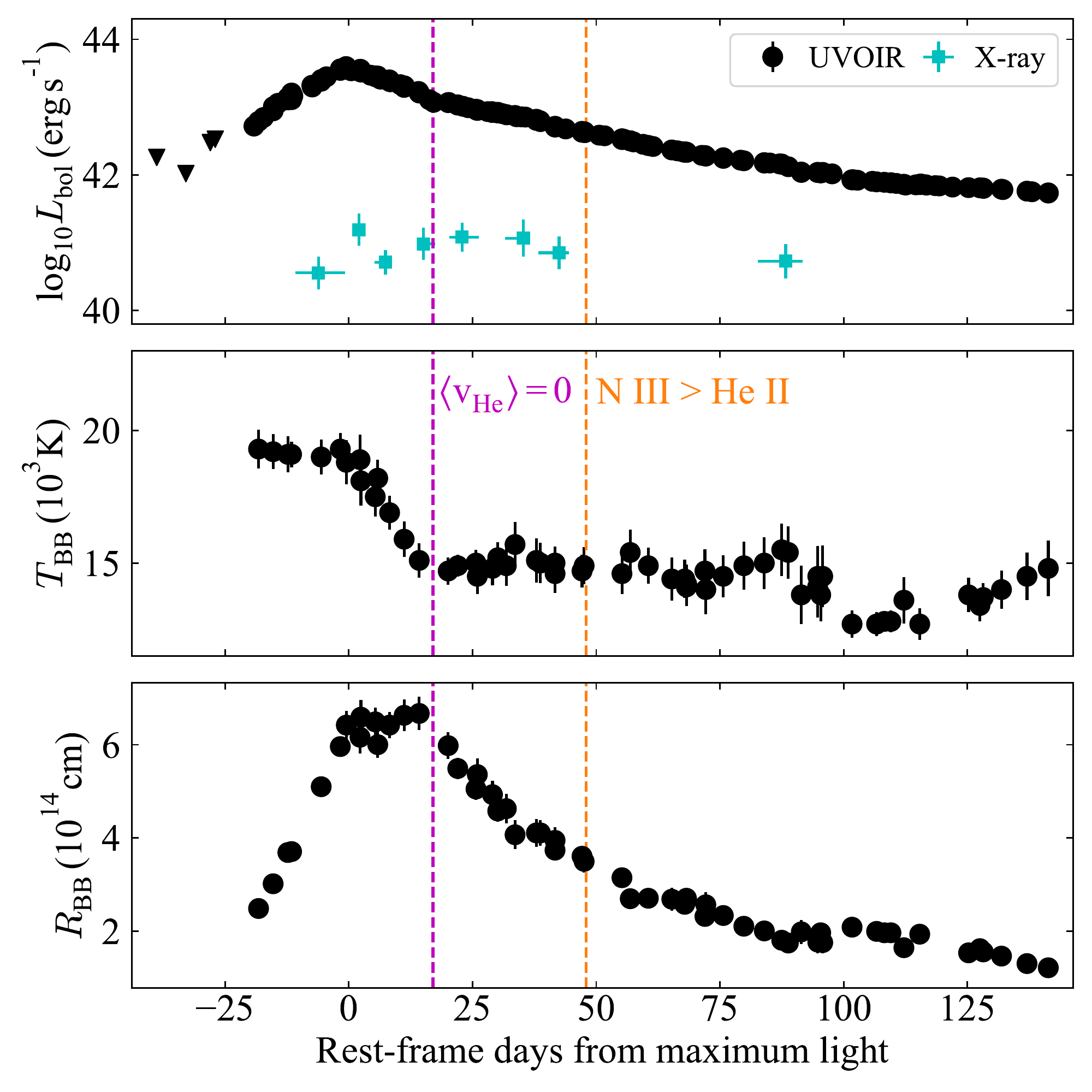}
  \caption{Top: Bolometric light curve of AT2019qiz derived from UV and optical photometry. The X-ray light curve is also plotted, which is $\approx10^3$ times fainter than the optical luminosity at peak but slower to fade. Middle: Temperature evolution. Bottom: Evolution of the blackbody radius. The temperature and radius are only shown for epochs covered without extrapolation by data in at least three photometric bands.
  Vertical lines indicate epochs of transition in the TDE spectrum, when the net blueshift of He~II goes to zero and when N~III becomes prominent (section \ref{sec:spec}).
  }
  \label{fig:bol}
\end{figure}

Ratios of these lines are used in the Baldwin-Phillips-Terlovich (BPT) diagram \citep{Baldwin1981} to probe the ionization mechanism of the gas. The ratios we measure for the host of AT2019qiz (Table \ref{tab:bpt}) lie intermediate between the main sequence of star-forming galaxies and galaxies with ionization dominated by an active galactic nucleus (AGN). This could be evidence of a weak AGN, or another source of ionization such as supernova shocks or evolved stars \citep{Kewley2001}. Several other TDE hosts lie in a similar region of the BPT parameter space \citep{Wevers2019,French2020rev}, while a number show direct evidence of AGN ionisation \citep{Prieto2016}. We note the caveat that if the lines in our spectra are excited by AGN activity, the calibrations used to estimate the metallicity may not always be valid.

To test the AGN scenario, we look at the mid-infrared colours. \citet{Stern2012} identify a colour cut $W1-W2>0.8$\,Vega mag to select AGN from WISE data. For the host of AT2019qiz, we find $W1-W2\approx0$\,Vega mag. At most a few percent of AGN have such a blue $W1-W2$ colour \citep{Assef2013}. \citet{Wright2010} employ a two-dimensional cut using the $W1-W2$ and $W2-W3$. The host of AT2019qiz has $W2-W3=1.8$\,mag, consistent with other spiral galaxies. Thus the emission detected by WISE is dominated by the galaxy, not an AGN.

The ratio of X-ray to [O~III] luminosity can also be used as an AGN diagnostic. Converting the X-rays to the 2-20\,keV band using our best-fit power-law, we measure a mean $L_X/L_{\rm [O~III]}=2.4\pm0.2$, which is consistent with a typical AGN \citep{Heckman2005}. However, the X-ray luminosity is only 0.03\% of the Eddington luminosity for a SMBH of $10^6$\,\M. Moreover, the temporal variation in the luminosity and hardness of the X-rays during the flare suggests a significant fraction of this emission comes from the TDE itself, rather than an existing AGN. In particular, the softening of the X-rays could indicate that as time increases, more of the emission is coming from the TDE flare, relative to an underlying AGN with a harder spectrum. Taking into account the BPT diagram, WISE colours, X-rays, and the morphology of the nucleus (section \ref{sec:morph}), we conclude that there is some support for a weak AGN, but that the galaxy is dominated by stellar light.

\section{Photometric analysis}\label{sec:phot}

\subsection{Bolometric light curve}

We construct the bolometric light curve of AT2019qiz by interpolating our photometry in each band to any epoch with data in the $g$, $r$ or $o$ bands, using \textsc{superbol} \citep{Nicholl2018}. We then integrate under the spectral energy distribution inferred from the multi-colour data at each epoch, and fit a blackbody function to estimate the temperature, radius, and missing energy outside of the observed wavelength range. A blackbody is an excellent approximation of the UV and optical emission from TDEs \citep[e.g][]{vanVelzen2020}. However, we note that the radius is computed under the assumption of spherical symmetry, which may not reflect the potentially complex geometry in TDEs. We include foreground extinction, but do not correct for the uncertain extinction within the host galaxy (formally, this makes our inferred luminosity and temperature curves lower limits). The bolometric light curve, temperature and radius evolution are plotted in Figure \ref{fig:bol}.

\begin{figure}
  \centering
  \includegraphics[width=\columnwidth]{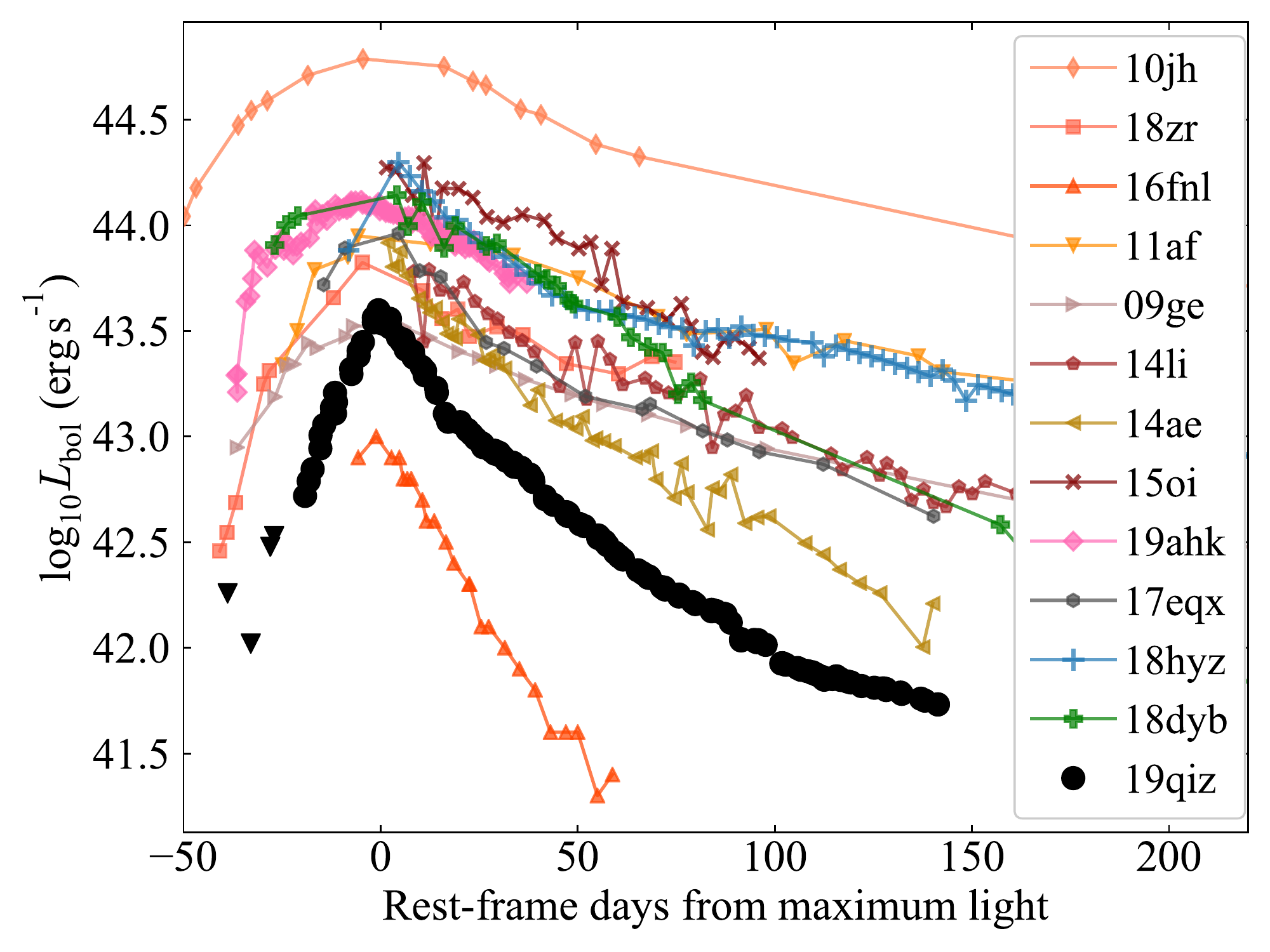}
  \caption{Comparison of the bolometric light curve to other TDEs from the literature  \citep{vanVelzen2019,Holoien2016b,Gezari2012,Chornock2014,Arcavi2014,Holoien2014,Holoien2016,Blagorodnova2017,Nicholl2019,Gomez2020,Leloudas2019}.
  }
  \label{fig:bolcomp}
\end{figure}


From the light curve we derive a peak date of MJD $58764\pm1$ (2019-10-08 UT)\footnote{The UV peaks slightly earlier (MJD 58764) than the optical (MJD 56766).}, a peak luminosity of $L=3.6\times10^{43}$\,\ergs, and integrated emitted energy of $E_{\rm rad}=1.0\times10^{50}$\,erg. Taking the black hole mass derived in section \ref{sec:host}, the peak luminosity corresponds to $\sim0.2 L_{\rm Edd}$, where $L_{\rm Edd}$ is the Eddington luminosity. We also plot the X-ray light curve to highlight the X-ray to optical ratio, which is $10^{-2.8}$ before peak. Since the X-rays appear to rise after the optical emission starts to fade, this ratio increases to $\approx 10^{-2.0}-10^{-1.8}$ between $20-50$ days after bolometric peak, and reaches $\approx 10^{-1.4}$ beyond 50 days.

We plot the bolometric light curve compared to other well-observed TDEs in Figure \ref{fig:bolcomp}. The fast rise (and decline), and low peak luminosity place it intermediate between the bulk of the TDE population and the original `faint and fast' TDE, iPTF16fnl \citep{Blagorodnova2017,Brown2018,Onori2019}.

\begin{figure}
  \centering
  \includegraphics[width=\columnwidth]{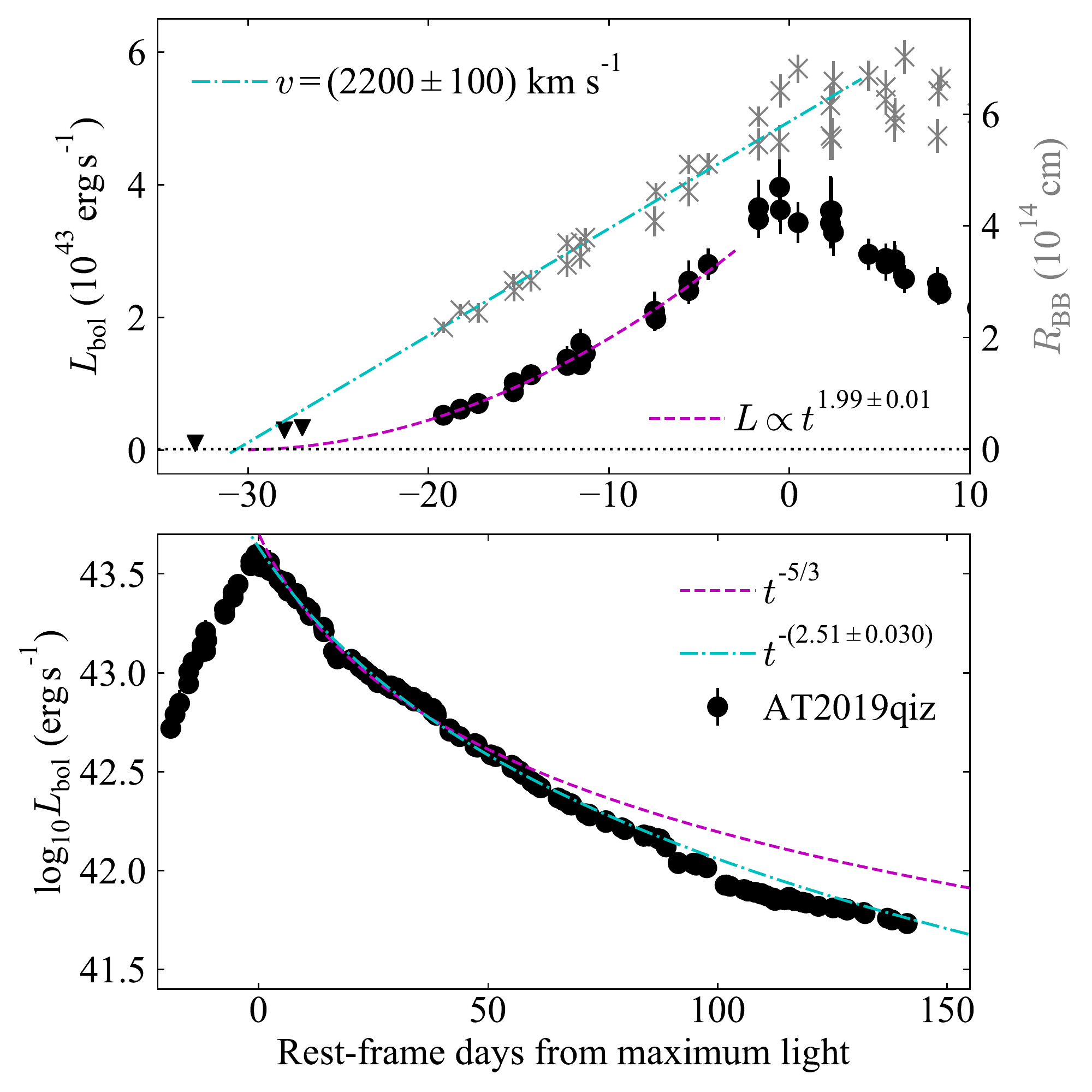}
  \caption{Power-law fits to the light curve. {Top: The photospheric radius (grey crosses) before maximum light grows linearly with a velocity $v\approx2200$\,\kms, while the luminosity (black circles) is best-fit as $L\propto t^2$}. Bottom: The declining light curve is steeper than the canonical $t^{-5/3}$.}
  \label{fig:pl}
\end{figure}


{The pre-maximum-light photospheric radius and temperature are unusually well constrained in AT2019qiz due to the early detections and multi-colour photometry.} The temperature is initially constant at $\approx 20,000$\,K before the light curve peaks, and then suddenly declines to $\approx 15,000$\,K over a period of $\sim 20$ days. It stays constant for the remainder of our observations, barring a possible slight dip around day 100 (though at this phase the UV data are noisier due to the large fractional host contribution that has been subtracted).


{The blackbody radius grows linearly up to maximum light, with a best-fit velocity of 2200\,\kms\ (Figure \ref{fig:pl}). Extrapolating back to radius $R=0$ implies a time of disruption 30.6 days before peak.} The radius then remains constant during the cooling phase identified in the temperature curve, before decreasing smoothly at constant temperature.

We fit the rising light curve with a power-law of the form $L = L_0 ((t-t_0)/\tau)^{\alpha}$ using the \textsc{curve\_fit} function in \textsc{scipy}. {We fix the initial time $t_0=30.6$ days, as inferred from the expanding photosphere (we find a near-identical fit to the light curve even if $t_0$ is left free). The best fit has a rise timescale $\tau=10.9$ days and $\alpha=1.99\pm0.01$. We plot this fit alongside the fit to the radius in Figure \ref{fig:pl}.} \citet{Holoien2019} modelled the rise of the TDE AT2019ahk, detected very soon after disruption, and also found a power-law consistent with $\alpha\approx2$.

We fit the declining light curve with a power-law function of the same form. As shown in the lower panel of Figure \ref{fig:pl}, we find a best-fit $\alpha=-2.54$, which is steeper than the canonical $L \propto t^{-5/3}$ predicted by simple fallback arguments \citep{Rees1988}. However this is not unusual among the diverse array of TDEs in the growing observed sample, and more recent theoretical work does not find a universal power-law slope for the mass return rate, nor that the light curve exactly tracks this fallback rate \citep[e.g.][]{Guillochon2013,Gafton2019}.

\subsection{TDE model fit}

To derive physical parameters of the disruption, we fit our multiband light curves using the Modular Open Source Fitter for Transients \citep[\textsc{mosfit};][]{Guillochon2018} with the TDE model from \citet{Mockler2019}. This model assumes a mass fallback rate derived from simulated disruptions of polytropic stars by a SMBH of $10^6$\,\M\ \citep{Guillochon2014}, and uses scaling relations and interpolations for a range of black hole masses, star masses, and impact parameters. The free parameters of the model, as defined by \citet{Mockler2019}, are the masses of the black hole, $M_\bullet$, and star, $M_*$; the scaled impact parameter $b$; the efficiency $\epsilon$ of converting accreted mass to energy; the normalisation and power-law index, $R_{\rm ph,0}$ and $l_{\rm ph}$, connecting the radius to the instantaneous luminosity; the viscous delay time $T_\nu$ (the time taken for matter to circularise and/or move through the accretion disk) which acts approximately as a low pass filter on the light curve; the time of first fallback, $t_0$; the extinction, proportional to the hydrogen column density $n_{\rm H}$ in the host galaxy; and a white noise parameter, $\sigma$. {The priors follow those used by \citet{Mockler2019}, and reflect the range of SMBH masses where optically-bright TDEs are expected \citep[e.g.][]{vanVelzen2018b}, the range of impact parameters covering both full and partial disruptions, accretion efficiencies for non-rotating to maximally-rotating black holes, and a broad range of possible photospheres and viscous timescales (see \citealt{Mockler2019} for details).}

\begin{figure}
  \centering
  \includegraphics[width=\columnwidth]{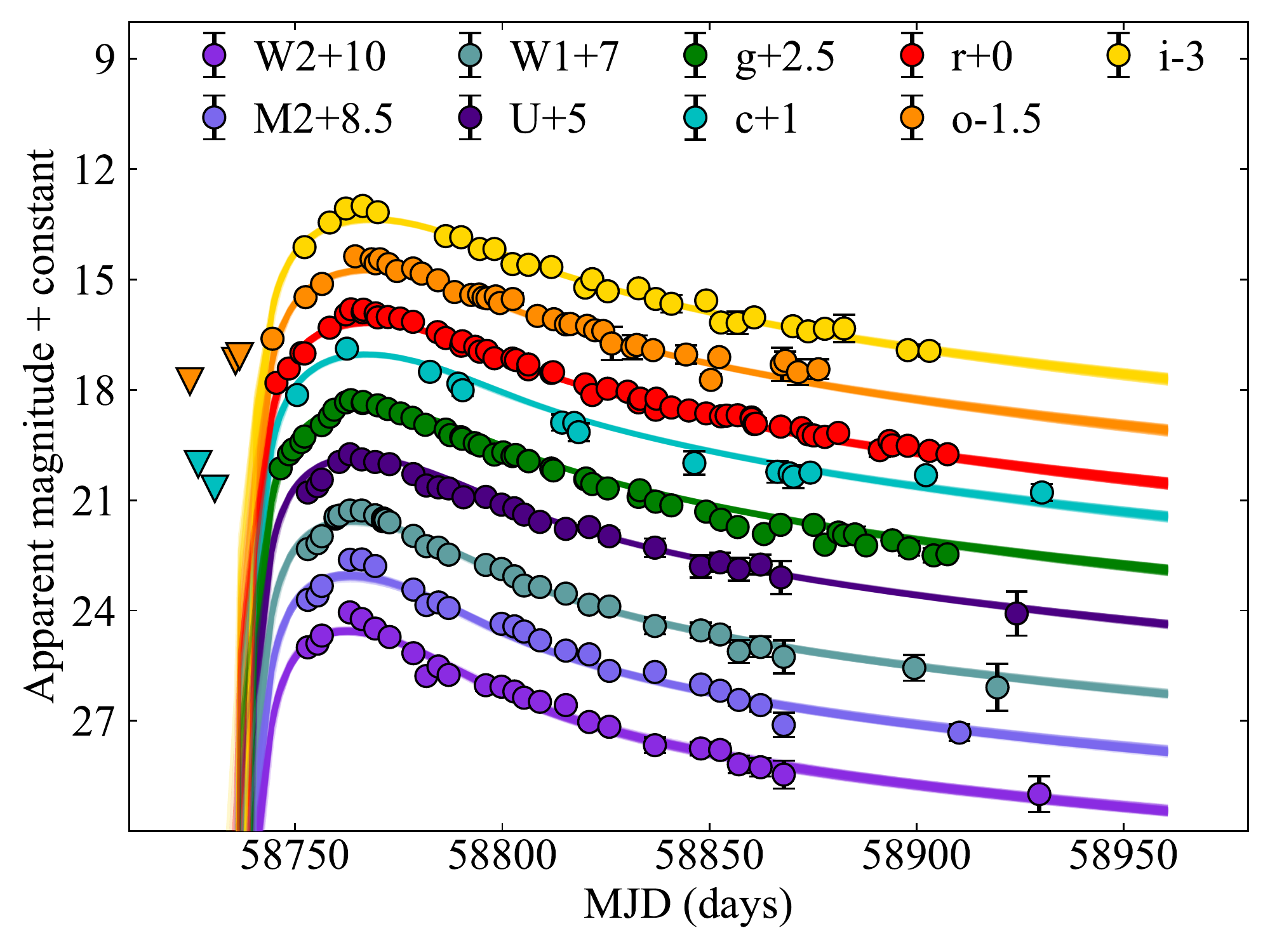}
  \caption{Fits to the multicolour light curve using the TDE model in \textsc{mosfit} \citep{Guillochon2018,Mockler2019}.}
  \label{fig:mosfit}
\end{figure}

\begin{table}
  \centering
  \begin{tabular}{cccc}
  \hline
  Parameter & Prior & Posterior & Units\\
  \hline
$ \log{(M_\bullet )}$ & $[5, 8]$ & $ 5.89^{+0.05}_{-0.06} $ & M$_\odot$ \\
$ M_*$ & $[0.01, 100]$ & $ 0.97 \pm 0.04 $ & M$_\odot$ \\
$ b$ & $[0, 2]$ & $ 0.22^{+0.02}_{-0.03}$ &   \\
$ \log(\epsilon) $ & $[-2.3, -0.4] $ & $ -2.23 ^{+0.14}_{-0.05}$ &   \\
$ \log{(R_{\rm ph,0} )} $ & $[-4, 4] $ & $ 1.12 \pm 0.06 $ &   \\
$ l_{\rm ph}$ & $[0, 4]$ & $ 0.66 \pm 0.03 $ &   \\
$ \log{(T_v )} $ & $[-3, 3] $ & $ 0.74^{+0.05}_{-0.06} $ & days  \\
$ t_0 $ & $[-50, 0]$ & $  -7.04 ^{+0.52}_{-0.60}$ & days  \\
$ \log{(n_{\rm H,host})}$ & $[19, 23]$ & $ 20.03 ^{+0.26}_{-0.48}$ & cm$^{-2}$ \\
$ \log{\sigma} $ & $[-4, 2] $ & $ -0.72 \pm 0.02$ &   \\
  \hline
\end{tabular}
  \caption{Priors and marginalised posteriors for the \textsc{mosfit} TDE model. Priors are flat within the stated ranges, except for $M_*$, which uses a Kroupa initial mass function. The quoted results are the median of each distribution, and error bars are the 16th and 84th percentiles. These errors are purely statistical; \citet{Mockler2019} provide estimates of the systematic uncertainty.}
  \label{tab:mosfit}
\end{table}

The fits are applied using a Markov Chain Monte Carlo (MCMC) method implemented in \textsc{emcee} \citep{Foreman2013} using the formalism of \citet{Goodman2010}. We burn in the chain for 10,000 steps, and then continue to run our simulation until the potential scale reduction factor (PSRF) is $<1.1$, indicating that the fit has converged. We plot 100 realisations of the Markov Chain in the space of our light curve data in Figure \ref{fig:mosfit}. The model provides a good fit to the optical bands, but struggles slightly to resolve the sharp peak present in the UV bands.

From this fit we derive the posterior probability distributions of the parameters, listed in Table \ref{tab:mosfit}, with two-dimensional posteriors plotted in the appendix. 
The inferred $t_0$ is {MJD $58737\pm1$, i.e.~$27\pm2$ days} before peak, consistent with the simpler power-law models. This suggests that the first detection of AT2019qiz is about a week after {the beginning of the flare}. The physical parameters point to the disruption of a roughly solar mass main sequence star by a black hole of {mass $10^{5.9}$\,\M.} This is consistent with the lower end of the SMBH mass range estimated from spectroscopy and the $M_\bullet-\sigma$ relation. In this case, the peak luminosity corresponds {to $0.36 L_{\rm Edd}$ (the Eddington luminosity). This is consistent with the typical Eddington ratios} measured for a sample of TDEs with well-constrained SMBH masses \citep{Wevers2019}.

The scaled impact parameter, {$b=0.22\pm0.02$, corresponds to a physical impact parameter $\beta \equiv R_t / R_p = 0.86\pm0.03$,} where $R_t$ is the tidal radius and $R_p$ the orbital pericentre. For the inferred SMBH mass, $R_t = 23 R_S$, where $R_S$ is the Schwarzschild radius. Using the remnant mass versus $\beta$ curve from \citet{Ryu2020} for a 1\,\M\ star, {up to $\sim 25\%$ of the star could have survived this encounter}. Interestingly, \citet{Ryu2020} predict a mass fallback rate proportional to $t^{-8/3}$ in this case (which they call a `severe partial disruption'), which is remarkably close to our best-fit power-law decline, $t^{-2.54}$.

\section{Spectroscopic analysis}\label{sec:spec}


\begin{figure}
  \centering
  \includegraphics[width=\columnwidth]{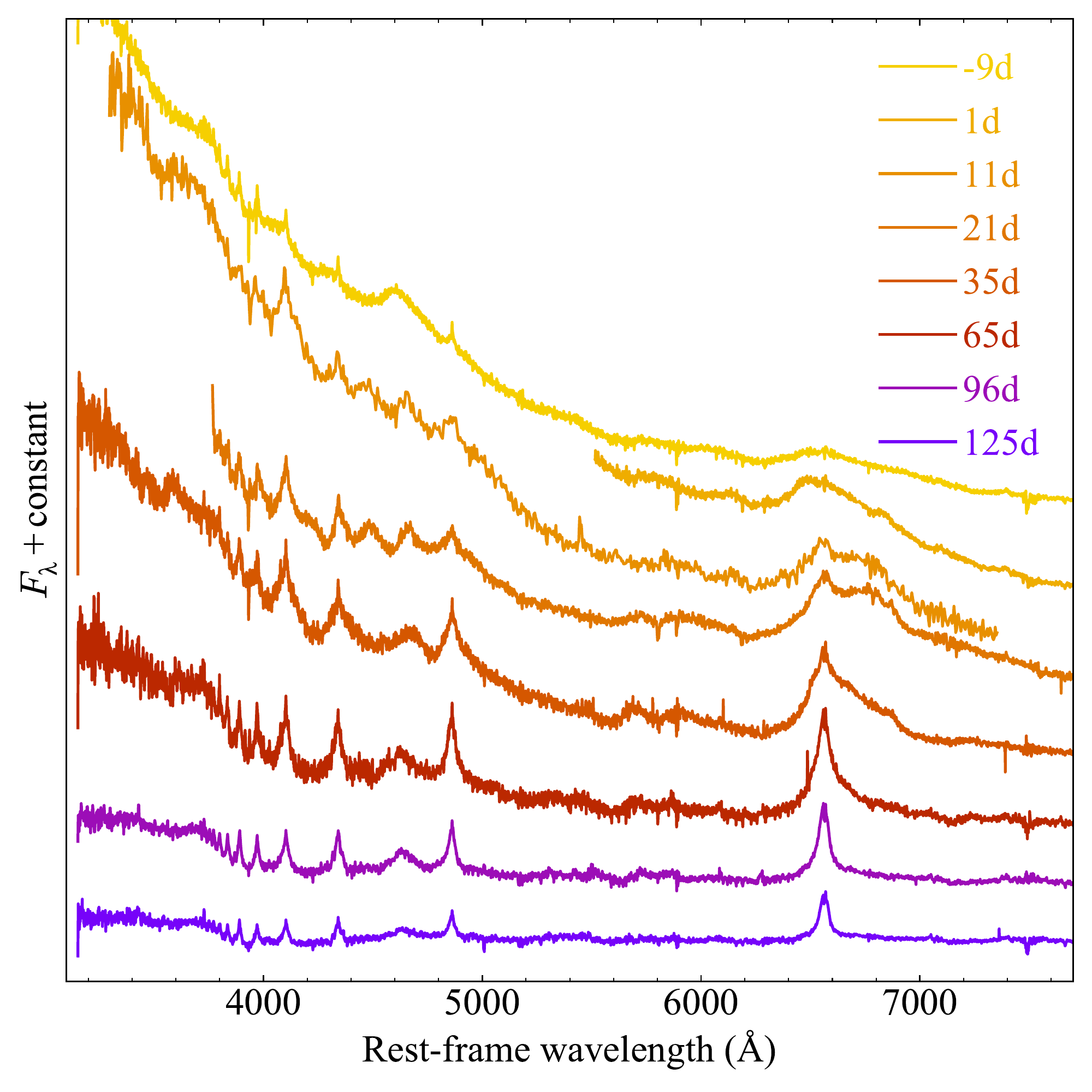}
  \includegraphics[width=\columnwidth]{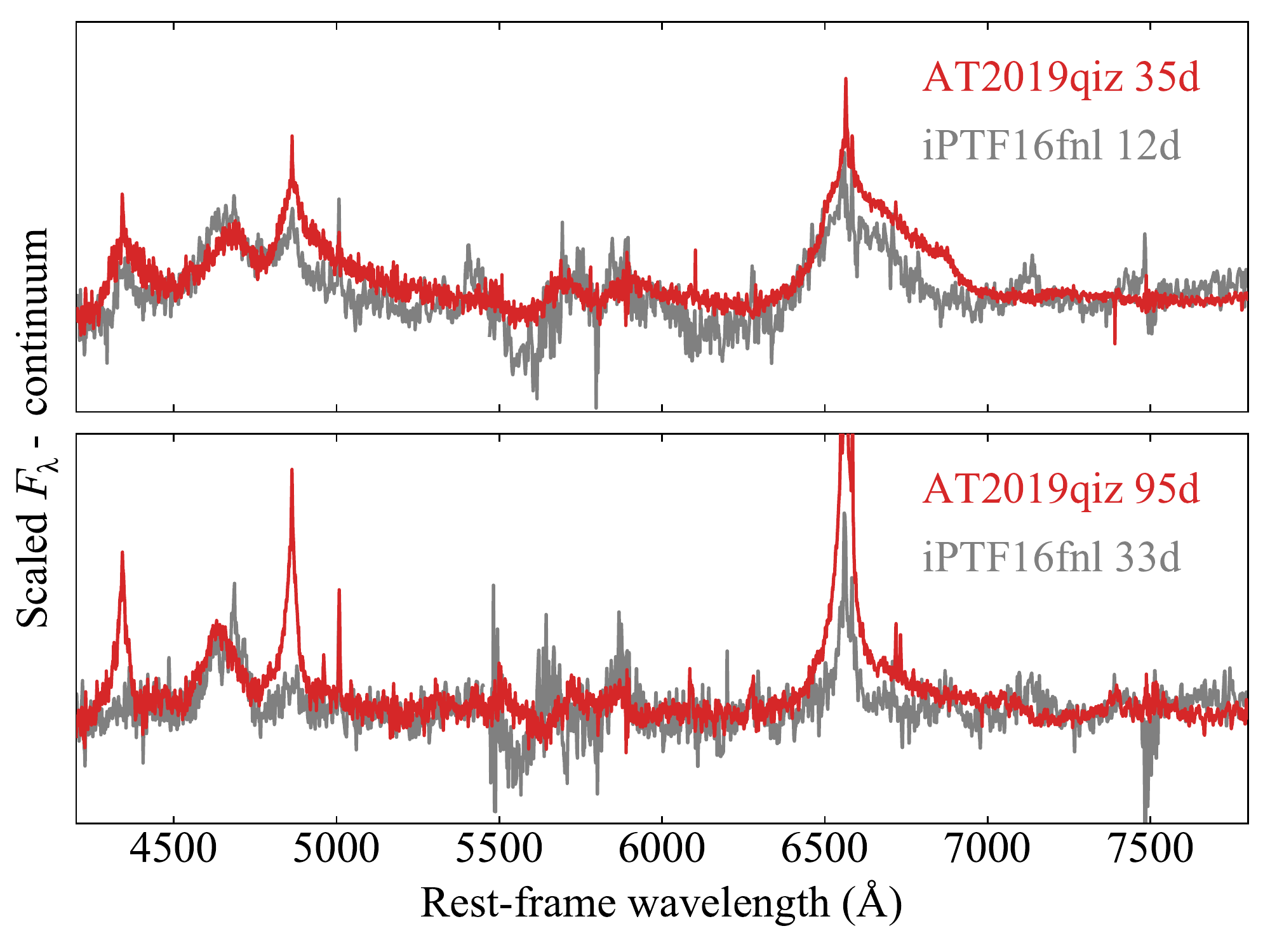}
  \caption{Top: Selected spectra from X-shooter, EFOSC2 and Binospec after subtraction of the host galaxy model (following the method outlined in the appendix). We have applied the subtraction procedure to all spectra, but here show only this subset (spanning the full range of observed phases) for clarity.
  Bottom: Comparison of AT2019qiz to X-shooter spectra of iPTF16fnl \citep[from][]{Onori2019}, the only TDE with a faster light curve evolution than AT2019qiz. The continuum has been removed using polynomial fits. The spectra shortly after maximum light are quite similar for these two events, though the Balmer lines are much weaker at late times in iPTF16fnl.}
  \label{fig:select}
\end{figure}

The early spectra are dominated by a steep blue continuum indicative of the high photospheric temperature (15,000-20,000\,K), superposed with broad emission bumps. As the spectra evolve and the continuum fades, the emission lines become more sharply peaked, while the host contribution becomes more prominent. In all the analysis that follows, we first subtract the host galaxy light using the model SED from \textsc{prospector} (section \ref{sec:host}). The full set of host-subtracted spectra, along with further details of the subtraction process, are shown in the appendix. In Figure \ref{fig:select}, we plot a subset of high signal-to-noise ratio, host-subtracted spectra spanning the evolution from before peak to more than 100 days after.

\begin{figure*}
  \centering
  \includegraphics[width=\textwidth]{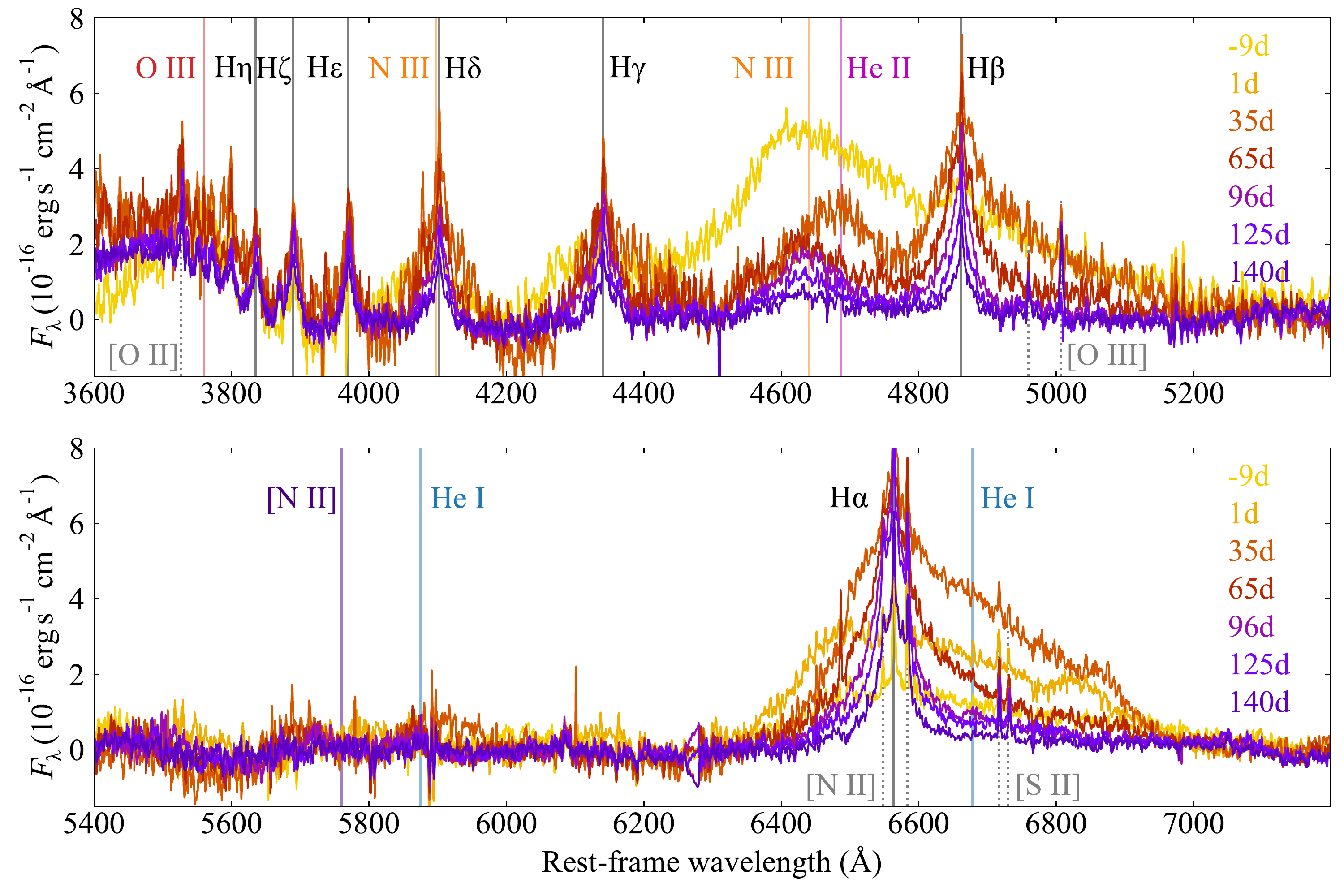}
  \caption{Host-subtracted X-shooter spectra after fitting and subtracting the continuum to highlight the emission lines. All strong lines from the TDE (solid coloured lines) and nebular emission in the host galaxy (grey dotted lines) are labelled. Note that the higher-order Balmer lines (H$\epsilon$, etc) and continuum bluewards of $\sim 4000$\,\AA\ may not be reliable, as we lack pre-disruption photometry to constrain the host galaxy model at these wavelengths.}
  \label{fig:id}
\end{figure*}

\subsection{Line identification}

To focus on the line evolution, we subtract the continuum using a 6th-order polynomial, with sigma-clipping to reject the line-dominated regions during the fit. The host- and continuum-subtracted spectra obtained with X-shooter are shown in Figure \ref{fig:id} (only this subset is shown for clarity of presentation). We identify and label the strong emission lines from both the TDE and the host galaxy. \textsc{prospector} allows to the user to turn nebular emission lines on and off; for the bulk of our analysis we use the predictions from \textsc{prospector} to subtract nebular lines, but in Figure \ref{fig:id} we leave the nebular emission in our data for completeness.
Balmer emission lines are at all times visible, with both a broad TDE component and a narrow host component. The other strong host lines are those used for the BPT analysis in section \ref{sec:host}.

As well as hydrogen, we also identify broad emission lines of He~II $\lambda4686$, the Bowen fluorescence lines of N~III $\lambda4100$ and $\lambda4640$ and likely O III $\lambda3670$ \citep{Bowen1935,Blagorodnova2018,Leloudas2019}, and possible weak emission of He~I~$\lambda5876$. The combination of hydrogen lines with the He~II/Bowen blend at around 4600\,\AA\ is common in TDE spectra, and qualify AT2019qiz as a TDE-Bowen in the recent classification scheme proposed by \citet{vanVelzen2020}, or an N-rich TDE in the terminology of \citet{Leloudas2019}.

\begin{figure*}
  \centering
  \includegraphics[width=5.8cm]{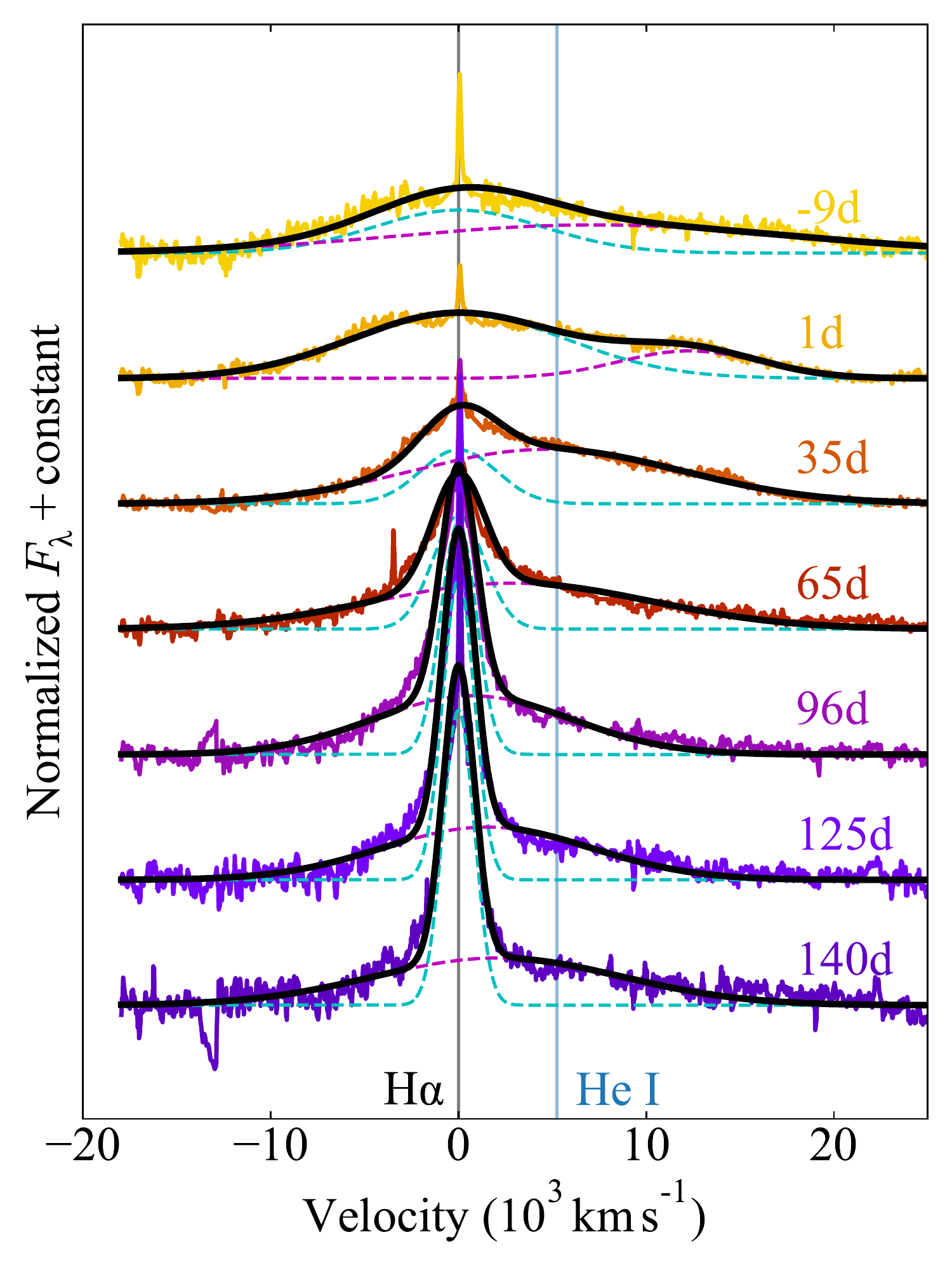}
  \includegraphics[width=5.8cm]{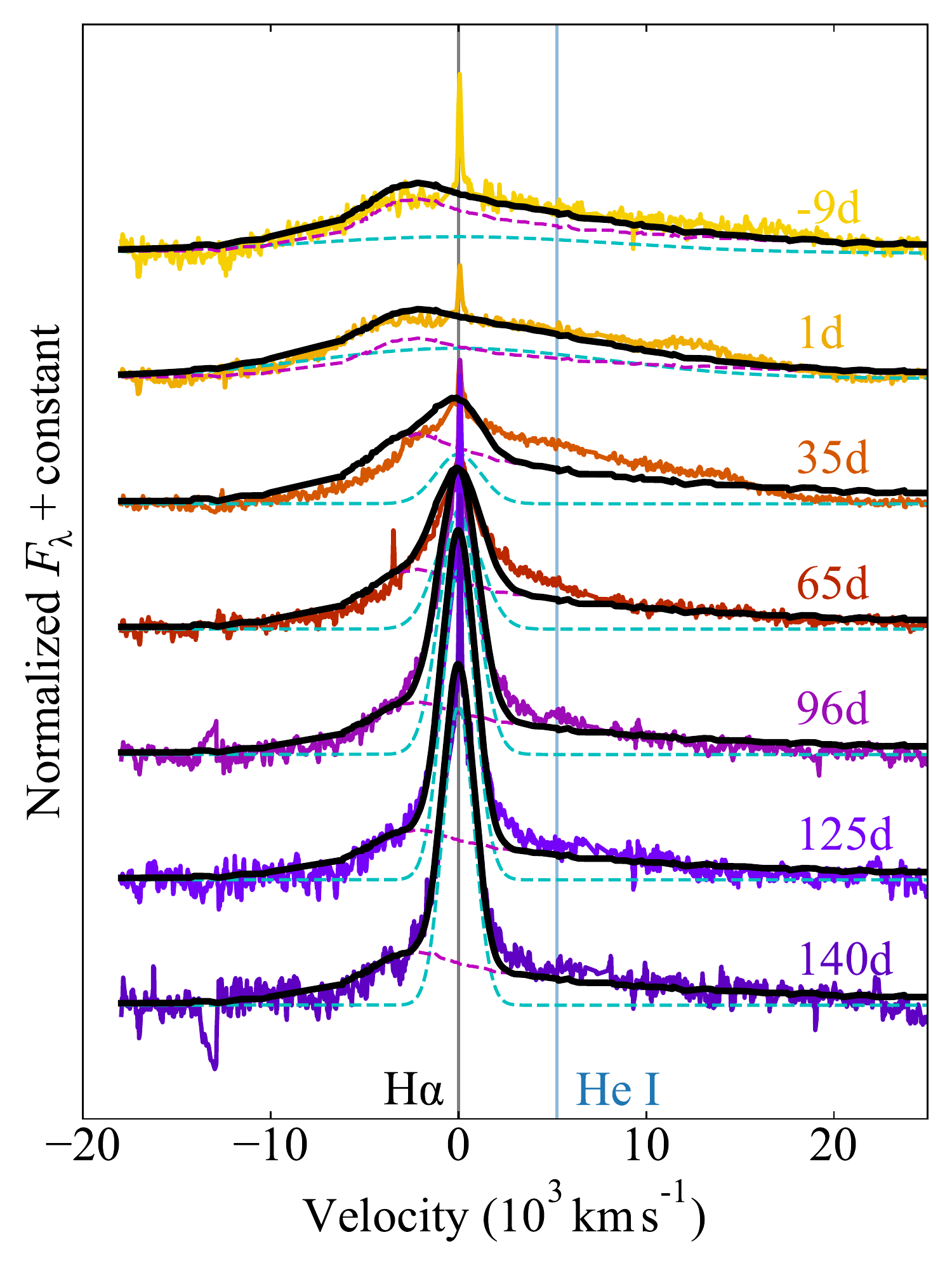}
  \includegraphics[width=5.8cm]{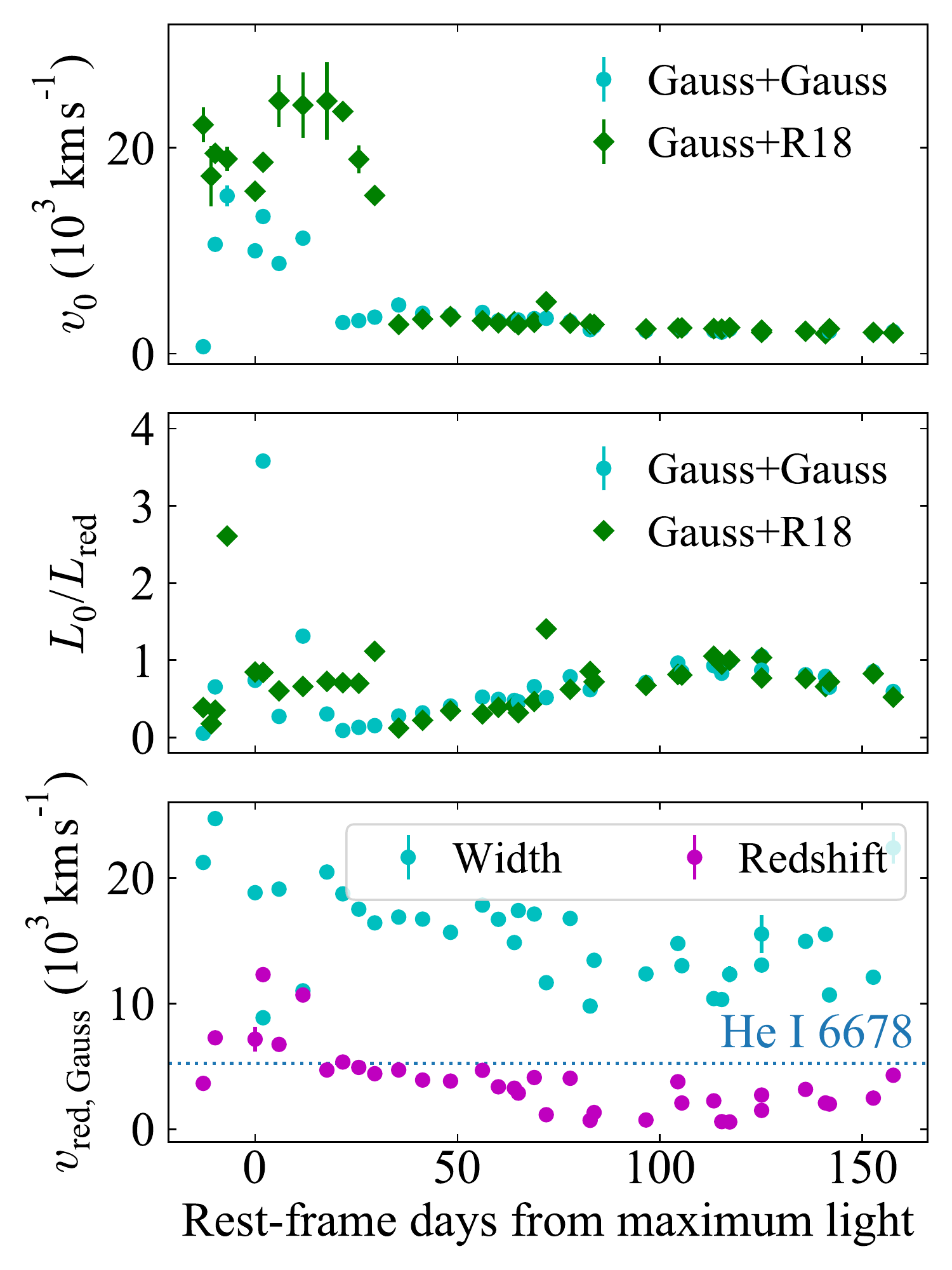}
  \caption{Left: Double-Gaussian fits to the continuum-subtracted H$\alpha$ line profile (only X-shooter spectra shown for clarity). One component is fixed around zero velocity with a variable width, while the other can vary in both width and red/blue-shift. The summed profile is shown as a solid line, while the fixed-centre and shifted profiles are shown as dotted lines.
  Middle: Same as left, but the offset component is now replaced with a 5000\,\kms\ outflow profile from \citet{Roth2018}.
  Right: Velocity and flux measurements from fits. The component with a fixed centre at zero velocity exhibits a similar evolution in both the double-Gaussian (left) and Gaussian+outflow (middle) models. In the former, this component is at all times narrower than the redshifted component (bottom panels). The broad component is not consistent with He~I~$\lambda6678$ (at various times it is centered at either too high or too low velocity). The flux ratio of the narrow component to the offset/outflow component increases over time from $\sim 0.5$ to $\sim 1$ for both models.}
  \label{fig:gauss}
\end{figure*}

In our light curve comparison, we found that AT2019qiz appeared fainter and faster than nearly any other TDE except for iPTF16fnl {(see also section \ref{sec:faint} for a more quantitative discussion)}. We plot a spectroscopic comparison in the lower panel of Figure \ref{fig:select}. The spectrum of AT2019qiz at about a month after maximum shows several similarities to iPTf16fnl at a slightly earlier phase of 12 days, particularly in the blue wing of H$\alpha$, though AT2019qiz exhibits a broader red wing. The ratio of H$\alpha$ compared to He~II is also quite consistent between these two events, modulo the slower evolution in AT2019qiz. At later times, the Balmer lines become weaker in iPTF16fnl, though H$\alpha$ narrows and becomes more symmetric, as we see in AT2019qiz. In fact, the ratios and velocity profiles of these lines evolve substantially with time, as we saw in Figure \ref{fig:id}. We will now investigate this in detail in the following sections.

\subsection{The H$\alpha$ profile}
\label{sec:ha}

We look first at the H$\alpha$ line. This region of the spectrum is plotted in velocity coordinates in Figure \ref{fig:gauss}. We have subtracted the continuum locally using a linear fit to line free regions at either side of the line (6170-6270\,\AA, 7100-7150\AA). The H$\alpha$ profile is initially asymmetric and shallow, with a blueshifted peak and a broad red shoulder. The red side may include some contribution from He~I~$\lambda6678$, but this is likely not a major contributor as we see only very weak He~I~$\lambda5876$. We initially fit the profile as the sum of two Gaussians whose normalisations and velocity widths vary independently, with one centroid fixed at zero velocity and the other free to vary. The three velocities (two widths and one offset) are plotted in Figure \ref{fig:gauss}. 

We find that the zero-velocity component is at all times narrower than the offset component, and over time decreases in width as the line becomes sharply peaked. The asymptotic velocity full-width at half-maximum (FWHM) is $\approx2000$\,\kms. The broader component is always redshifted, though this shift decreases as the red shoulder becomes less prominent. This component maintains a width of $\sim 15,000$\,\kms, though the scatter in measuring this component is quite large at later times when the shoulder is less prominent. We confirm that this feature is not a blend with He~I, as the velocity offset does not match the wavelength of that line (and varies over time).

If the broadening of the redshifted component is due to rotation, the implied radius of the emitting material is $\approx 200 R_S \approx 5-10 R_p$. This is consistent with the size of TDE accretion disks in the simulations of \citet{Bonnerot2020}. {Disk profiles in TDE emission lines have been claimed in PT09djl \citep{Arcavi2014,Liu2017} and AT2018hyz \citep{Short2020,Hung2020}. However the difficulty in interpreting these line profiles is illustrated by the case of AT2018zr (also called PS18kh), which had flat-topped Balmer lines argued by \citet{Holoien2018} to originate in an ellipical disk and by \citet{Hung2019} to instead come from an outflow. AT2019qiz does not show a classic flat-topped or `double-horned' disk profile at early or late phases, though it is interesting to note that around 20 days after maximum (e.g. Figure \ref{fig:select}), the red shoulder temporarily resembles a second peak, with the blue peak close to rest wavelength. 
}

{To produce this profile with a disk model would require a highly elliptical disk, viewed close to edge on and with a near-vertical orientation of the pericenter with respect to the observer, as was suggested to be the case for PT09djl by \citet{Liu2017}. If we were to interpret the day 20 H$\alpha$ profile of AT2019qiz as a disk, it would be surprising to find the same highly specific geometry in two out of four TDEs with claimed disk signatures. On the other hand the transient appearance of a double-peaked profile at this phase after peak would be reminiscent of AT2018hyz. The transience of these signatures could be due to either optical depth effects \citep{Short2020,Gomez2020} or contamination by an additional emission component \citep{Hung2020}.  Detailed time-series modelling with disk profiles will be needed to confirm if this scenario is compatible with AT2019qiz.
}

\begin{figure*}
  \centering
  \includegraphics[width=\columnwidth]{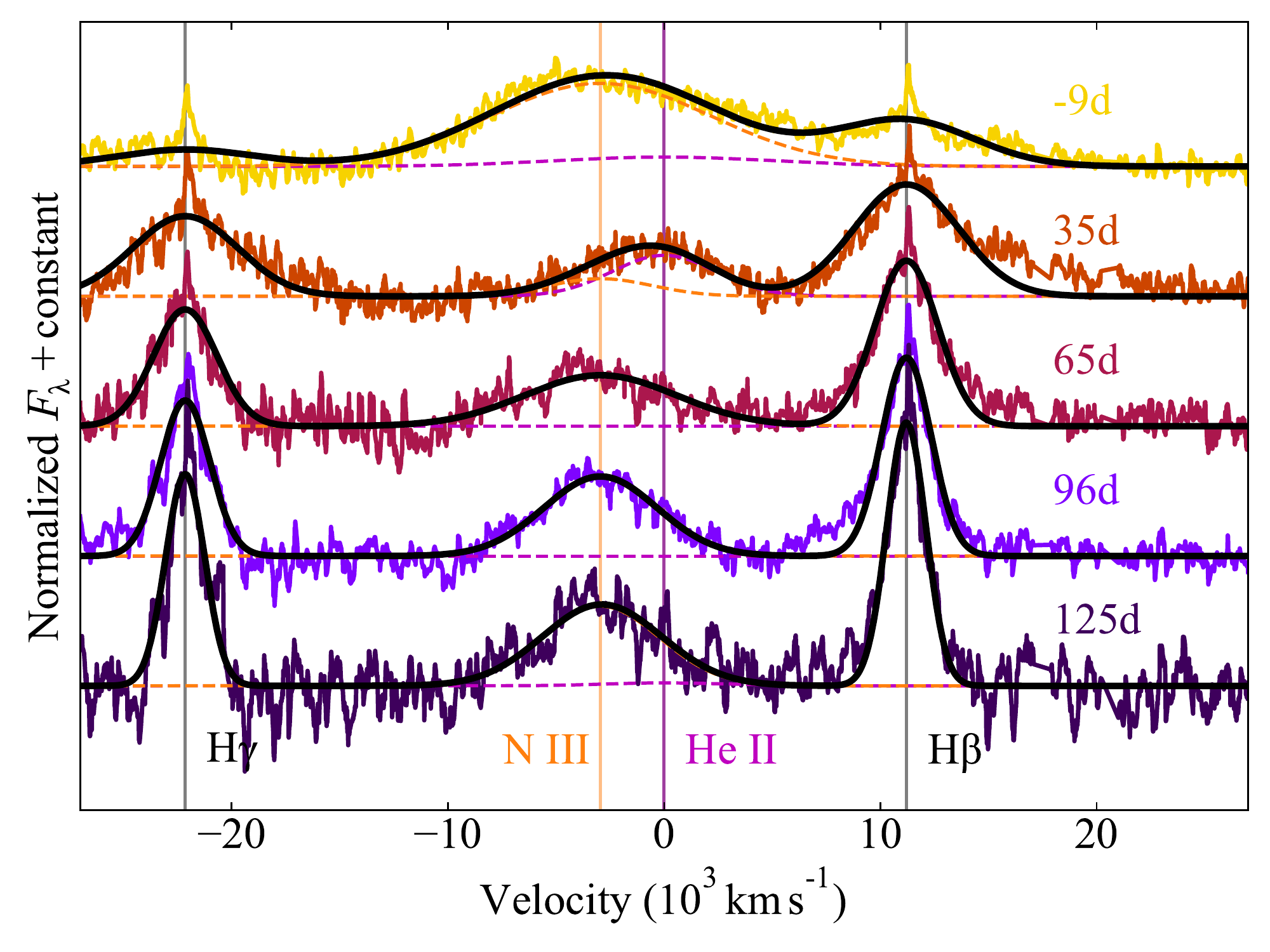}
   \includegraphics[width=\columnwidth]{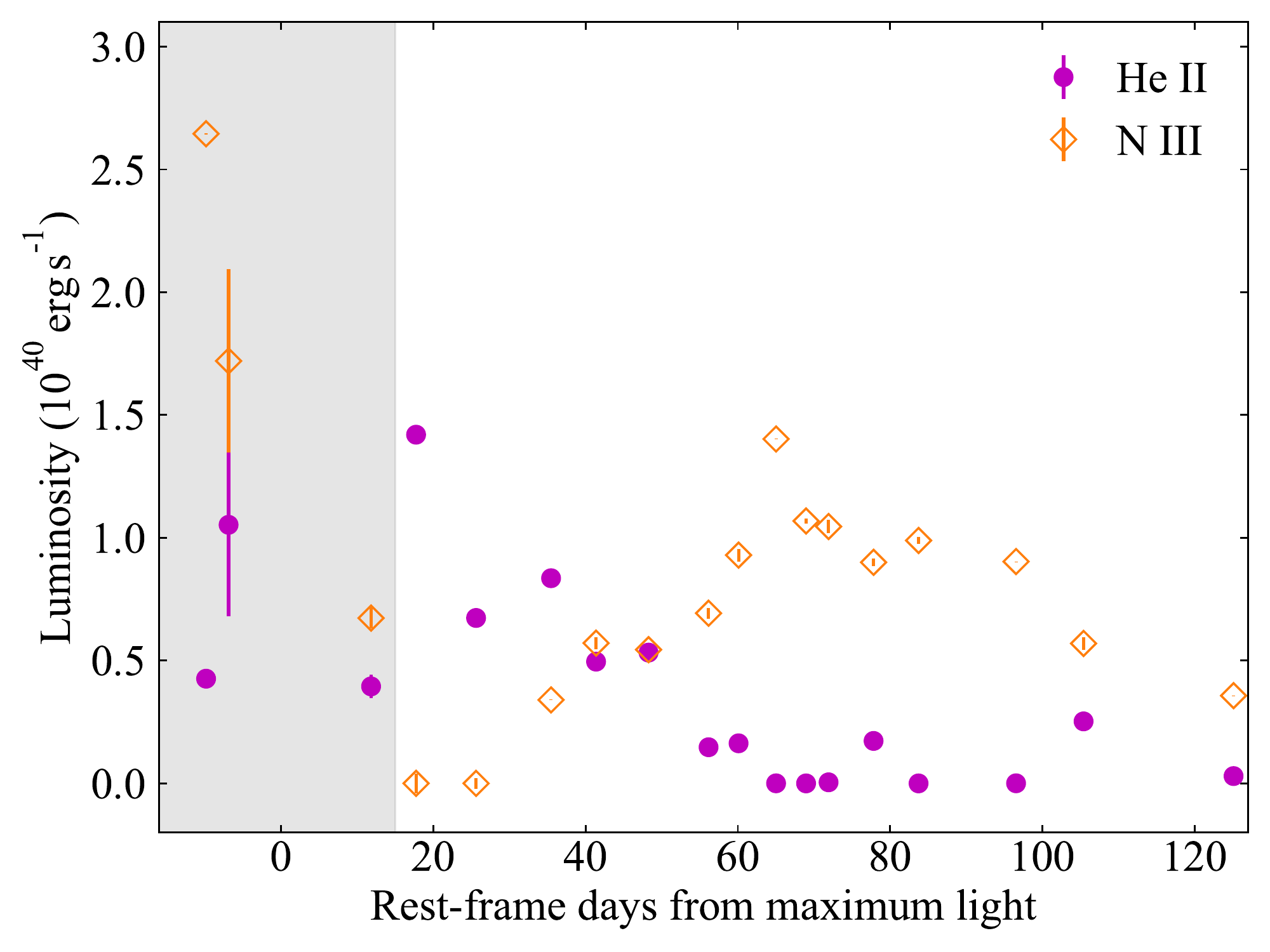}
 \caption{
 Left: Gaussian fits to the continuum-subtracted He~II, N~III, H$\beta$ and H$\gamma$ line profiles (only X-shooter spectra shown for clarity). All lines are centred at their rest position. The velocity width of the two Balmer lines is constrained to be the same. We also impose that the widths of He~II and N~III must match each other to help the fit converge. The summed profile is shown as a solid line, while the He~II and N~III profiles are shown as dotted lines.
 Right: Deblended luminosity evolution of He~II and N~III lines measured from Gaussian fits. The grey shaded region is where the peak of the blended feature is bluewards of N~III, i.e. Gaussian fits do not well describe the blueshifted feature.}
  \label{fig:gaussHe}
\end{figure*}

Alternatively, this redshifted component could correspond to emission from the receding part of an outflow, and a Gaussian or double-Gaussian profile may be an oversimplification. \citet{Roth2018} (hereafter, RK18) calculated line profiles including the effects of electron scattering above a hot photosphere in an outflowing gas. Qualitatively, their models show properties similar to AT2019qiz: a blueshifted peak (seen here only at early times) and a broad red shoulder. A decreasing optical depth or velocity in these models leads to a narrow core. We compare the observed H$\alpha$ profiles in Figure \ref{fig:gauss} to models from RK18, but we find that the implied velocity from the data is lower than any of the available models, apart from in the earliest epochs where a model with $v=5000$\,\kms\ gives an acceptable match. We will return to the early-time line profile in detail in section \ref{sec:outflow}.

We fit the H$\alpha$ profiles again, this time as a sum of a Gaussian centred at zero velocity (as before) and the $v=5000$\,\kms\ outflow model from RK18. The zero-velocity component could correspond to emission from pre-existing gas, since the RK18 profile should account for the TDE emission self-consistently. The only free parameter for the outflow component is the normalisation of the RK18 spectrum. This gives a good fit at early ($\lesssim0$ days) and late ($\gtrsim50$ days) times, but gives an inferior fit around 30 days compared to the double-Gaussian model. We note that the published model assumed a photospheric radius of $2.7\times10^{14}$\,cm, similar to the blackbody radius of AT2019qiz well before and after peak, whereas at peak the radius of AT2019qiz is a factor of $\gtrsim2$ larger, which may explain this discrepancy. As is shown in Figure \ref{fig:gauss}, the velocity of the Gaussian core, and the ratio of luminosity between the broad component and the zero-velocity component, are comparable between the double-Gaussian and Gaussian+RK18 fits.

In both models, when the narrow core of the line is revealed at late time we measure $v\approx 2000$\,\kms. Rather than line emission from the TDE itself, an alternative interpretation of the line profile is a pre-existing broad line region (BLR) illuminated by the TDE (recalling that this galaxy shows evidence for hosting an AGN; section \ref{sec:host}). Interpreting the width of the narrow component as a Keplerian velocity would yield an orbital radius $\approx 5\times 10^{15}$\,cm ($\approx 10^4$ Schwarzschild radii, $R_{\rm S}$). We note that an outflow from the TDE can reach this distance and interact with a BLR within 100 days if the expansion velocity is $\gtrsim 6000$\,\kms.

At least one previous TDE in a galaxy hosting an AGN has shown evidence of lighting up an existing BLR (PS16dtm; \citealt{Blanchard2017}), and further candidates have been discovered \citep{Kankare2017}.
{Although the narrow core of the Balmer lines is therefore quite plausibly associated with a BLR, it is difficult to interpret the entire line evolution in this way. The red wing does not decrease significantly in velocity over time, as it should if the emission is coming from material progressively further out (i.e.~with a lower orbital velocity). Moreover, AGN BLRs do not produce the very strong, broad, asymmetrical He~II and Bowen lines that co-exist with the broad early component of H~I, and which we now discuss.}

\subsection{The 4650\,\AA\ He-Bowen blend}
\label{sec:he}

Next we examine the He~II region of the spectrum. This is complicated by a blend of not only He~II~$\lambda4686$ and N~III~$\lambda4640$, but also H$\beta$ and H$\gamma$. In Figure \ref{fig:gaussHe}, we set zero velocity at the rest-frame wavelength of He~II. The earliest spectra before maximum light show a single broad bump with a peak that is bluewards of both He~II and N~III. After maximum, the H$\beta$ line becomes much more prominent, with a sharp profile similar to H$\alpha$, while the broad bump fades and by $\approx20$ days after maximum is centered at zero velocity. This indicates the early emission is dominated by He~II rather than N~III. However as the spectra evolve this line moves back to the blue, and by $\sim70$ days is centered at the rest wavelength of N~III. This is where it remains over the rest of our observations.

We quantify this by fitting this region with a sum of four Gaussians (He~II, N~III, H$\beta$ and H$\gamma$). All profiles are centred at zero velocity. The two Balmer lines are constrained to have the same width, and to make the problem tractable we also impose a further condition that the widths of He~II and N~III match each other. The fits are overlaid on Figure \ref{fig:gaussHe}, where we also show the evolution in luminosity of the He~II and N~III components. The luminosities of the two components are poorly constrained pre-peak, but by $\sim20$ days after peak the He~II component is clearly dominant, with almost no contribution from N~III. Soon afterwards, the N~III luminosity increases while that of He~II drops, and by 50 days N~III is the dominant component, with no significant He~II flux detectable after $\gtrsim 60$ days.

Other TDEs have shown separate resolved components of He~II and N~III \citep{Blagorodnova2018,Leloudas2019}, however this transition from almost fully He~II dominated to fully N~III dominated is remarkable, particularly given that the Bowen mechanism is triggered by the recombination of ionised He II. \citet{Leloudas2019} measured the He~II/N~III ratio for four TDEs with confirmed Bowen features, of which the two with events with $>2$ epochs of measurement showed an increasing He~II/N~III up to at least 50 days, opposite to what we observe in AT2019qiz. One of these events, AT2018dyb, may have shown a turnover after 60 days, but this is based on only one epoch, and ASASSN-14li continued to increase its He~II/N~III ratio for at least 80 days after peak. Based on two epochs measured for iPTF16axa, it may have shown a decrease in He~II/N~III as in AT2019qiz.

One other TDE to date has shown a clear increase in N~III while He~II fades, as we see in AT2019qiz. \citet{Onori2019} analysed a series of X-shooter spectra of iPTF16fnl, and found that the ratio of He~II/N~III decreased from $>5$ to $<1$ over a period of $\sim 50$ days, similar to the magnitude and timescale of the evolution in AT2019qiz. Interestingly, we note that the measured He~II/N~III ratios are in tension with the simplest theoretical predictions for \emph{all} TDEs in which Bowen lines have been detected. \citet{Netzer1985} calculated this ratio for a range of physical conditions appropriate for AGN, and found in all cases that $L_{\rm N~III}/L_{\rm He~II}\approx0.4-0.85$. The reversal in this ratio in AT2019qiz, as well as iPTF16fnl \citep{Onori2019}, AT2018dyb \citep{Leloudas2019}, and iPTF15af \citep{Blagorodnova2018}, is a puzzle that requires the application of detailed line transfer calculations.

\begin{figure}
  \centering
  \includegraphics[width=\columnwidth]{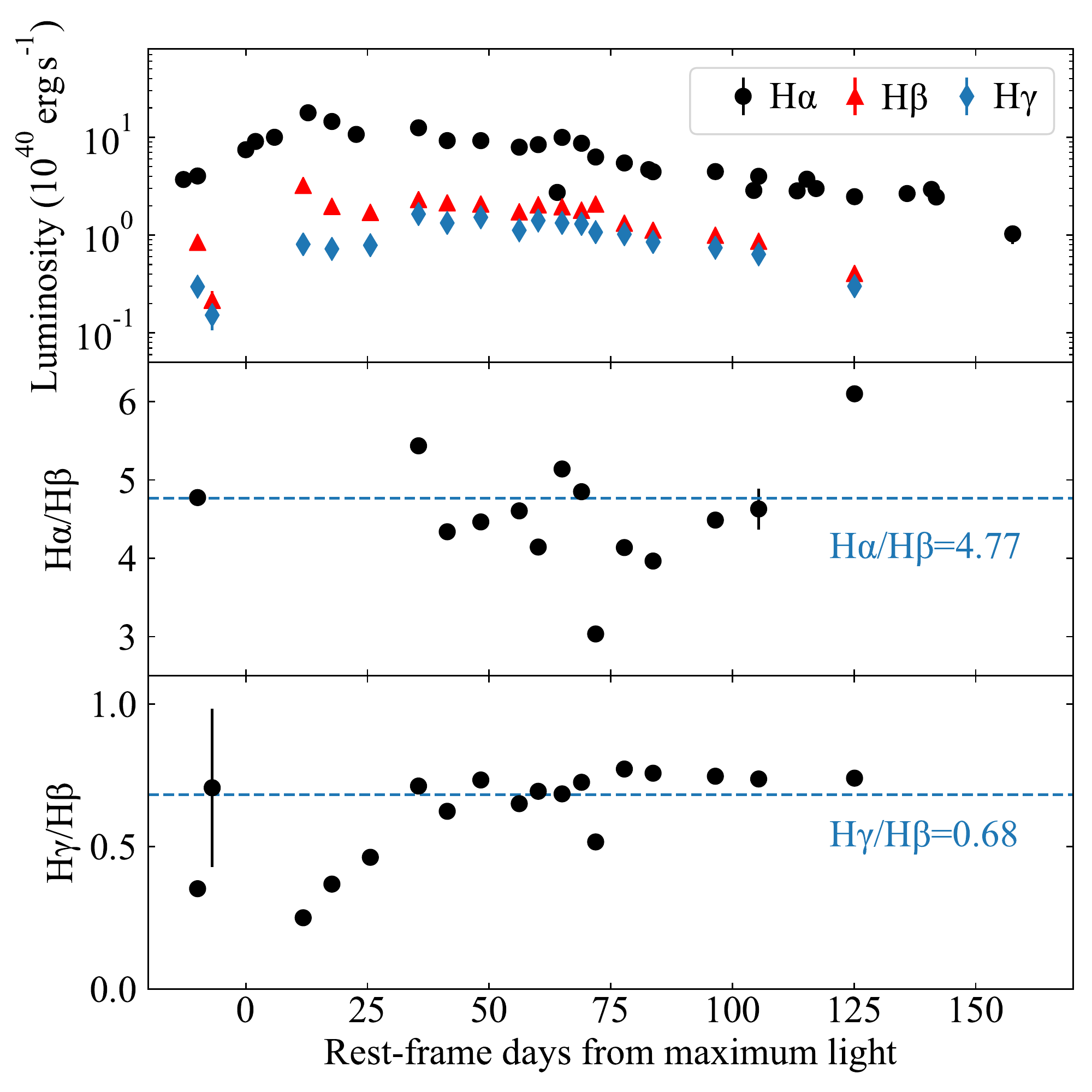}
  \caption{Luminosities and ratios of the Balmer emission lines as a function of time, measured using Gaussian fits. Dashed horizontal lines show the mean ratios. These have not been corrected for the (uncertain) extinction in the host galaxy.}
  \label{fig:ratio}
\end{figure}

\subsection{Balmer line ratios}

Using the fits to this part of the spectrum and to H$\alpha$, we calculate the ratios between the Balmer lines. For the H$\alpha$ luminosity we include both components in measuring the total flux (and we confirm that the luminosity derived from the fits matches that obtained by direct integration). For H$\beta$ and H$\gamma$ the fits include only one component, though at early times there may be a weak red shoulder visible in these lines too. The ratios are plotted in Figure \ref{fig:ratio}. There is no strong evidence for an evolution in the line ratio with time. We find H$\alpha$/H$\beta\approx4.7$ and H$\gamma/$H$\beta\approx0.7$. These ratios are marginally consistent with photoionisation and recombination \citep{Osterbrock2006}. This is what one would expect, for example, if the line emission is dominated by an AGN BLR.
A moderate internal dust extinction of $E(B-V)\approx 0.5$\,mag would bring the H$\alpha$/H$\beta$ ratio into agreement with the theoretical Case B value of 2.86, though we note that the light curve analysis supports a lower extinction. The host galaxy modelling also favours a low (spatially-averaged) extinction, as does the rather weak Na~I~D absorption. The equivalent widths of each line in the doublet (D1$\approx$D2$\approx0.1$) correspond to $E(B-V)=0.02$ in the calibration of \citet{Poznanski2012}. However, the extinction could be significantly higher in the nucleus.

It has recently been pointed out that a number of TDEs show much flatter Balmer decrements (H$\alpha$/H$\beta\approx1$; \citealt{Short2020,Leloudas2019}), which may indicate collisional excitation of these lines, e.g. in a disk chromosphere \citep{Short2020}. This ratio can be difficult to measure in TDEs due to line blending, but the excellent data quality available for AT2019qiz confirms a higher Balmer ratio in this event. If we assume a collisional origin for the Balmer lines in AT2019qiz, the implied temperature of the line-forming region is $\lesssim3000$\,K, following \citet{Short2020}. Alternatively, a large Balmer decrement may indicate shock powering, as is seen in many Type IIn supernovae \citep[e.g.][]{Smith2010}.

Finally, we note that the spectra in Figure \ref{fig:id} appear to show a positive Balmer jump (i.e.~enhanced flux below the Balmer break at 3646\,\AA), though we caution again that this may be due to an unreliable host subtraction at short wavelengths. If the positive jump is real, this could signify an extended atmosphere, where the additional bound-free opacity below the Balmer break means that we observe a $\tau\sim1$ surface at a larger radius, and hence see a larger effective emitting surface.

\subsection{Outflow signatures in the pre-maximum spectrum}\label{sec:outflow}

\begin{figure}
  \centering
  \includegraphics[width=\columnwidth]{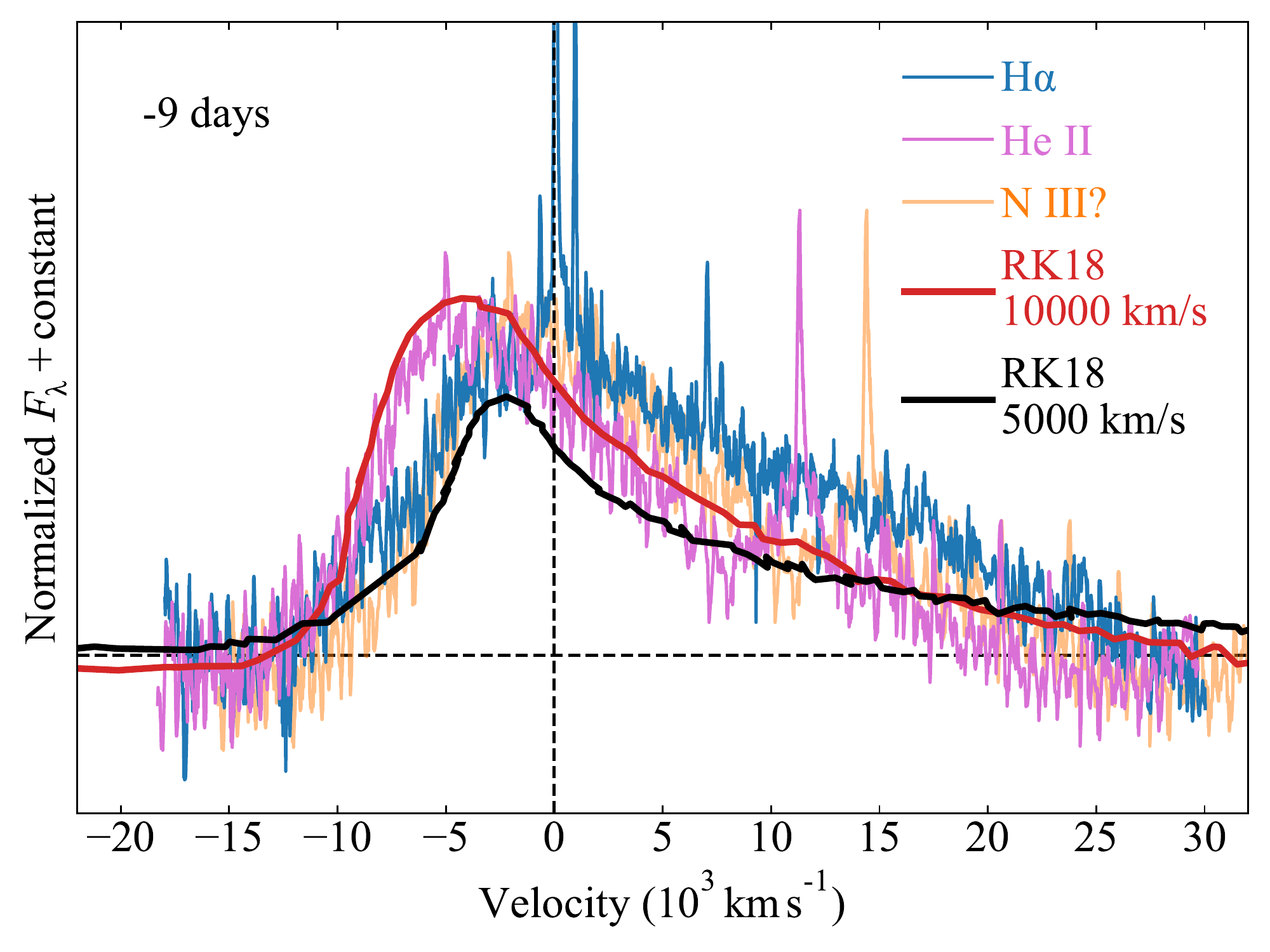}
  \caption{Velocity profiles of the strongest emission lines in the X-shooter spectrum at 9 days before peak, during the early rise/expansion phase. We show  H$\alpha$, and the broad He~II/Bowen bump for two different cases: (i) assuming that the line is dominated by He~II (pink) or (ii) that it is dominated by N~III (orange). Although the latter case gives a profile quite similar to H$\alpha$, the evolution over the next 20-30 days suggests that He~II is more likely the dominant component.
  Overplotted are model line profiles from \citet{Roth2018}. Assuming the blend is dominated by He~II, an excellent match is found for a 10,000\,\kms\ outflow. H$\alpha$ comes from slower material with velocity $\lesssim 5000$\,\kms.
}
  \label{fig:blueshift}
\end{figure}

The analysis in sections \ref{sec:ha} and \ref{sec:he} suggested the early blueshifted line profiles may be indicative of an outflow. We make this more explicit here by comparing the H$\alpha$ and He~II/Bowen line profiles in our earliest X-shooter spectrum, obtained 9 days before maximum light. Figure \ref{fig:blueshift} shows these lines in velocity coordinates. We reiterate that at early times, the 4650\,\AA\ blend seems to be dominated by He~II rather than N~III (Figure \ref{fig:gaussHe} and section \ref{sec:he}), but we plot the profile for both possible cases for completeness.

Both lines exhibit a peak blueshifted from their rest wavelengths (whether the 4650\,\AA\ feature is He~II or N~III). Assuming that the He~II line is strongest at early times, we see that He~II shows a larger blueshift than H$\alpha$. Both lines have a similar broad and smooth red shoulder. We quantify the velocity difference using the electron-scattering outflow models from RK18, previously applied only to H$\alpha$ in section \ref{sec:ha}. Line broadening in these models includes both the expansion and thermal broadening. We note the important caveat that these profiles were calculated explicitly only for H$\alpha$, but RK18 suggest that the qualitative results should apply to the other optical lines too. We proceed here under that assumption.

The fit to H$\alpha$ with a 5000\,\kms\ model (the lowest velocity published model) is indicative of an outflow but does not fully capture the shape of the red side of the line. Interpolating/extrapolating by eye between the parameters explored by RK18, we suggest that a larger optical depth and a slightly lower velocity would likely produce a closer match to the observed profile. Alternatively, a two-component model as explored in Figure \ref{fig:gauss} can produce a satisfactory fit.

The He~II line shows even stronger evidence for an outflow. This profile gives an excellent match to a 10,000\,\kms\ outflow model from RK18 (with other parameters the same as the 5000\,\kms\ model). If this line profile was instead N~III dominated, the inferred velocity would be closer to that of H$\alpha$.

In our photometric analysis, we found that the blackbody photosphere of AT2019qiz initially expanded at a velocity $\gtrsim 2000$\,\kms\ (Figure \ref{fig:pl}). This suggests a velocity gradient or homologously expanding outflow, where the line forming region above has a greater velocity than the continuum photosphere below. {This is consistent with the RK18 models, which assume a homologous velocity profile.
We will discuss such a profile in detail in the next section.}

In this context it is somewhat surprising that He~II exhibits a larger velocity than H$\alpha$, as \citet{Roth2016} found that H$\alpha$ should be emitted further out in radial coordinates than He~II due to its greater self-absorption optical depth. In this case He~II, emitted deeper in the debris, would experience a larger electron scattering optical depth. We speculate that degeneracies between electron scattering optical depth and velocity may be responsible for the apparent contradiction. Alternatively, this may indicate that N~III is a better identification for this feature at both early and late times (though not at the phases where we clearly do see He~II at zero velocity). This could be the case if the density in the line emitting region is high enough for efficient operation of the Bowen mechanism at early and at late times, but not at intermediate phases $\approx$20-40 days after peak.

\begin{figure}
  \centering
  \includegraphics[width=\columnwidth]{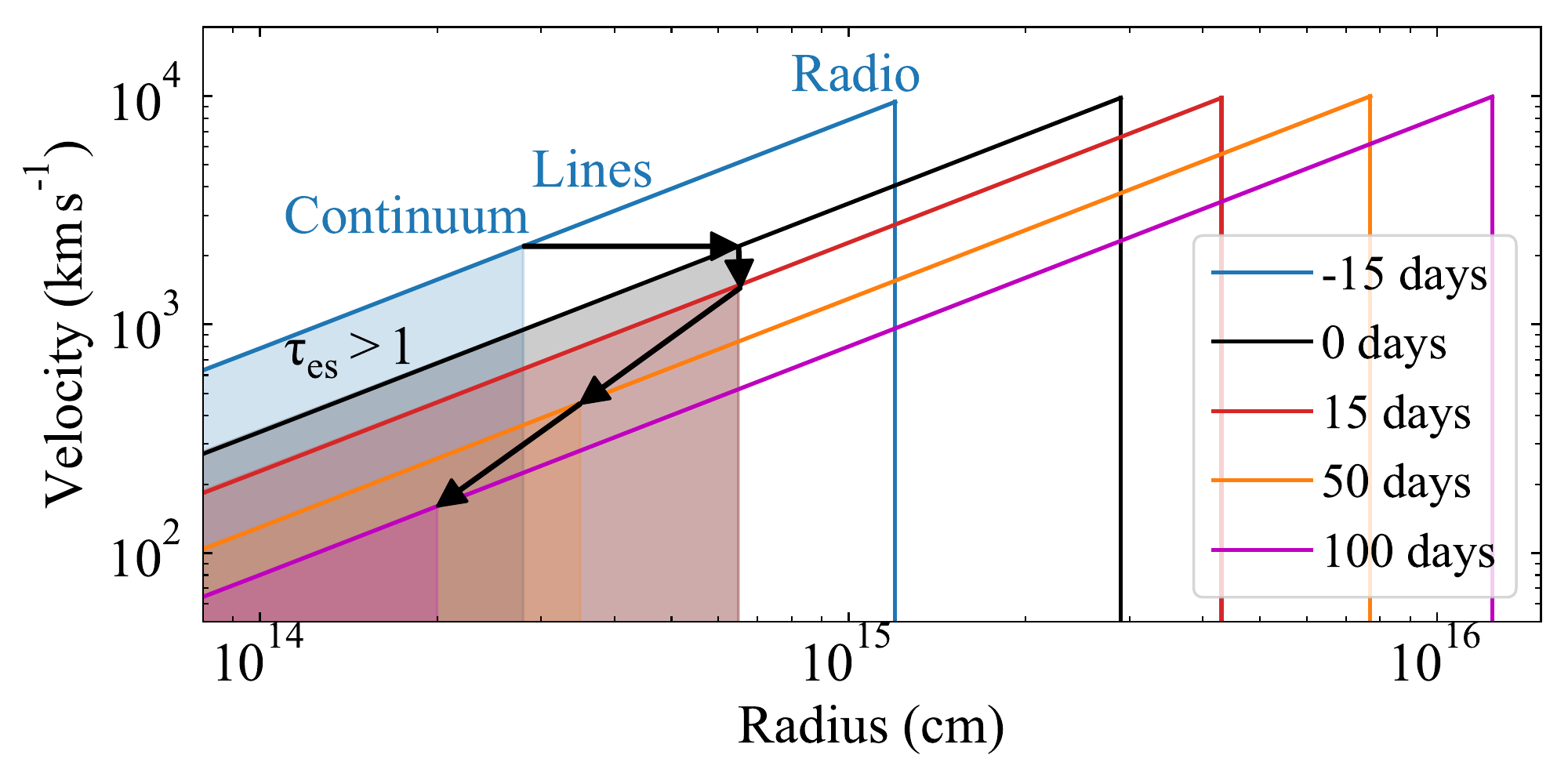}
  \caption{
  {Representative snapshots in time of a homologous velocity profile that simultaneously describes our photometry, spectroscopy, and the radio detection. The slope is tightly constrained at phases $t\leq0$ by the evolution of the photospheric radius (Figure \ref{fig:pl}), which has $v_{\rm ph}=2200$\,\kms\ and $R=6.5\times10^{14}$\,cm at maximum light. The scale (maximum) velocity of $v_{\rm sc}\approx 10,000$\,\kms\ is derived from fitting the spectral lines. After maximum light, we assume that the ejecta continue to expand with this profile. We mark the \emph{observed} radius of the photosphere (Figure \ref{fig:bol}) at each epoch; the region below the photosphere (the reprocessing layer) is shaded. Line emission comes from the unshaded region, which has a characteristic size of $\sim 10^{15}$\,cm, consistent with the parameters used by RK18 when modelling the line profiles. Radio emission comes from the leading edge of the ejecta shocking the ISM (K.~D.~Alexander in preparation). The arrows highlight the movement of the photosphere: as it recedes to a smaller radius after maximum light, the bulk of the line-emitting region is at lower velocity. The X-ray emitting region resides at a much smaller radius ($\sim R_t\lesssim10^{13}$\,cm) not visible on this plot; X-rays must escape either through gaps in the reprocessing layer or a polar funnel \citep{Dai2018}.}
}
  \label{fig:schem}
\end{figure}

\begin{figure*}
  \centering
  \includegraphics[width=10cm]{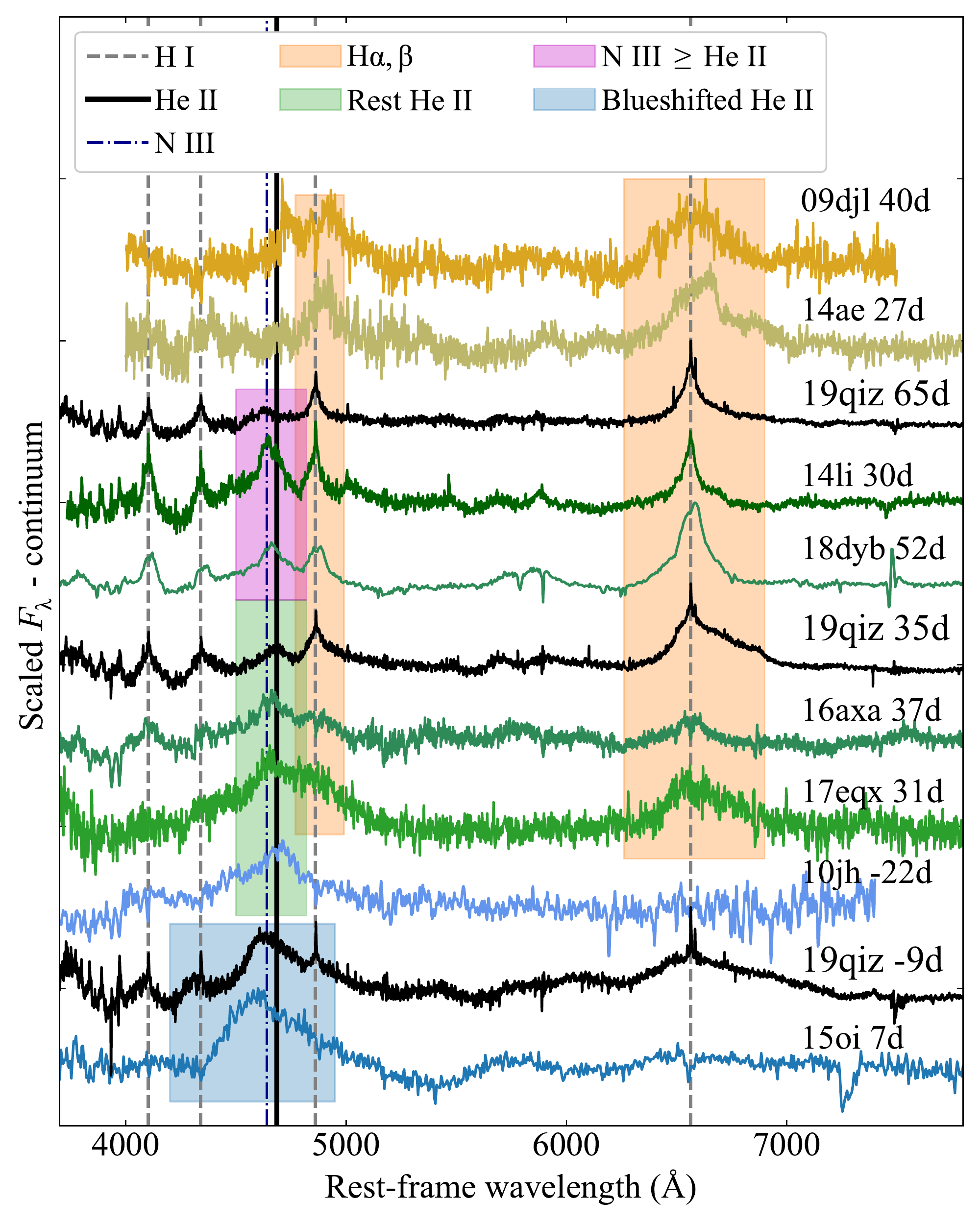}
  \caption{AT2019qiz in the context of the H-Bowen-He sequence of TDE spectra \citep{Arcavi2014,vanVelzen2020,Leloudas2019,Nicholl2019}.
  Although this TDE is best classified as Bowen type, at various phases the line profile in the Bowen/He region of the spectrum most closely resembles events with blueshifted He II (e.g.~ASASSN-15oi; \citealt{Holoien2016b}), rest-frame He II (e.g.~iPTF16axa; \citealt{Hung2017}), or strong N III (e.g.~AT2018dyb, ASASSN-14li; \citealt{Leloudas2019,Holoien2014}). The Balmer line profiles are most similar to these latter events. }
  \label{fig:speccomp}
\end{figure*}

\section{Discussion}\label{sec:dis}

\subsection{The outflow-dominated early evolution and the onset of accretion}

{As discussed in previous sections, the early optical (photospheric and emission line) evolution of AT2019qiz can be interpreted as a roughly homologous outflow. Figure \ref{fig:schem} illustrates such a profile, with the velocity gradient fixed at early times by the tight constraints on the expansion rate of the photosphere (section \ref{sec:phot}), and the maximum velocity set by modelling of the optical emission lines ($v_{\rm sc}\approx10,000$\,\kms). Although this profile is likely a simplification of the messy geometry in a real TDE, it will serve as a useful reference in the discussion to follow.}

The detection of blueshifted emission lines is not unique to AT2019qiz, with other TDEs exhibiting either blueshifted He~II or asymmetric H$\alpha$ profiles, as shown in a comparison with other TDE spectra in Figure \ref{fig:speccomp}. However, due to its proximity allowing detection in the radio, we can unambiguously say that AT2019qiz launched an outflow. Preliminary modelling of the radio light curve (K.~D.~Alexander et al., in preparation) indicates that the velocity of radio emitting material is $v_{\rm radio}\approx0.03c\approx10,000$\,\kms. 
This {consistency between optical- and radio-derived velocities strongly suggests that the optical line broadening is indeed a consequence of expansion, and therefore supports} the picture that the complex line profiles in TDEs do -- at least in some cases -- form in outflowing gas (RK18).

The dense time-series spectroscopy of AT2019qiz and very early photometric data allow us to study how the outflow influences the optical rise of the TDE and determine the dominant power sources over the course of the light curve evolution. We now combine our results from the optical, X-rays, and spectroscopy to examine each phase of the transient.

\subsubsection{The constant-velocity phase}

The earliest detections of AT2019qiz are characterised by a power-law like rise to maximum light over a period of $\approx 35$ days. During this time, the blackbody radius increases at a roughly constant velocity of $\gtrsim 2000$\,\kms, with no significant temperature evolution (Figure \ref{fig:pl}). Blueshifted emission lines, with $v\sim 5000-10,000$\,\kms\ (Figure \ref{fig:blueshift}), and a radio detection confirm the rapid expansion. 

Generally, such an outflow would cool, but in a TDE we may have a complex trade-off between cooling through adiabatic expansion and heating through continuous energy injection. A quasi-spherical expansion or wind at a perfectly constant velocity and temperature would lead to a luminosity evolution $L\propto t^2$, consistent with the data. We also note that the mass return rate at early times may influence the shape of the rising light curve, and recent simulations find $\dot{M}\propto t^2$ during the second periapsis passage (D.~Liptai et al., in preparation). A steeper power-law would indicate a non-spherical outflow: for example, the ballistic ejection of bipolar blobs which then undergo additional expansion due to their own internal pressure. This has been used to explain some observations of classical novae \citep{Shore2018}, but to our knowledge this geometry of mass ejection is not predicted by TDE simulations.

Several models predict a quasi-spherical outflow in TDEs, with $v\sim 10,000$\,\kms, consistent with the observed velocity. These fall into two broad classes (much like models for the TDE luminosity), which have been revealed through detailed TDE simulations. Collisions between debris streams can launch material on unbound trajectories \citep{Jiang2016}, or drive an outflow via shocks \citep{Lu2020}. Alternatively, radiatively-inefficient accretion can lead to most of the energy from the fallback going into mechanical outflows. \citet{Metzger2016} described this process analytically, and recent work by D.~Liptai (private communication) has demonstrated that this does occur in first-principles numerical simulations.

Our early observations of AT2019qiz at all wavelengths help to break the degeneracy in these models. The detection of X-rays during this early phase {could} indicate that accretion began promptly: stream collisions are not predicted to be X-ray sources \citep{Piran2015}, whereas X-ray observations of TDEs to date are generally consistent with disk models \citep{Jonker2020,Mummery2020,Wevers2020}. The X-rays are not likely to arise from the outflow, as the luminosity is orders of magnitude greater than that predicted by modelling the radio SED (K.~D.~Alexander et al., in preparation). Moreover, we see indications of Bowen fluorescence (N~III and likely O~III) in the early spectra, and these require a far-UV source to photoionise He~II (the Bowen lines are pumped by He~II Ly$\alpha$; \citealt{Bowen1935}). \citet{Leloudas2019} also interpreted Bowen lines in several TDEs as a signature of obscured accretion.

We also note the UV excess in the light curve at peak, relative to the TDE model that best fits the optical bands (Figure \ref{fig:mosfit}). The UV excess could be explained as a consequence of multiple energy sources contributing to the light curve, i.e. accretion and collisions. In other TDEs, a UV excess has been seen at late times, even in the form of a second maximum in the light curve \citep{Leloudas2016,Wevers2019b}, with the first peak interpreted as stream collisions and the second as the formation of a disk. If accretion begins sufficiently early, we may see both processes at once, giving a light curve at peak that cannot be easily fit with a one-component model.

\subsubsection{The constant-radius phase}

At the point when its bolometric light curve peaks, AT2019qiz transitions to a rather different behaviour: rapid cooling at constant radius (Figure \ref{fig:bol}). In the context of an outflow model, this may be a consequence of the optical depth in the ejecta. \citet{Metzger2016} argue that the wind is initially optically thick, and that photons are advected outwards to a characteristic trapping radius where they can escape. This advection is responsible for downgrading the X-rays released by accretion to UV/optical photons. 

Using our inferred SMBH mass and impact factor in equation 24 from \citet{Metzger2016}, we estimate a trapping radius for AT2019qiz of $\sim 2\times10^{14}$\,cm, within a factor of a few of our measured photospheric radius at maximum light. Therefore we suggest that the photosphere follows the debris outwards, unable to cool efficiently until the expansion reaches this trapping radius, at which point photons can free-stream from the outer ejecta and the effective photosphere stays frozen, despite continued expansion of the outflow.

The matter at the photosphere is now able to cool. From photon diffusion \citep{Metzger2016}, this occurs on a timescale
\begin{equation}
    t_{\rm cool} \sim \tau_{\rm es} R_{\rm ej}/c \approx 3 \kappa_{\rm es} M_{\rm ej} / 4 \pi R c,
\end{equation}
which is $\sim10$ days if we assume $R\sim R_{\rm ph}=6\times10^{14}$\,cm, with an ejected mass $M_{\rm ej}\sim 0.1$\,\M\ above the photosphere and an electron scattering opacity $\kappa_{\rm es}=0.34$\,cm$^2$\,g$^{-1}$ for hydrogen-rich matter. This is reasonably well matched to the duration of this phase of the evolution.

During this short-lived phase, the line profiles change dramatically: by around 20 days after maximum light, the He~II profile is roughly symmetrical and shows no evidence of a net velocity offset. We interpret this as supporting evidence for a model where the expansion and eventual cooling of the outflow lead to a reduction in the optical depth, and hence the disappearance of the broadened, blueshifted profiles visible during the rising phase.

\subsubsection{The constant-temperature phase}

At the point when the blueshifts vanish from the line profiles, the behaviour of the photosphere changes again: the cooling stops, the temperature settling at $\approx 15,000$\,K, and the radius now begins a smooth decrease. This is more akin to the typical behaviour seen in other optical TDEs, which generally show constant temperatures and decreasing blackbody radii.

At the same time, we observe a rise in the X-ray light curve (Figure \ref{fig:hr}), indicating that as the {luminosity emitted from the outflowing photosphere} fades, at this phase we are seeing more directly the contribution of accretion. The X-ray/optical ratio increases, but remains low ($\lesssim10^{-2}$; Figure \ref{fig:bol}), meaning that most of the accretion power is still reprocessed by the optically thick inner part of the outflow acting as the Eddington envelope \citep{Loeb1997}. The spectral evolution confirms this, as the presence of strong Bowen lines requires the absorption of significant far-UV/X-ray flux by the ejecta.

The X-ray and optical luminosity both decline from here, consistent with a decreasing accretion rate, as also inferred from the \textsc{mosfit} model fit. As the accretion power decreases, so too will the energy injected into the mechanical outflow, leading to a decreasing velocity and therefore possibly a higher density (for the same degree of mass-loading) in the inner regions. This may explain the increasing prominence of N~III at this phase, as the Bowen mechanism is more efficient at high density \citep{Bowen1935,Netzer1985}. Assuming $\sim 0.1$\,\M\ of hydrogen-dominated debris in the reprocessing region and taking the photospheric radius of $\approx3\times10^{14}$\,cm at this phase, we find an electron density $N_e \approx 10^{12} f_{\rm ion}$\,cm$^{-3}$, where $f_{\rm ion}$ is the ionisation fraction, which for even modest ionisation is consistent with the density regime, $N_e>10^{10}$\,cm$^{-3}$, where Bowen lines are expected to be strong \citep{Netzer1985}.

We note however that other TDEs have shown a decreasing N~III/He~II ratio at late times \citep{Leloudas2019}, which is harder to explain with an increasing density in the reprocessing layer. We posit that this could arise from a viewing angle effect in a geometry similar to that proposed by \citet{Nicholl2019} to explain the late onset of blueshifted He~II in AT2017eqx. If a TDE has a quasi-spherical photosphere but a faster outflow along the polar direction, and is viewed off-axis (see their Figure 13), more low-density material (emitting He~II but not N~III) will become visible along the dominant axis over time.

The corollary to this picture is that a TDE with an increasing N~III/He~II ratio (like AT2019qiz) must have been viewed more directly `face-on'. Such a scenario is supported by the detection of X-rays, which may be visible only for viewing angles close to the pole in the unified model of \citet{Dai2018}. In fact, consistency is also found with \citet{Nicholl2019}, who suggested that TDEs with outflow signatures in their early spectra should also exhibit X-ray emission, since both features are more likely for a polar viewing angle if the outflow is not spherical.

It is interesting to note that the X-rays in AT2019qiz are among the faintest observed in a TDE to date, and would not have been detectable at the greater distances where most TDEs have been found. Deeper X-ray observations of a larger TDE sample are still needed to test for any correlations between the X-ray properties and those of the optical spectral lines. An X-ray to optical ratio that increases with time has been seen in other TDEs \citep{Gezari2017,Wevers2019b,Jonker2020}, and may be a common feature even for TDEs with a low X-ray luminosity at peak.

\subsubsection{Overall energetics}

One persistent question pertaining to TDEs is the so-called `missing energy' problem \citep{Lu2018}. This states that the observed luminosity, totalling $\lesssim 10^{51}$\,erg for most TDEs, is orders of magnitude below the available energy from the accretion of any significant fraction of a solar mass onto the SMBH. Examining our \textsc{mosfit} parameters for AT2019qiz, we have a bound mass of $\approx 0.5$\,\M, and a radiative efficiency of $\approx 1\%$ \citep[typical of observed TDEs;][]{Mockler2019}. This gives a total available energy of $\approx 10^{52}$\,erg, whereas direct integration of the bolometric light curve yields only $10^{50}$\,erg.

In comparison, a homologously expanding outflow of similar mass with a scale velocity $\sim 10,000$\,\kms\ carries a total energy of $3\times10^{50}$\,erg. This indicates that the fraction of energy going into the mechanical outflow is greater than that released as radiation. However, this is not enough to account for the apparently missing energy.  Particularly at early times, accretion can be very inefficient (at producing radiation or driving outflows), and if the infall is spherically symmetric much of the energy can be simply advected across the event horizon.

\citet{Lu2018} suggest that most of the energy is radiated promptly in the far-UV. This seems unlikely for AT2019qiz, as the total energy in soft X-rays over the time of our observations is only $\sim 1\%$ of the near-UV/optical total, and neither the modest blackbody temperature ($\sim 15,000-20,000$\,K) nor the shallow power-law in the X-rays (Figure \ref{fig:xspec}) point towards a large far-UV excess. Still, this scenario can be tested by future searches for a mid-infrared echo on timescales of $\sim$ years \citep{Lu2018,JiangN2016}.

Alternatively, a large fraction of the energy can be released in the UV and X-rays not promptly, but rather over $\sim$ decade-long timescales via ongoing accretion in a stable viscous disk. Indications of such behaviour have been observed for a number of nearby optical TDEs via late-time flattening in the UV light curves \citep{vanVelzen2018} and X-ray observations \citep{Jonker2020}. At our estimated accretion efficiency, AT2019qiz could radiate the rest of its expected total energy ($\sim10^{52}$\,erg) by accreting a few percent of a solar mass per year. Recently, \citet{Wen2020} modeled the X-ray spectra of TDEs and found that a combination of a `slimming disk' with a very slowly declining accretion rate, and the energy directly lost into the black hole without radiating, appeared to solve the missing energy problem for the systems they studied. The proximity of AT2019qiz makes it an ideal source to test these scenarios with continued monitoring.

\subsection{The nature of faint and fast TDEs}
\label{sec:faint}

Based on the light curve comparisons in section \ref{sec:phot}, AT2019qiz appears to be the second faintest and fastest among known TDEs. To better quantify this statement, we examine the comprehensive TDE sample from \citet{vanVelzen2020}, and define a `faint and fast' TDE as one with a peak blackbody luminosity $\log (L/{\rm erg}\,{\rm s}^{-1})<43.5$ and exponential rise time $t_{\rm r}<15$\,d. Four events meet these criteria: AT2019qiz, iPTF16fnl, PTF09axc, and AT2019eve. However, \citet{vanVelzen2020} also find that there is no correlation between the rise and decline timescales of TDEs (unlike SNe; \citealt{Nicholl2015}). AT2019eve is one of the slowest-fading TDEs, and so appears to be qualitatively different from AT2019qiz and iPTF16fnl, which evolve rapidly in both their rise and decline phases. The remaining event, PTF09axc, has little data available after peak to measure the decline timescale, but appears to be more symmetrical than AT2019eve.

Of all optical TDEs with host galaxy mass measurements, iPTF16fnl and AT2019eve are among only four with host masses $\log(M_*/M_\odot)<9.5$ \citep{vanVelzen2020}, the others being AT2017eqx and AT2018lna, which have more typical light curves \citep{Nicholl2019,vanVelzen2020}. This suggests that the fastest events may be linked to low mass SMBHs, as proposed by \citet{Blagorodnova2017}. However, the case of AT2019qiz demonstrates that similar events can occur in more massive galaxies, and the SMBH mass we measure, $\sim 10^6$\,\M, is typical of known TDEs \citep{Wevers2019,Mockler2019}.

We also found that AT2019qiz is similar to iPTF16fnl in its spectroscopic properties at early times (though the strong narrow component in the Balmer lines at later times, possibly associated with an existing BLR, was not observed in iPTF16fnl; \citealt{Blagorodnova2017,Onori2019}). Both are classified as TDE-Bowen by \citet{vanVelzen2020}, and have characteristically small blackbody radii at maximum light. However, the rise and decay timescales for the TDE-Bowen class as a whole span a similar range to the TDE-H class, so a small radius alone is not sufficient to lead to a fast light curve. Moreover, AT2019eve and PTF09axc are both of type TDE-H.

As the early phase when AT2019qiz and iPTF16fnl are most similar is also the time when the dynamics of the outflow control the light curve evolution of AT2019qiz, it is possible that the properties of the outflow are most important for causing fast TDEs. In this case, such events may indeed be more common at lower SMBH mass, as the escape velocity from the tidal radius scales as $v_{\rm esc}\propto M_\bullet^{1/3}$, meaning an outflow can escape more easily at low $M_\bullet$, but are not limited to low mass black holes as other factors, such as the impact parameter, can also have an effect. Low SMBH mass also lends itself to faint TDEs, if the population have similar Eddington ratios, such that many of these fast TDEs should also be faint. AT2019qiz is only somewhat underluminous, consistent with its more typical SMBH mass.

While iPTF16fnl does show early-time line profiles consistent with an outflow (blueshifted H$\alpha$), it was not detected to deep limits in the radio, despite being at a comparable distance to AT2019qiz. While analysis of the radio data for AT2019qiz is deferred to a forthcoming work, we suggest that a pre-existing AGN in this galaxy may lead to a higher ambient density around the SMBH, producing a more luminous radio light curve as the TDE outflow expands into this medium. This is consistent with the overrepresentation of pre-existing AGNs within the sample of radio-detected TDEs \citep{Alexander2020}. Note that the outflows discussed here are not relativistic jets, which could be detectable at any plausible nuclear density \citep{Generozov2017}.

\section{Conclusions}\label{sec:conc}

We have presented extensive optical, UV and X-ray observations of AT2019qiz, which is the closest optical TDE to date at only 65.6\,Mpc, and a comprehensive analysis of its photometric and spectroscopic evolution. We summarise our main findings here.

\begin{itemize}
    \item AT2019qiz occurred in a galaxy likely hosting a weak AGN, as indicated by BPT line ratios and a nuclear point source. The galaxy has a high central concentration, as seen in other TDE hosts, and a strong quenching of the star formation within the last $\sim1$\,Gyr.
    \item The SMBH mass, measured using the $M_\bullet-\sigma$ relation, is $10^{5.75}-10^{6.52}$\,\M, depending on the calibration used. The mass inferred from light curve modelling, $10^{5.9}$\,\M, is within this interval.
    \item The bolometric light curve shows a rise in luminosity $L\propto t^{2}$, consistent with expansion at constant temperature and velocity $\approx 2200$\,\kms.
    \item The decay is steeper than the canonical $L\propto t^{-5/3}$ but close to the $t^{-8/3}$ predicted for a partial disruption \citep{Ryu2020}. The best fit light curve model with \textsc{mosfit} also suggests a partial disruption, of a $\approx 1$\,\M\ star, with about 75\% of the star stripped during the encounter.
    \item The peak luminosity, $L=3.6\times10^{43}$\,\ergs, and integrated emission, $E=1.0\times10^{50}$\,erg, are both among the lowest measured for a TDE to date.
    \item The early spectra show broad emission lines of H and most likely He~II, with blueshifted peaks and asymmetric red wings, consistent with electron scattering in an expanding medium with $v\approx 3000-10,000$\,\kms.
    \item Around maximum light, the temperature suddenly drops while the radius stays constant. This can be explained as trapped photons advected by the outflow suddenly escaping when it reaches the radius at which the optical depth is below unity.
    \item After peak, the lines become much more symmetrical, and He~II is gradually replaced by N~III, indicating efficient Bowen fluorescence and thus a source of far-UV photons. The late-time H lines show an unusually strong peak which may be from a pre-existing BLR.
    \item The time-varying X-ray emission (i.e.~from the TDE rather than the pre-existing AGN) and Bowen lines suggest that accretion commenced early in this event.
\end{itemize}

The detection of this event at radio wavelengths, likely enabled by its fortuitous proximity and possibly a high ambient density, confirms a (non-relativistic; K.~D.~Alexander et al., in preparation) expansion. This removes ambiguity in interpreting the line profiles of AT2019qiz as arising in an outflow, and confirms the velocities we have inferred from its optical properties. By extrapolation, outflows (even if undetected in the radio in the general case) are likely important in the many other TDEs with similar line profiles to AT2019qiz.

The properties of the outflow in this case (size, density, optical depth) are consistent with the reprocessing layer needed to explain the low X-ray luminosities of most optical TDEs, and in particular to provide the high-density conditions required for Bowen fluorescence in the inner regions. Thus, AT2019qiz {offers perhaps the strongest support to date for} the long-standing picture that outflows are responsible for the `Eddington envelope' hypothesised to do this reprocessing \citep{Loeb1997,Strubbe2009,Guillochon2014,Metzger2016}. The evidence for accretion early in this event, including X-ray detections before maximum light, suggest that the outflow was in this case powered by radiatively inefficient accretion (\citealt{Metzger2016}, D.~Liptai et al., in preparation), rather than stream collisions.

The exquisite data presented here will make AT2019qiz a Rosetta stone for interpreting future TDE observations in the era of large samples expected from ZTF \citep{vanVelzen2020}, the Rubin Observatory \citep{Lsst2009} and other new and ongoing time-domain surveys.


\section*{Acknowledgements}

We thank the anonymous referee for their many suggestions that improved this paper.
We thank Miguel P\'erez-Torres for helpful discussions.
MN is supported by a Royal Astronomical Society Research Fellowship.
TW is funded in part by European Research Council grant 320360 and by European Commission grant 730980. PGJ and GC acknowledge support from European Research Council Consolidator Grant 647208.
GL and PC are supported by a research grant (19054) from VILLUM FONDEN.
MG is supported by the Polish NCN MAESTRO grant 2014/14/A/ST9/00121.
NI is partially supported by Polish NCN DAINA grant No. 2017/27/L/ST9/03221.
IA is a CIFAR Azrieli Global Scholar in the Gravity and the Extreme Universe Program and acknowledges support from that program, from the Israel Science Foundation (grant numbers 2108/18 and 2752/19), from the United States - Israel Binational Science Foundation (BSF), and from the Israeli Council for Higher Education Alon Fellowship.
JB, DH, and CP were supported by NASA grant 80NSSC18K0577.
TWC acknowledges the EU Funding under Marie Sk\l{}odowska-Curie grant agreement No 842471.
LG was funded by the European Union's Horizon 2020 research and innovation programme under the Marie Sk\l{}odowska-Curie grant agreement No. 839090. This work has been partially supported by the Spanish grant PGC2018-095317-B-C21 within the European Funds for Regional Development (FEDER).
SGG acknowledges support by FCT under Project CRISP PTDC/FIS-AST-31546 and UIDB/00099/2020.
IM is a recipient of the Australian Research Council Future Fellowship FT190100574.
TMB was funded by the CONICYT PFCHA / DOCTORADOBECAS CHILE/2017-72180113.
KDA acknowledges support provided by NASA through the NASA Hubble Fellowship grant HST-HF2-51403.001-A awarded by the Space Telescope Science Institute, which is operated by the Association of Universities for Research in Astronomy, Inc., for NASA, under contract NAS5-26555.
This work is based on data collected at the European Organisation for Astronomical Research in the Southern Hemisphere, Chile, under ESO programmes 1103.D-0328 and 0104.B-0709 and as part of ePESSTO+ (the advanced Public ESO Spectroscopic Survey for Transient Objects Survey), observations from the Las Cumbres Observatory network, the 6.5 meter Magellan Telescopes located at Las Campanas Observatory, Chile, the MMT Observatory, a joint facility of the University of Arizona and the Smithsonian Institution and the William Herschel Telescope (programme W19B/P7). \textit{Swift} data were supplied by the UK Swift Science Data Centre at the University of Leicester. The Liverpool Telescope and William Herschel Telescope are operated on the island of La Palma by Liverpool John Moores University in the Spanish Observatorio del Roque de los Muchachos of the Instituto de Astrofisica de Canarias with financial support from the UK Science and Technology Facilities Council.
\\
\\
\textit{
$^{1}$\bham\\
$^{2}$\edinburgh\\
$^{3}$\cambridge\\
$^{4}$\northwestern\\
$^{5}$\dtu\\
$^{6}$\iaps\\
$^{7}$\mpi\\
$^{8}$\stockholm\\
$^{9}$\cfa\\
$^{10}$\brera\\
$^{11}$\telaviv\\
$^{12}$\cifar\\
$^{13}$\warsaw\\
$^{14}$\radboud\\
$^{15}$\sron\\
$^{16}$\monash\\
$^{17}$\ozgrav\\
$^{18}$\weizmann\\
$^{19}$\lco\\
$^{20}$\ucsb\\
$^{21}$\anu\\
$^{22}$\eso\\
$^{23}$\iap\\
$^{24}$\granada\\
$^{25}$\lisbon\\
$^{26}$\cardiff\\
$^{27}$\mssl\\
$^{28}$\soton\\
$^{29}$\bello\\
$^{30}$\iucaa\\
$^{31}$\qub
}
\\
\\
\textbf{Data Availability} All data in this paper will be made publicly available via WISeREP and the Open TDE Catalog.




\bibliographystyle{mnras}
\bibliography{refs} 




\appendix

\section{Host galaxy subtraction}

\subsection{UV photometry}

\begin{figure}
  \centering
  \includegraphics[width=8cm]{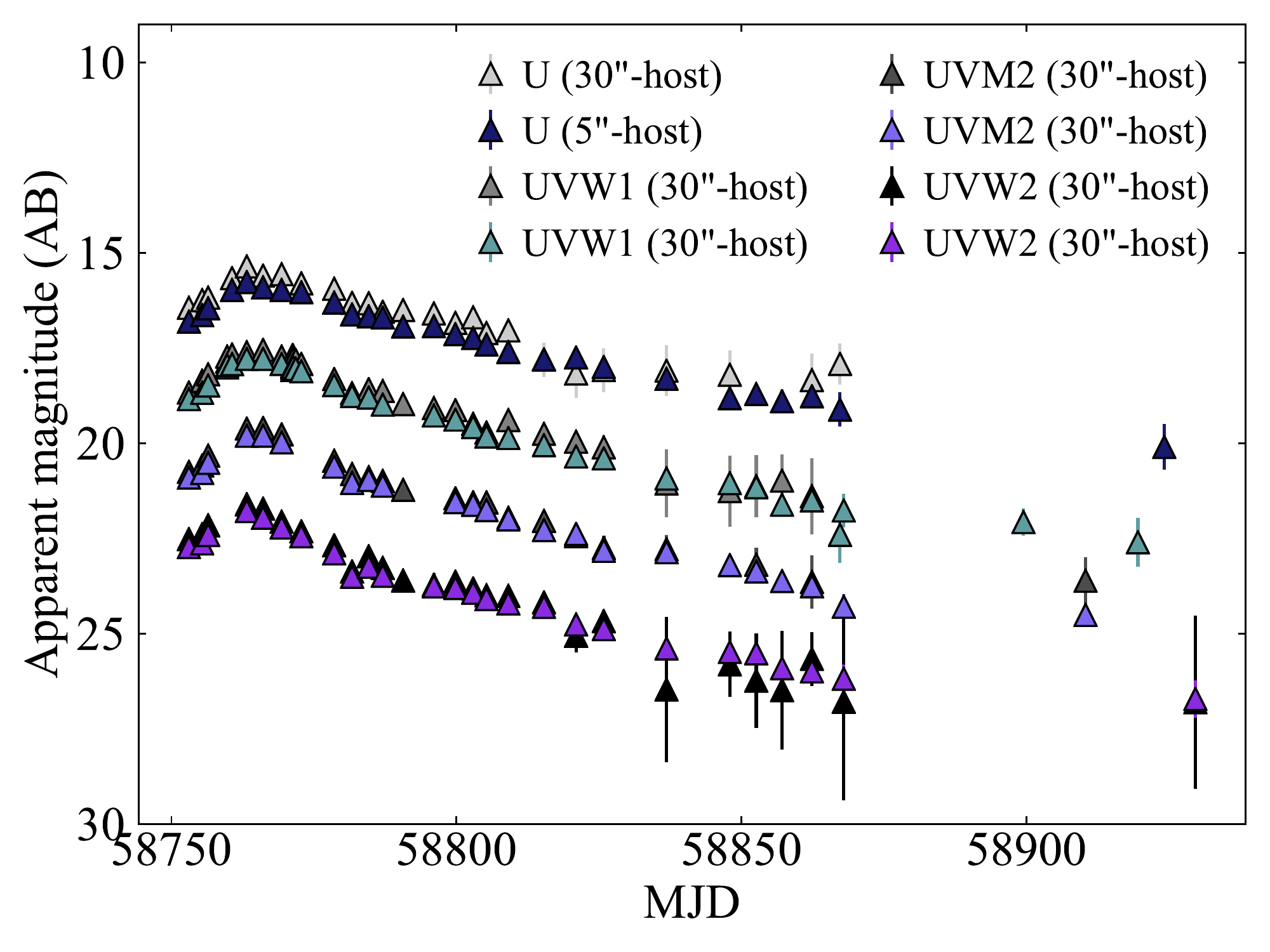}
  \caption{Host subtraction from UVOT magnitudes. The grey points show photometry measured in a large 30" aperture to fully capture the transient and host flux, with the entire host flux from the \textsc{prospector} model in each band subtracted. The coloured triangles show the UVOT photometry measure in a 5" aperture, where the calibration is more reliable and the host contamination lower, with 10-20\% of the total host flux subtracted.
  }
  \label{fig:uvhost}
\end{figure}

{Figure \ref{fig:uvhost} shows the difference between the host-subtracted UVOT light curves using a $5''$ aperture compared to a 30$''$ aperture, as described in section \ref{sec:uvot}.
}

\subsection{Spectroscopy}

\begin{figure*}
  \centering
  \includegraphics[width=5.8cm]{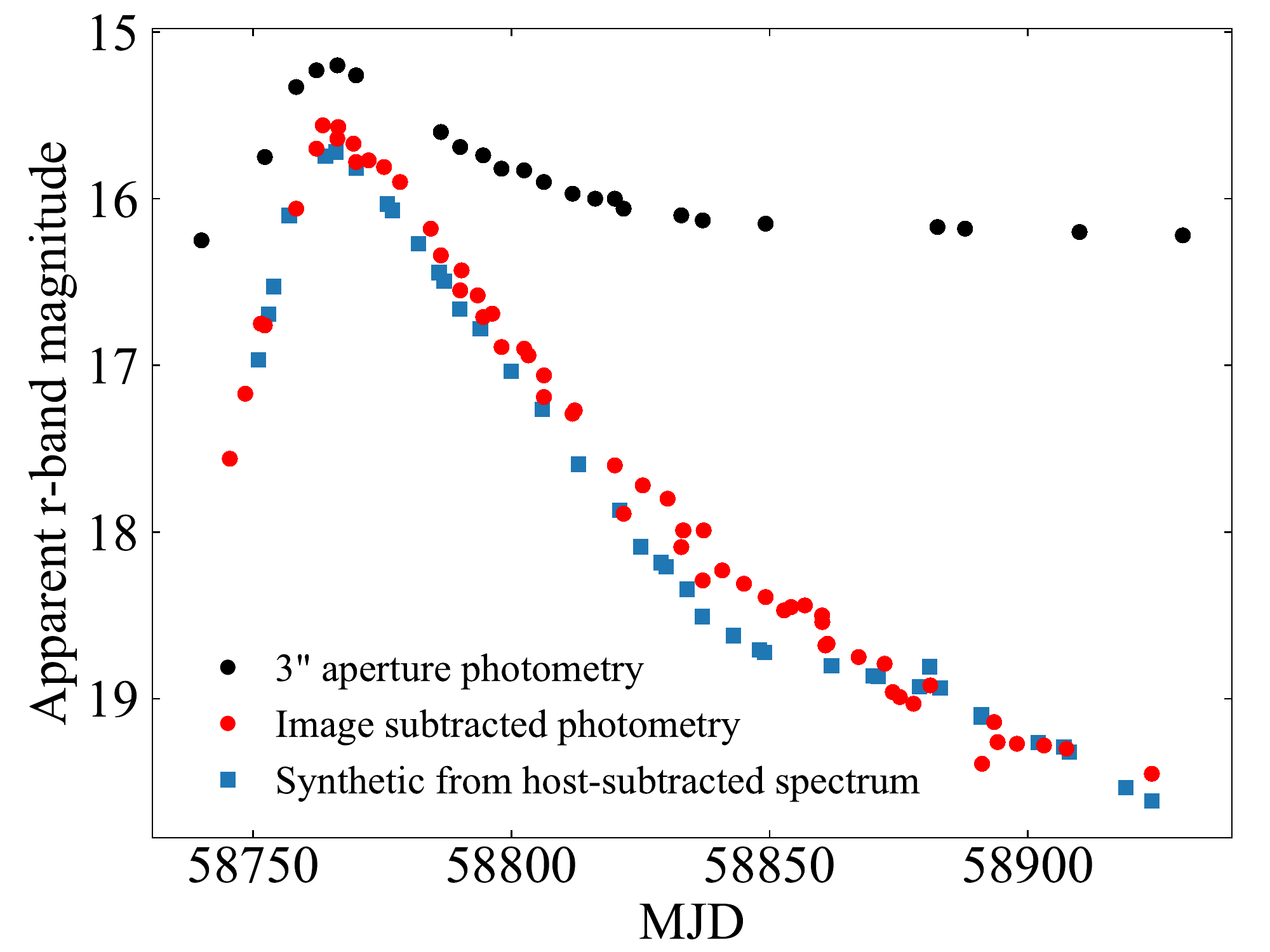}
  \includegraphics[width=5.8cm]{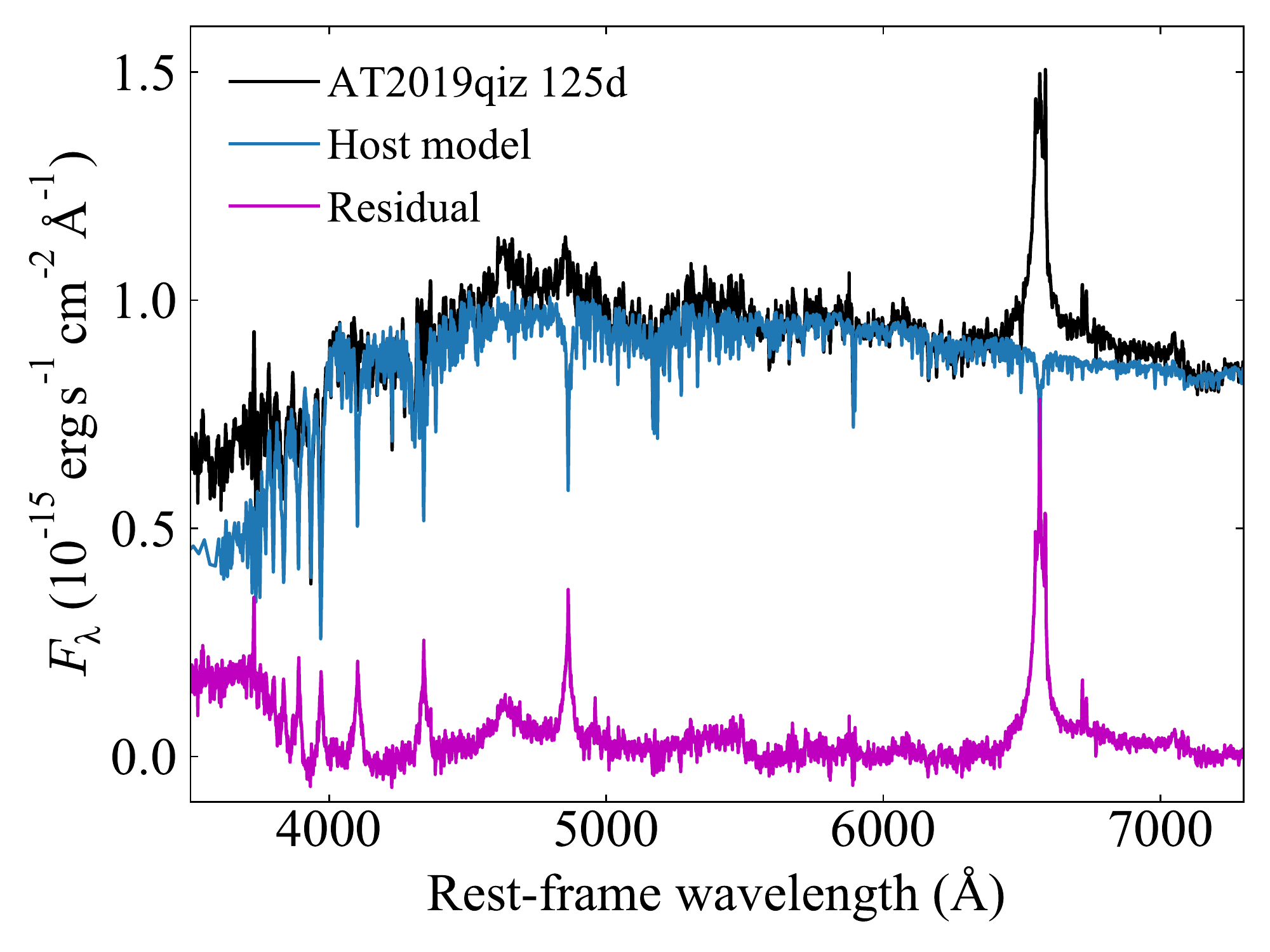}
  \includegraphics[width=5.8cm]{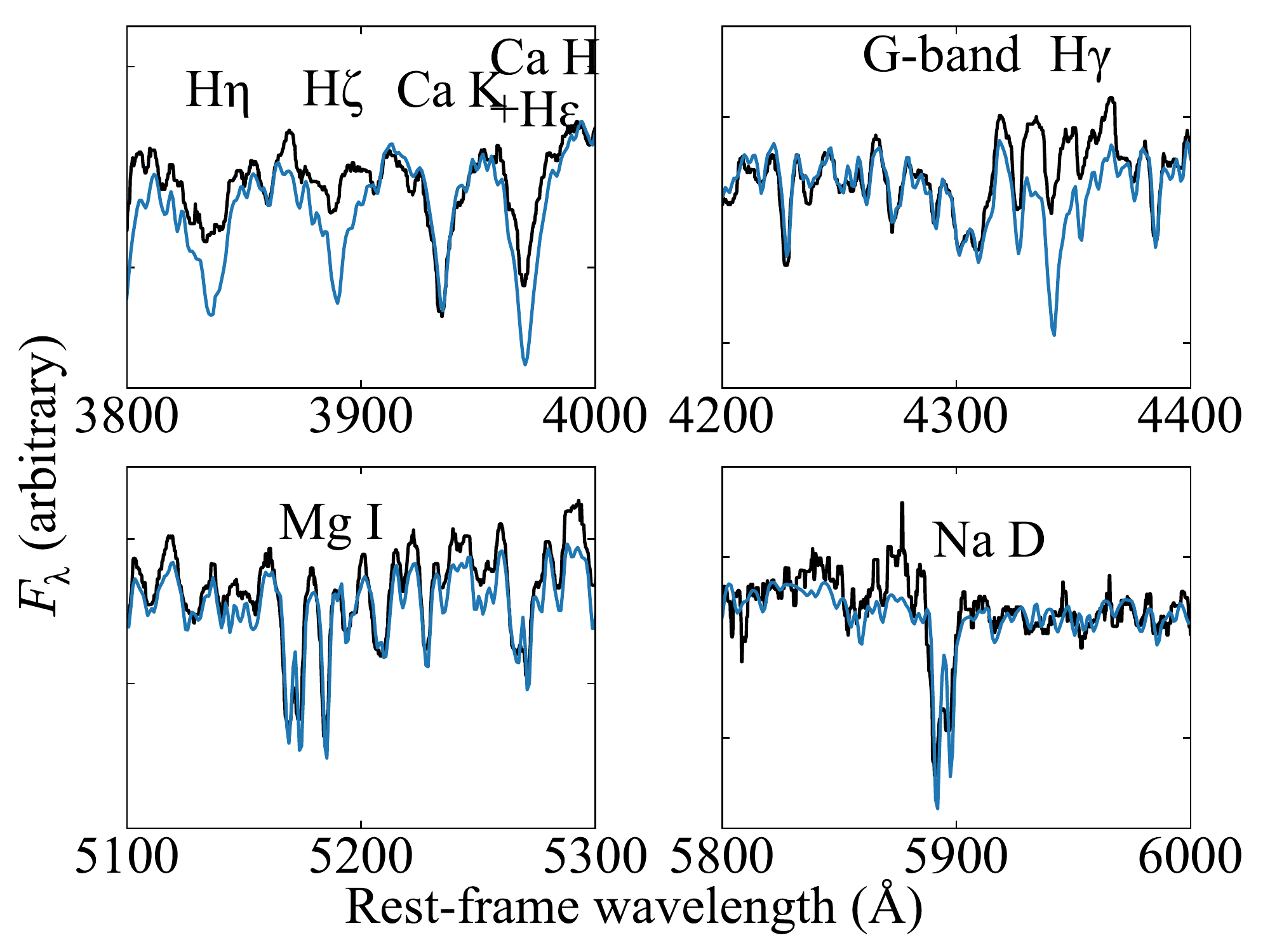}
  \caption{Removing host galaxy light from the TDE spectra. Left: Raw aperture photometry in the central $3''$ (includes host emission) vs image-subtracted PSF-fitting photometry (no host). Each TDE spectrum is scaled to the former (interpolated to epochs with spectra) before subtraction, and the synthetic magnitude after subtracting the host is verified to closely match the latter. Middle: A late-time spectrum of AT2019qiz compared to the host model, which has been scaled and convolved to match the data. Subtraction residuals (i.e.~the TDE-only spectrum) are also shown. Right: close-ups of strongest host galaxy absorption lines. We obtain an excellent match between the observed and \textsc{prospector}-predicted equivalent widths of the metal lines. The model predicts Balmer absorptions that are not observed in the data due to filling by the TDE emission lines.}
  \label{fig:hostprocess}
\end{figure*}

The \textsc{prospector} model spectrum allows us to remove stellar continuum from our spectroscopic data to better study emission from the TDE. Figure \ref{fig:hostprocess} illustrates the method. We first construct an $r$-band light curve measured from LCO data (without any image subtraction) in an aperture of radius $3''$, chosen to match the typical aperture size used for extracting the spectrum. This unsubtracted light curve plateaus when the TDE light falls below that of the host. We then scale each spectrum (remembering that it also contains both host and TDE light) to match this light curve; synthetic photometry is calculated on the spectrum using \textsc{pysynphot}. 

The model host spectrum from \textsc{prospector} is scaled to $m_r=16.47$\,mag, measured in a matching $3''$ aperture in the PanSTARRS $r$-band image. At each epoch, this model spectrum is interpolated to the same wavelength grid as the data, convolved with a Gaussian function to match the instrument-specific resolution, and subtracted. We verify that the fraction of flux removed in this way is reasonable by performing synthetic photometry on the subtracted spectrum, and find that it matches the \emph{host-subtracted} photometry to better than 0.5\,mag at all times. As a final step, we apply a small scaling to the subtracted spectrum to correct these $\lesssim {\rm few} \times 0.1$\,mag discrepancies with the subtracted light curve.

Figure \ref{fig:hostprocess} also shows the spectrum of AT2019qiz at 125 days after the light curve peak (i.e.~when host light is the dominant component) compared to the host model, after scaling to the observed magnitudes and convolving to match the resolution, as described above. The middle panel shows the full optical range, showing the good match to the continuum shape, with some TDE emission lines clearly visible above the host level. The host model shows deep Balmer absorption lines (common in TDE host galaxies) while the subtraction residuals (i.e.~the TDE-only spectrum, also plotted) exhibits prominent Balmer emission. To verify that these are real features rather than oversubtraction, we show close-ups around several of the main host absorption lines. The model gives an excellent match to the Ca II, Mg I, Na I and G-band absorptions present in the data, giving us confidence that the TDE H emission is real (and cancels out the host absorptions in the bluer Balmer lines). Moreover, the TDE shows a clear H$\alpha$ (and weak H$\beta$) emission line even before subtraction, with a profile similar to the emission lines in the subtracted spectrum. Finally, we find that the emission line fluxes decrease with time, which would not be true in the case of over-subtraction.

\begin{figure*}
  \centering
  \includegraphics[width=\textwidth]{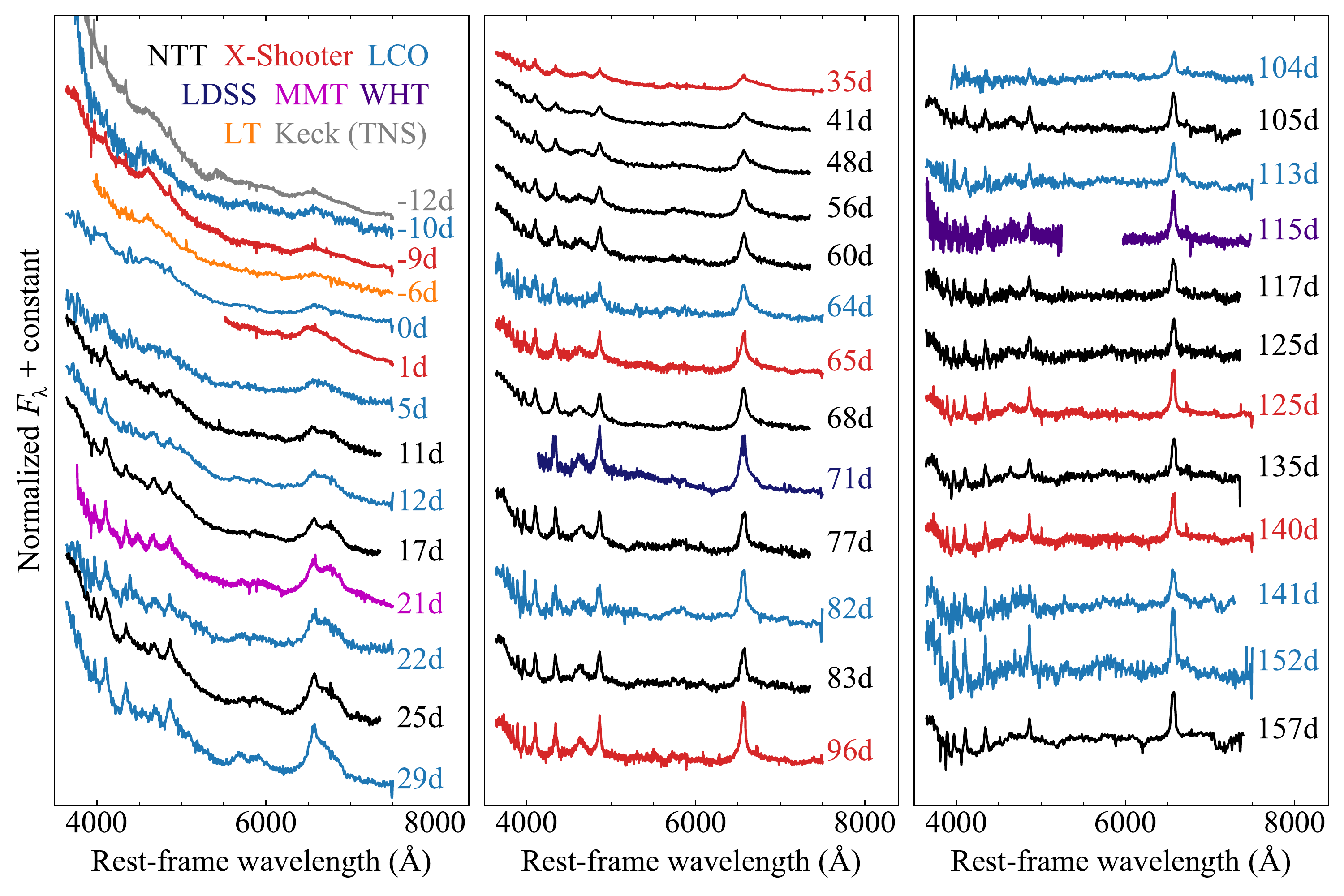}
  \caption{Same as Figure \ref{fig:spec} but with host contribution removed.}
  \label{fig:specsub}
\end{figure*}

Figure \ref{fig:specsub} shows all spectra of AT2019qiz as in Figure \ref{fig:spec}, but in this case after applying the host subtraction process.

\section{Mosfit Posteriors}

\begin{figure*}
  \centering
  \includegraphics[width=15cm]{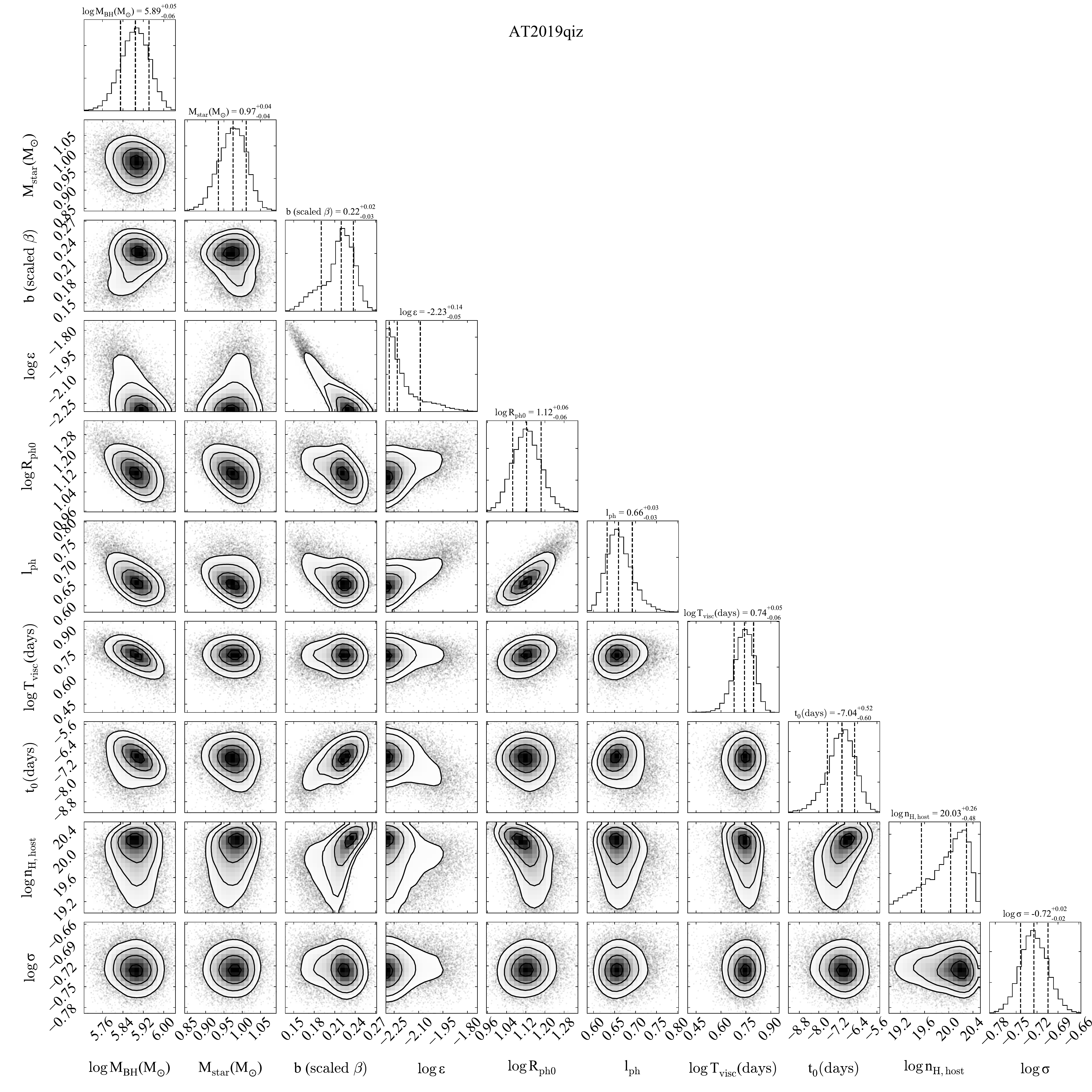}
  \caption{Posterior probability density functions for the free parameters of the model light curves in Figure \ref{fig:mosfit}.}
  \label{fig:post}
\end{figure*}

The priors and marginalised posteriors of our \textsc{mosfit} TDE model were listed in Table \ref{tab:mosfit}. In Figure \ref{fig:post}, we plot the full two-dimensional posteriors, which show some degeneracies between $\epsilon$ and $\beta$, and $l_{\rm ph}$ and $R_{\rm ph,0}$.

\bsp	
\label{lastpage}
\end{document}